\documentclass[epj]{svjour}
\usepackage{graphicx,psfrag}% Include figure files
\usepackage{dcolumn}% Align table columns on decimal point
\usepackage{bm}% bold math
\usepackage{amsmath}
\usepackage{amssymb}

\usepackage{ulem} %% for strike-through
\usepackage[usenames]{color}

\def\ba{\begin{eqnarray}}
\def\ea{\end{eqnarray}}
\def\be{\begin{equation}}
\def\ee{\end{equation}}
\def\bc{\begin{center}}
\def\ec{\end{center}}

\def\prt{\partial}

\newcommand{\Ima}{\text{Im}}

\newcommand{\mev}{\text{ MeV}}
\newcommand{\gev}{\text{ GeV}}
\newcommand{\fm}{\text{ fm}}

\begin{document}
\title{Theoretical study of incoherent $\phi$ photoproduction 
on a deuteron target}

%\subtitle{Do you have a subtitle?\\ If so, write it here}

\author{T. Sekihara\inst{1,2} 
\and A. Mart\'{i}nez Torres\inst{2}
\and D. Jido\inst{2}
\and E. Oset\inst{3}
% \thanks is optional - remove next line if not needed
%\thanks{\emph{Present address:} Insert the address here if needed}%
}                     % Do not remove
%
%\offprints{}          % Insert a name or remove this line
%
\institute{Department of Physics, Graduate School of Science, Kyoto
  University, Kyoto, 606-8502, Japan %
  \and %
  Yukawa Institute for Theoretical Physics, Kyoto University, Kyoto
  606-8502, Japan %
  \and %
  Departamento de F\'{\i}sica Te\'orica and IFIC, Centro Mixto
  Universidad de Valencia-CSIC, Institutos de Investigaci\'on de
  Paterna, Aptdo. 22085, 46071 Valencia, Spain}
\date{Received: date / Revised version: date}
% The correct dates will be entered by Springer
%
\abstract{ 
  We study the photoproduction of $\phi$ mesons in deuteron, paying
  attention to the modification of the cross section from bound
  protons to the free ones.  
  For this purpose we take into account Fermi motion in single
  scattering and rescattering of $\phi$ to account for $\phi$
  absorption on a second nucleon as well as the rescattering of the
  proton on the neutron.  
  We find that the contribution of the double scattering for $\phi$ is
  much smaller than the typical cross section of $\gamma p \to \phi p$
  in free space, which implies a very small screening of the $\phi$
  production in deuteron.
  The contribution from the proton rescattering, on the other hand, is
  found to be not negligible compared to the cross section of $\gamma
  p \to \phi p$ in free space, and leads to a moderate reduction of
  the $\phi$ photoproduction cross section on a deuteron at forward
  angles if the LEPS set up is taken into account.
  The Fermi motion allows contribution of the single scattering in
  regions forbidden by phase space in the free case.  In particular,
  we find that for momentum transferred squared close to the maximum
  value, the Fermi motion changes drastically the shape of $d \sigma
  /dt$, to the point that the ratio of this cross section to the free
  one becomes very sensitive to the precise value of $t$ chosen, or
  the size of the bin used in an experimental analysis.
  Hence, this particular region of $t$ does not seem the most
  indicated to find effects of a possible $\phi$ absorption in the
  deuteron.  This reaction is studied theoretically as a function of
  $t$ and the results are contrasted with recent experiments at LEPS
  and Jefferson Lab.  The effect of the experimental angular cuts at
  LEPS is also discussed, providing guidelines for future experimental
  analyses of the reaction.
   % end of PACS codes
} %end of abstract
\maketitle
\section{Introduction}
\label{intro}

The photoproduction of $\phi$ mesons on nucleons has attracted much
attention, both
experimentally~\cite{Ballam:1972eq,Besch:1974rp,Behrend:1978ik,Barber:1981fj,Anciant:2000az,Barth:2003bq}
and
theoretically~\cite{Bauer:1977iq,Titov:1996bg,Titov:1997qz,Titov:1998bw,Titov:1998tx,Titov:1999eu,Kisslinger:1999jk,felipe}.
Renewed recent efforts at LEPS~\cite{Mibe:2005er} have stimulated also
theoretical work~\cite{Titov:2008zz}. The determination of the
strangeness content of the nucleon has been one of the motivations for
these studies.  Tests on Pomeron exchange in this reaction have also
been another one of the motivations. Experimental work on the deuteron
has been done at LEPS~\cite{Chang:2009yq} looking for $d\sigma
  /dt$ close to $t_{\text{max}}$ and related theoretical work on near
threshold $\phi$ photoproduction on the deuteron has been done
in~\cite{Titov:2007fc}.  Photoproduction of $\phi$ mesons in nuclei
has also been addressed in~\cite{Ishikawa:2004id}, looking at the
transparency ratio~\cite{Hernandez:1992rv}, deducing from there an
enhanced $\phi N$ cross section in nuclei with respect to the one on a
free proton. This issue is of relevance to theories on vector
modification in a nuclear medium~\cite{Rapp:1999ej,Hayano:2008vn}. A
theoretical calculation of $\phi$ photoproduction in nuclei has been
performed in~\cite{Magas:2004eb}, and compared with the experimental
results of~\cite{Ishikawa:2004id}.  Very recently there has been
  further experimental research concerning $\phi$ production in
  deuterium.  In Ref.~\cite{haiyan} $\phi$ photoproduction near
  threshold from a deuterium target is studied, concluding that the
  extracted $d\sigma /dt$ is consistent with predictions based on a
  quasifree mechanism, in contradiction with the claims done at LEPS
  in a different momentum transfer region. Our theoretical results
  support the findings of Ref.~\cite{haiyan} and shed light on the
  different results claimed for the LEPS
  experiment~\cite{Chang:2009yq}.  

In the present work we address the problem of $\phi$ photoproduction
in the smallest nucleus, the deuteron, contrast our finding with
  those of Ref.~\cite{haiyan}, and point out missing experimental
information for a proper comparison of our results with the
recent measurements of Ref.~\cite{Chang:2009yq}. By analogy to the
photoproduction in nuclei we also have here effects of Fermi motion
and of $\phi$ and proton rescatterings, which are studied here in
detail and compared to the data. We find small effects of double
scattering for $\phi$ compared to single scattering in consonance
  with the findings of Ref.~\cite{haiyan} except in regions of phase
space forbidden to the scattering on free nucleons.  The proton
rescattering effect, on the other hand, is found to be not negligible
at the forward angles compared to the single scattering contribution.
The Fermi motion effects are also moderate, but of course they are
extremely important in the regions forbidden to the scattering on free
nucleons. This is the case particularly around the $t_{\text{max}}$
region of the free proton, where Fermi motion distorts drastically the
shape of the distribution, making this region not well suited to
investigate other possible two body mechanisms. The findings of the
paper for different values of $t$ and the effect of the angular cuts
of the LEPS set up are shown, opening a window for further reanalysis
of the reaction of~\cite{Chang:2009yq} in regions better suited to
extract relevant information.

\section{Formulation}
\label{sec:form}

In this section, we explain our approach to calculate the cross section 
of the $\gamma p \to \phi p$ and $\gamma d \to \phi p n$ reactions.

\subsection{$\bm{\gamma p \to \phi p}$ reaction}
\label{subsec:gamma-p}

\subsubsection{Kinematics}
\label{subsubsec:kin1}

\begin{figure}[t]
  \centering
  \includegraphics[scale=0.18]{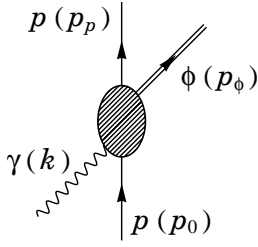}
  \caption{Kinematics for the $\gamma p \to \phi p$ reaction. }
  \label{fig:gamma-p} 
\end{figure}

Let us first provide the formulation for the $\gamma p \to \phi p$ reaction. 
The momenta of the particles in the initial and final states are shown in 
Fig.~\ref{fig:gamma-p}. In terms of these variables the cross section is 
defined as, 
\begin{align}
& \sigma _{\gamma p \to \phi p} (E_{\gamma}^{\text{lab}} ) 
\nonumber \\
& = \frac{2 M_{p}}{4 \sqrt{(k \cdot p_{0})^{2} - M_{\gamma}^{2} M_{p}^{2}}} 
\int \frac{d^{3} p_{\phi}}{(2 \pi )^{3}} \frac{1}{2 \omega _{\phi}} 
\int \frac{d^{3} p_{p}}{(2 \pi )^{3}} \frac{2 M_{p}}{2 E_{p}} 
\nonumber \\
& \phantom{=} \times \overline{\sum _{\lambda}} \sum _{\lambda}
| T_{\gamma p \to \phi p} |^{2} 
(2 \pi )^{4} \delta ^{4} (k + p_{0} - p_{\phi} - p_{p}) , 
\end{align}
where $k$, $p_{0}$, $p_{\phi}$, and $p_{p}$ are momenta of initial photon and 
proton and final $\phi$ and proton, respectively, and $T_{\gamma p \to \phi p}$ is the
scattering amplitude for the $\gamma p \to \phi p$ reaction. 
By means of the two summation symbols, the sum and average of $|T|^{2}$ for the 
polarizations of $\gamma$, $\phi$, and initial and final protons are done. 
The cross section is a function of the initial photon energy 
$E_{\gamma}^{\text{lab}}$ at the laboratory frame 
where the initial proton is at rest. Using the relation 
$\sqrt{(k \cdot p_{0})^{2} - M_{\gamma}^{2} M_{p}^{2}} = M_{p} E_{\gamma}^{\text{lab}}$ 
and performing the phase-space integration in the center-of-mass frame, one 
can obtain, 
\be
\sigma _{\gamma p \to \phi p} 
= \frac{p_{\text{cm}}^{\prime} M_{p}}{16 \pi ^{2} E_{\gamma}^{\text{lab}} \sqrt{s}} 
\int d \Omega _{p} 
\overline{\sum _{\lambda}} \sum _{\lambda} | T_{\gamma p \to \phi p} |^{2} , 
\label{eq:2}
\ee
with, 
\be
p_{\text{cm}} = \frac{\lambda ^{1/2} (s, \, M_{\gamma}^{2}, \, M_{p}^{2} )}
{2 \sqrt{s}} , 
\quad 
p_{\text{cm}}^{\prime} = \frac{\lambda ^{1/2} (s, \, M_{\phi}^{2}, \, M_{p}^{2} )}
{2 \sqrt{s}} , 
\ee
where $p_{\text{cm}}$ ($p_{\text{cm}}^{\prime}$) is the initial (final) state momenta in 
the center-of-mass 
frame and $s$ the Mandelstam variable $(k + p_{0})^{2}$. 
In Eq.~(\ref{eq:2}), $\Omega _{p}$ is the solid angle for the final 
proton in the center-of-mass frame. 

For the $\phi$ photoproduction, the differential cross section 
$d \sigma /dt$, with Mandelstam variable $t = (p_{\phi} - k)^{2}$, is an important 
observable in the experiments~\cite{Mibe:2005er,Chang:2009yq}. In the 
center-of-mass frame of the $\gamma p \to \phi p$ reaction, $t$ is written as 
\be
t = M_{\phi}^{2} 
- 2 p_{\text{cm}} (\omega _{\phi}^{\text{cm}} - p_{\text{cm}}^{\prime} \cos \theta _{p}) , 
\ee
with $\theta _{p}$ being the angle between the incident photon and the $\phi$ 
meson momenta, and $\omega _{\phi}^{\text{cm}} = 
\sqrt{M_{\phi}^{2} + p_{\text{cm}}^{\prime}}$ the $\phi$ energy. 
The maximum and minimum values of $t$, $t_{\text{max}}$ and $t_{\text{min}}$, 
are, 
\be
t_{\text{max}} (s) = 
M_{\phi}^{2} - 2 p_{\text{cm}} (\omega _{\phi}^{\text{cm}} - p_{\text{cm}}^{\prime}) , 
\label{eq:tmax}
\ee
\be
t_{\text{min}} (s) = 
M_{\phi}^{2} - 2 p_{\text{cm}} (\omega _{\phi}^{\text{cm}} + p_{\text{cm}}^{\prime}) , 
\label{eq:tmin}
\ee
respectively. Now using the relation, 
\be
d t = 2 p_{\text{cm}} p_{\text{cm}}^{\prime} d \cos \theta _{p} , 
\ee
Eq.~(\ref{eq:2}) can be written as follows: 
\begin{align}
\frac{d \sigma _{\gamma p \to \phi p}}{d t} 
& = \frac{M_{p}}{16 \pi p_{\text{cm}} E_{\gamma}^{\text{lab}} \sqrt{s}} 
\overline{\sum _{\lambda}} \sum _{\lambda} | T_{\gamma p \to \phi p} |^{2} , 
\end{align}
where we have performed the azimuthal angle integration. 

\subsubsection{Scattering amplitude}
\label{subsubsec:amp1}

In this section we give details on the scattering amplitude for the $\gamma p \to \phi p$ 
reaction. 
In Ref.~\cite{Mibe:2005er} it was shown that $d\sigma _{\gamma p \to \phi p}/dt$ 
has an exponential dependence as a function of $t$. To take this into account 
we use a phenomenological amplitude given by, 
\be
T_{\gamma p \to \phi p} 
= a_{p} (s) \exp (b \tilde{t} / 2) 
\times \epsilon _{\mu} (\gamma ) \epsilon ^{\mu} (\phi ) , 
\label{eq:Amp-phenom}
\ee
with $\tilde{t} = t - t_{\text{max}}$. 
Here $a_{p}$ is a factor which determines the strength of the total cross 
section. 
Based on Ref.~\cite{Rogers:2005bt}, we take the following $s$ dependence 
for the factor $a_{p}$: 
\be
a_{p} (s) = \alpha \left ( \frac{s}{\text{GeV}^{2}} \right )^{\beta} 
\left [ 
1 + R_{a} e ^{- R_{b} (E_{\gamma}^{\text{lab}} (s) - R_{c})^{2}} 
\right ]^{1/2} , 
\ee
with parameters $\alpha = 0.0167 \gev ^{-1}$, $\beta = 2.29$, 
$R_{a}=0.71$, $R_{b}=16.5 \gev ^{-2}$, and $R_{c}=2 \gev$, so as 
to reproduce the experimental data~\cite{Mibe:2005er}, and the photon 
energy $E_{\gamma}^{\text{lab}}$ in $a_{p}$ is evaluated as 
$E_{\gamma}^{\text{lab}}=(s-M_{p}^{2})/2M_{p}$, as a function of $s$. 
We note that this form is not same as that in Ref.~\cite{Rogers:2005bt}, 
where the authors fit the differential cross section $d\sigma /dt$ rather 
than the scattering amplitude. 
The parameter $b$ in Eq.~(\ref{eq:Amp-phenom}) is taken from~\cite{Mibe:2005er} 
as $b=3.38 \gev ^{-2}$. On the other hand, 
$\epsilon _{\mu} (\gamma )$ and $\epsilon _{\mu} (\phi )$ are the photon and 
$\phi$ polarization vectors, respectively. In this study we 
take the Coulomb gauge for the electromagnetic interaction, hence, 
Eq~(\ref{eq:Amp-phenom}) is rewritten as, 
\be
T_{\gamma p \to \phi p} 
= - a_{p} (s) 
\exp (b \tilde{t} / 2) 
\times \vec{\epsilon} (\gamma ) \cdot \vec{\epsilon} (\phi ) . 
\label{eq:ampli}
\ee

For the sum over the polarizations, we have the following relations, 
\begin{align}
& \sum _{\lambda _{\gamma}} \epsilon ^{\ast i} (\gamma ) \epsilon ^{j} (\gamma ) 
= \delta ^{ij} 
- \frac{k^{i} k^{j}}{|\vec{k}|^{2}} ,
\label{eq:polarization-gamma}
\\
& \sum _{\lambda _{\phi}} \epsilon ^{\ast \mu} (\phi ) \epsilon ^{\nu} (\phi ) 
= - g ^{\mu \nu} + \frac{p_{\phi}^{\mu} p_{\phi}^{\nu}}{M_{\phi}^{2}} . 
\label{eq:polarization-phi}
\end{align}
By summing and averaging over the polarizations, we obtain, 
\begin{align}
& \overline{\sum _{\lambda}} \sum _{\lambda} | T_{\gamma p \to \phi p} |^{2} 
%\nonumber \\ & 
= |a_{p}|^{2} \exp (b \tilde{t}) \left [ 
1 + \frac{|\vec{p}_{\phi}|^{2}}{2 M_{\phi}^{2}} 
\sin ^{2} \theta _{p} 
\right ] , 
\end{align}
Note that the spin 
component of the proton does not appear in this phenomenological form of the 
cross section.

\subsection{$\bm{\gamma d \to \phi p n}$ reaction}
\label{subsec:gamma-d}

\subsubsection{Kinematics}
\label{subsubsec:kin2}

\begin{figure}[t]
  \centering
    \includegraphics[scale=0.18]{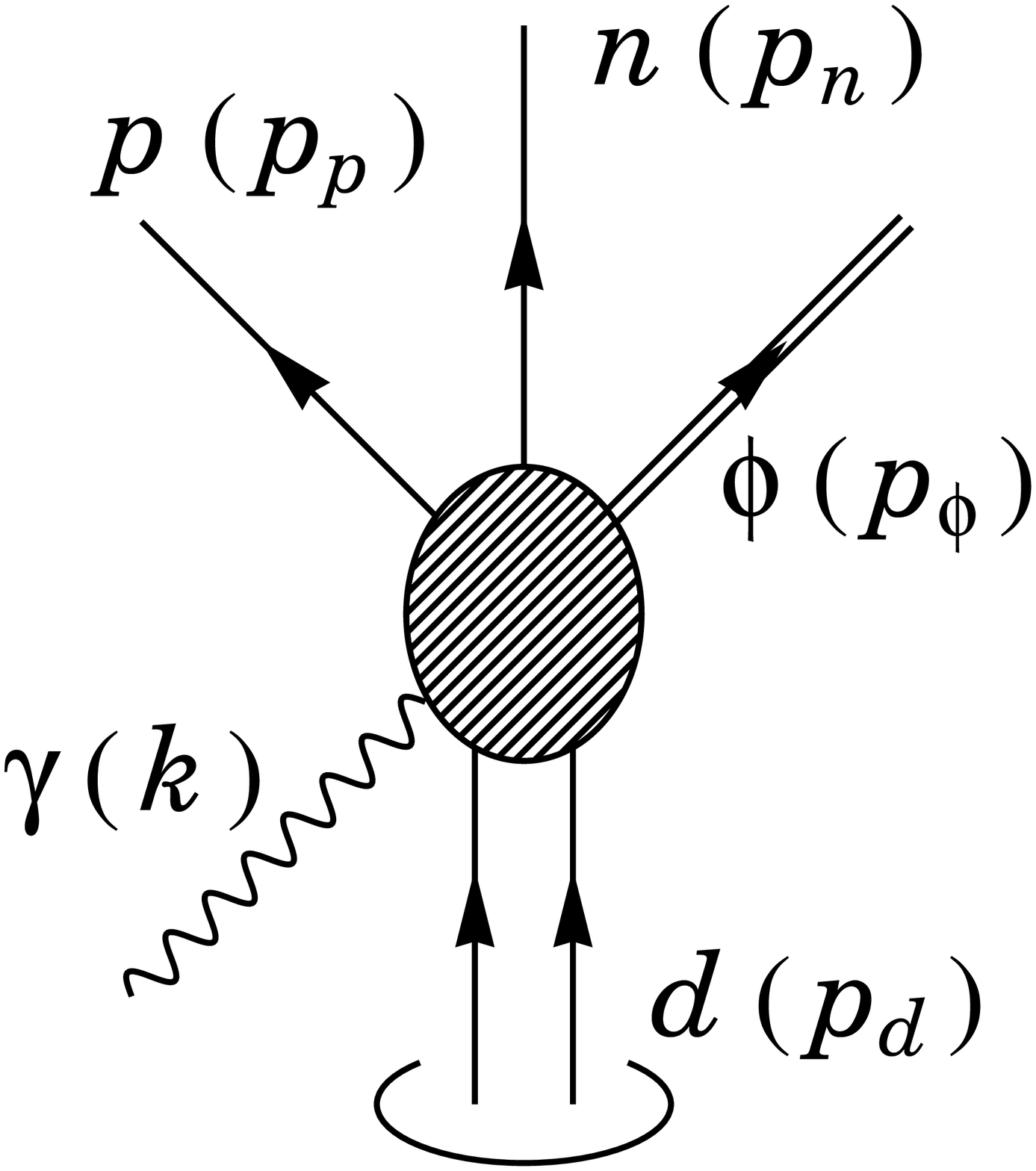} 
  \caption{Kinematics for the $\gamma d \to \phi p n$ reaction. }
  \label{fig:gamma-d}
\end{figure}

Now let us provide the formulation for the $\gamma d \to \phi p n$ reaction. 
The cross section for the three-body final state is written 
as~\cite{Amsler:2008zzb}, 
\begin{align}
& \sigma _{\gamma d \to \phi p n} (E_{\gamma}^{\text{lab}} ) 
\nonumber \\
& = \frac{M_{p} M_{n}}{4 E_{\gamma}^{\text{lab}} \sqrt{s_{\text{tot}}}} 
\frac{1}{(2 \pi )^{4}} \int d M_{p n} | \vec{p}_{\phi} | |\vec{p}_{p}^{\; \ast} | 
\int _{-1}^{1} d \cos \theta _{\phi} \int d \Omega _{p}^{\ast} 
\nonumber \\
& \phantom{=} \times \overline{\sum _{\lambda}} \sum _{\lambda} 
| T_{\gamma d \to \phi p n} |^{2} . 
\end{align}
Here $s_{\text{tot}}=(k+p_{d})^{2}$ is the Mandelstam variable with the initial 
photon and deuteron momenta, $k$ and $p_{d}$, respectively, $M_{pn}$ the 
$p$-$n$ invariant mass, $\vec{p}_{\phi}$ and $\theta _{\phi}$ the $\phi$ momentum 
and the scattering angle between the 
incident photon and the final $\phi$ in the 
total center-of-mass frame, respectively, and $\vec{p}_{p}^{\; \ast}$ and 
$\Omega_{p}^{\ast}$ the proton momentum and the proton solid angle in the $p$-$n$ 
center-of-mass frame, respectively. 

In order to make connection with the work of~\cite{Chang:2009yq} we change the 
integral variables from $M_{p n}$ and 
$\cos \theta _{\phi}$ to $t_{\phi}\equiv (p_{\phi}-k)^{2}$ and 
$u_{\phi}\equiv (p_{\phi}-p_{d})^{2}$ with the $\phi$ momentum $p_{\phi}$. 
For this purpose we use the following kinematical relations, 
\begin{align}
& s_{\text{tot}} + t_{\phi} + u_{\phi} = M_{d}^{2} + M_{\phi}^{2} + M_{p n}^{2} , 
\label{eq:Mandelstam-stu}
\\
& s_{\text{tot}} - 2 \omega _{\phi}^{\text{cm}} \sqrt{s_{\text{tot}}} + M_{\phi}^{2} 
= M_{p n}^{2} , 
\label{eq:Mpn}
\\
& t_{\phi} = M_{\phi}^{2} 
- 2 E_{\gamma}^{\text{cm}} (\omega _{\phi}^{\text{cm}} 
- |\vec{p}_{\phi} | \cos \theta _{\phi}) , 
\label{eq:tphi}
\end{align}
where $E_{\gamma}^{\text{cm}}$ is the photon energy in the total center-of-mass 
frame, and $\phi$ energy $\omega _{\phi}^{\text{cm}}=\sqrt{M_{\phi}^{2}+|\vec{p}_{\phi}|^{2}}$. 
Since $s_{\text{tot}}$ is fixed and the masses $M_{d}^{2}$ and $M_{\phi}^{2}$ are 
constant, we have from Eq.~(\ref{eq:Mandelstam-stu}), 
\be
\frac{\prt M_{p n}}{\prt t_{\phi}} = 
\frac{\prt M_{p n}}{\prt u_{\phi}} = 
\frac{1}{2 M_{p n}} . 
\ee
In addition, since $\omega _{\phi}^{\text{cm}}$ and $|\vec{p}_{\phi}|$ are 
functions of $t_{\phi}$ and $u_{\phi}$, we have from Eq.~(\ref{eq:tphi}), 
\begin{align}
\frac{\prt \cos \theta _{\phi}}{\prt t_{\phi}} 
& = \frac{\omega _{\phi}^{\text{cm}}}{4 E_{\gamma}^{\text{cm}} \sqrt{s_{\text{tot}}} 
|\vec{p}_{\phi}|^{3}} 
(t_{\phi} - M_{\phi}^{2} + 2 E_{\gamma}^{\text{cm}} \omega _{\phi}^{\text{cm}} )
\nonumber \\
& \phantom{=}
+ \frac{1}{2 E_{\gamma}^{\text{cm}} |\vec{p}_{\phi}|} 
\left ( 1 - \frac{E_{\gamma}^{\text{cm}}}{\sqrt{s_{\text{tot}}}} \right ) , 
\end{align}
\begin{align}
\frac{\prt \cos \theta _{\phi}}{\prt u_{\phi}} 
& = \frac{\omega _{\phi}^{\text{cm}}}
{4 E_{\gamma}^{\text{cm}} \sqrt{s_{\text{tot}}} |\vec{p}_{\phi}|^{3}} 
(t_{\phi} - M_{\phi}^{2} + 2 E_{\gamma}^{\text{cm}} \omega _{\phi}^{\text{cm}} ) 
\nonumber \\ &
\phantom{=} + \frac{1}{2 E_{\gamma}^{\text{cm}} |\vec{p}_{\phi}|}
\left ( - \frac{E_{\gamma}^{\text{cm}}}{\sqrt{s_{\text{tot}}}} \right
) .
\end{align}
Here we have used the relation, 
\be
\frac{\prt |\vec{p}_{\phi}|}{\prt t_{\phi}} = 
\frac{\prt |\vec{p}_{\phi}|}{\prt u_{\phi}} = 
\frac{\prt |\vec{p}_{\phi}|}{\prt \omega _{\phi}^{\text{cm}}}
\frac{\prt \omega _{\phi}^{\text{cm}}}{\prt M_{p n}} 
\frac{\prt M_{p n}}{\prt t_{\phi}} = 
- \frac{\omega _{\phi}^{\text{cm}} }{2 \sqrt{s_{\text{tot}}} |\vec{p}_{\phi}|} , 
\ee
where $\prt \omega _{\phi}^{\text{cm}} / \prt M_{p n} = - M_{p n} / 
\sqrt{s_{\text{tot}}}$ is evaluated from 
Eq.~(\ref{eq:Mpn}). As a consequence, we have, 
\be
d M_{p n} d \cos \theta _{\phi} 
= J (M_{p n} , \, \cos \theta _{\phi} ; \, t_{\phi}, \, u_{\phi})
d t_{\phi} d u_{\phi}
\ee
with the Jacobian, 
\begin{align}
J (M_{p n} , \, \cos \theta _{\phi} ; \, t_{\phi}, \, u_{\phi})
= \frac{1}{4 E_{\gamma}^{\text{cm}} M_{p n} |\vec{p}_{\phi} | } . 
\end{align}
Now we can write down the final form of the total cross section for the 
$\gamma d \to \phi p n$ reaction as, 
\begin{align}
\sigma _{\gamma d \to \phi p n} (E_{\gamma}^{\text{lab}} ) 
& = 
\frac{M_{p} M_{n}}{16 (E_{\gamma}^{\text{lab}})^{2} M_{d}}
\frac{1}{(2 \pi )^{4}} \int d t_{\phi} \int d u_{\phi} 
\frac{|\vec{p}_{p}^{\; \ast} |}{M_{p n}} 
\nonumber \\
& \phantom{=} \times 
\int d \Omega _{p}^{\ast} 
\; \overline{\sum _{\lambda}} \sum _{\lambda} | T_{\gamma d \to \phi p n} |^{2} , 
\label{eq:23}
\end{align}
where we have used $E_{\gamma}^{\text{cm}} \sqrt{s_{\text{tot}}} 
= E_{\gamma}^{\text{lab}} M_{d}$. Or, equivalently, we have, 
\begin{align}
\frac{d \sigma _{\gamma d \to \phi p n}}{d t_{\phi}} 
& = \frac{M_{p} M_{n}}{16 (E_{\gamma}^{\text{lab}})^{2} M_{d}}
\frac{1}{(2 \pi )^{4}} \int _{u_{\phi , \text{min}}}^{u_{\phi , \text{max}}} d u_{\phi} 
\frac{|\vec{p}_{p}^{\; \ast} |}{M_{p n}} 
\nonumber \\
& \phantom{=} 
\times \int d \Omega _{p}^{\ast} 
\; \overline{\sum _{\lambda}} \sum _{\lambda} | T_{\gamma d \to \phi p n} |^{2} , 
\end{align}
with $u_{\phi , \text{min}}$ and $u_{\phi , \text{max}}$ the minimum and maximum value 
of $u_{\phi}$ for fixed $t_{\phi}$, 
\begin{align}
& u_{\phi , \text{min}} 
= M_{\phi}^{2} + M_{d}^{2} + (M_{p} + M_{n})^{2} - s_{\text{tot}} - t_{\phi} , 
\\
& u_{\phi , \text{max}} 
= 2 M_{\phi}^{2} + M_{d}^{2} - 2 \sqrt{s_{\text{tot}}} \omega _{\phi}^{\prime} - t_{\phi} ,
\end{align}
\begin{align}
& \omega _{\phi}^{\prime} = \sqrt{p_{\phi}^{\prime 2} + M_{\phi}^{2}} , 
\quad 
p_{\phi}^{\prime} = \frac{M_{\phi}^{2} - t_{\phi}}{4 E_{\gamma}^{\text{cm}}} 
- \frac{M_{\phi}^{2} E_{\gamma}^{\text{cm}}}{M_{\phi}^{2} - t_{\phi}} , 
\end{align}
where $u_{\phi , \text{min}}$ ($u_{\phi , \text{max}}$) is achieved in the case 
that $M_{p n}$ has its minimum (maximum) value with fixed $t_{\phi}$ (see 
Eq.~(\ref{eq:Mandelstam-stu})). From the kinematics, $M_{pn}$ takes values 
between, 
\be
(M_{p} + M_{n})^{2} \le M_{p n}^{2} \le M_{\phi}^{2} + s_{\text{tot}} 
- 2 \sqrt{s_{\text{tot}}} \omega _{\phi}^{\prime} . 
\ee
We note that both $u_{\phi , \text{min}}$ and $u_{\phi , \text{max}}$ depend on 
$t_{\phi}$. We also 
write down the minimum and maximum $t_{\phi}$ of the $\gamma d \to \phi p n$ 
reaction, $t_{\phi , \text{min}}$ and $t_{\phi , \text{max}}$, which will be needed 
for Eq.~(\ref{eq:23}), as, 
\begin{align}
& t_{\phi , \text{min}} 
= M_{\phi}^{2} - 2 E_{\gamma}^{\text{cm}} ( \omega _{\text{max}} + p_{\text{max}} ) , 
\label{eq:tphimin}
\\
& t_{\phi , \text{max}} 
= M_{\phi}^{2} - 2 E_{\gamma}^{\text{cm}} ( \omega _{\text{max}} - p_{\text{max}} ) , 
\label{eq:tphimax}
\end{align}
with, 
\begin{align}
& \omega _{\text{max}} = \sqrt{p_{\text{max}}^{2} + M_{\phi}} , 
\\
& p_{\text{max}} = 
\frac{\lambda ^{1/2} (s_{\text{tot}}, \, M_{\phi}^{2}, \, (M_{p}+M_{n})^{2} )}
{2 \sqrt{s_{\text{tot}}}} . 
\end{align}
Here $p_{\text{max}}$ corresponds to the maximum momentum for the $\phi$ in the 
total center-of-mass frame, in which $M_{p n}=M_{p}+M_{n}$.

\subsubsection{Scattering amplitude}
\label{subsubsec:amp2}

\begin{figure}[t]
  \centering
  \begin{tabular*}{8cm}{@{\extracolsep{\fill}}cc}
    \includegraphics[scale=0.18]{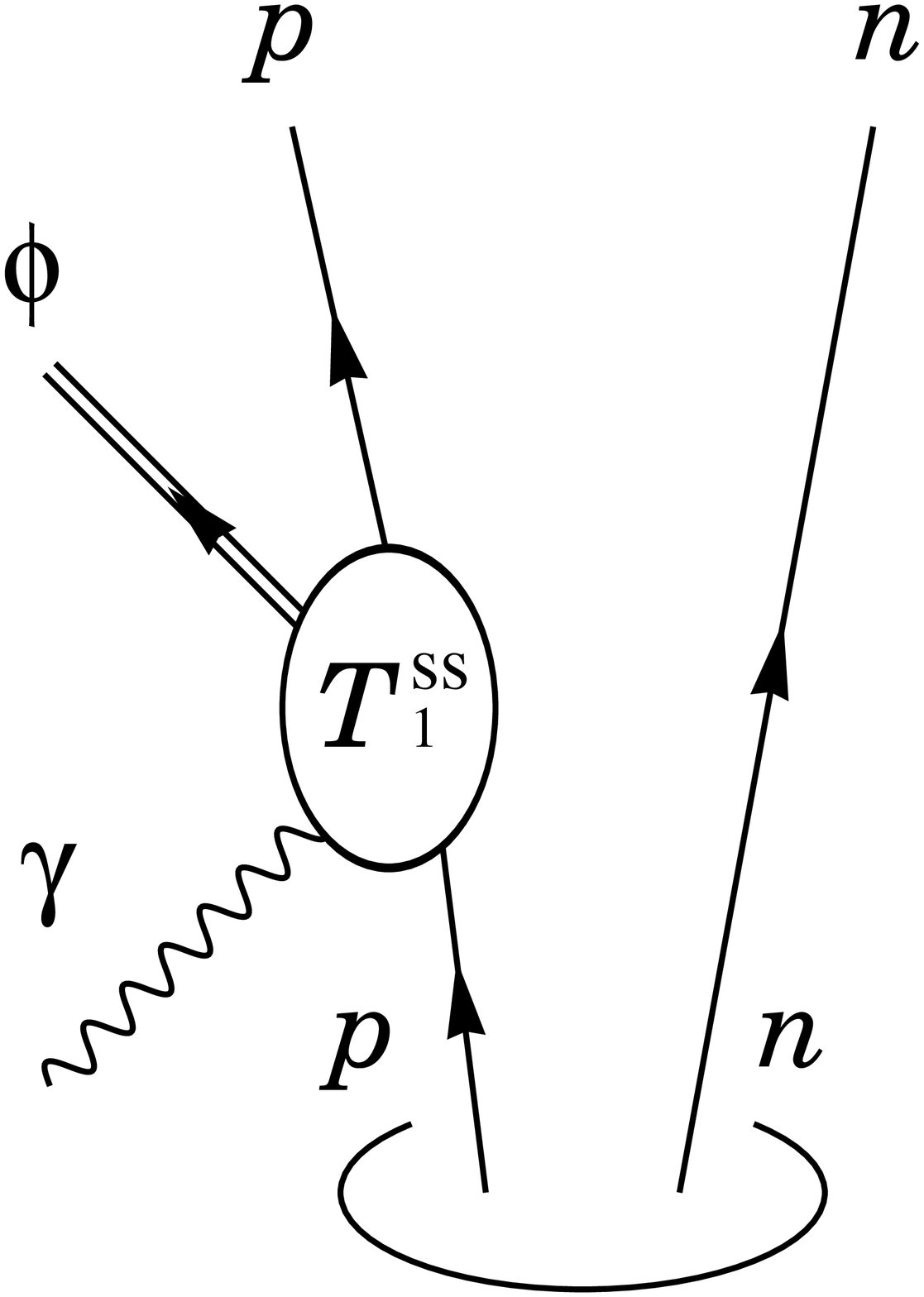} & 
    \includegraphics[scale=0.18]{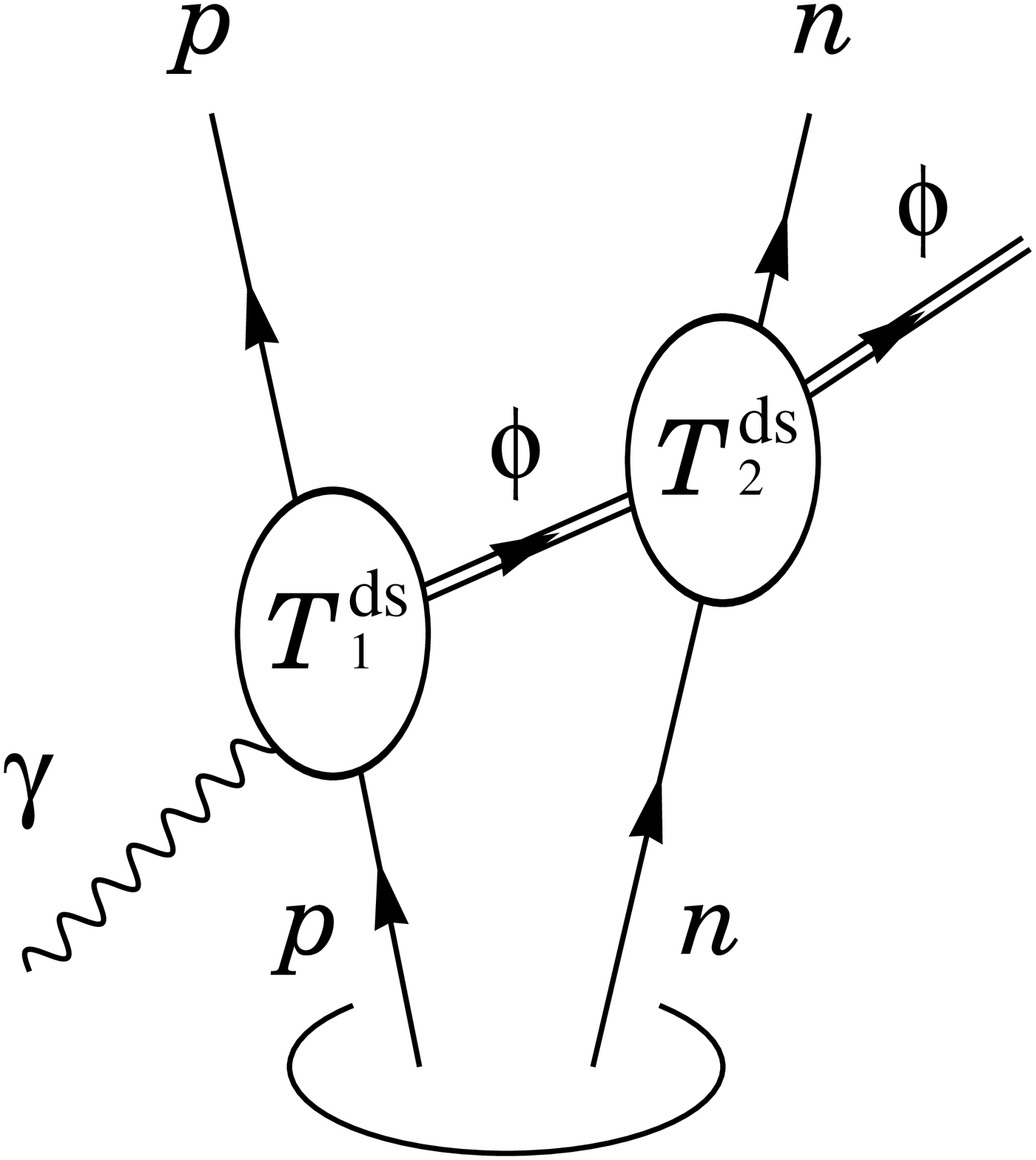} \vspace{5pt} \\
    (a)  $T^{\text{ss}}$ & (b)  $T^{\text{ds}}$ 
  \end{tabular*}
  \caption{Diagrams for the calculatin of $\gamma d \to \phi p n$ reaction. }
  \label{fig:gamma-d-each}
\end{figure}

In this section we develop the formalism to obtain the scattering amplitude for the $\gamma d \to \phi p n$
reaction. Since our aim is to compare our results
with~\cite{Chang:2009yq}, where the cross section on the proton of the
deuteron is singled out, we also select from the full model of the
$\gamma d \to \phi p n$ the terms where there is primary production of
the $\phi$ on the proton.  It is easy to extend this formulation to a
case of the neutron of the deuteron in the same way as the proton
case. The $\phi$ photoproduction amplitude from the proton of the
deuteron is obtained from the mechanisms depicted in
Fig.~\ref{fig:gamma-d-each}. The diagram of the left represents the
single scattering amplitude, $T^{\text{ss}}$. The diagram of the right
represents the double scattering amplitude,
$T^{\text{ds}}$. Modification of the $\phi$ photoproduction amplitude
on the proton of the deuteron with respect to that on a free proton
would be attributed to this double scattering amplitude, as well as to
Fermi motion and binding effects associated to the deuteron wave function.
We will see later that $T^{\text{ss}}$ and $T^{\text{ds}}$ are
correlated destructively with each other. Hence, we expect that
$T^{\text{ds}}$ decreases the cross section of the impulse
approximation.

The evaluation of the amplitude including the deuteron wave function is given
in~\cite{Jido:2009jf}, where the authors discussed the $K^{-} d \to \pi \Sigma
n$ reaction. Following~\cite{Jido:2009jf} we obtain for the impulse
approximation amplitude, 
\be T^{\text{ss}}
= T_{1}^{\text{ss}} \times \tilde{\varphi} (|\vec{p}_{n} -
\vec{p}_{d}/2|) , 
\label{eq:Tss}
\ee 
written in terms of the elementary $\gamma p \to \phi p$ amplitude $T_{1}^{\text{ss}}$ 
and the deuteron wave function
$\tilde{\varphi}$ in momentum space. The elementary $\gamma p \to \phi
p$ amplitude has already appeared in Eq.~(\ref{eq:ampli}), and
we rewrite it for the case of the single scattering in the deuteron target as, 
\be
T_{1}^{\text{ss}} = - a_{p} (M_{\phi p}^{2}) \exp (b
\tilde{t}^{\text{ss}} / 2) \times \vec{\epsilon} (\gamma ) \cdot
\vec{\epsilon} (\phi ) , 
\label{eq:T1ss}
\ee 
where, 
\be
\tilde{t}^{\text{ss}}=(p_{\phi} - k)^{2} - t_{\text{max}} ( M_{\phi p}^{2} ), 
\ee 
with $t_{\text{max}}$ defined in
Eq.~(\ref{eq:tmax}). In the case of the single scattering for the
$\gamma d \to \phi p n$ reaction, $a_{p}$ and $t_{\text{max}}$ are 
functions of $M_{\phi p}^{2}=(p_{\phi} + p_{p})^{2}$ instead of $s$ in the free 
$\gamma p \to \phi p$ reaction. 

For the deuteron wave function, we neglect the $d$-wave component 
and we use a parameterization of the $s$-wave component given by an analytic 
function~\cite{Lacombe:1981eg} as, 
\be
\tilde{\varphi} (p) = \sum _{j=1}^{11} \frac{C_{j}}{p^{2} + m_{j}^{2}} , 
\label{eq:deutWF}
\ee
with $C_{j}$ and $m_{j}$ determined in~\cite{Machleidt:2000ge}. 

Let us now consider the double scattering amplitude. 
Following~\cite{Jido:2009jf}, one can show that this amplitude is given by, 
\be
T^{\text{ds}} = \int \frac{d^{3} q_{\text{ex}}}{(2 \pi )^{3}} 
\frac{\tilde{\varphi} 
(|\vec{p}_{\phi} + \vec{p}_{n} - \vec{q}_{\text{ex}} - \vec{p}_{d}/2|)}
{q_{\text{ex}}^{2} - M_{\phi}^{2} + i M_{\phi} \Gamma _{\phi}} 
\times T_{1}^{\text{ds}} T_{2}^{\text{ds}} , 
\label{eq:Tds}
\ee
where $q_{\text{ex}}^{\mu}$ is the exchanged $\phi$ meson momentum, 
$T_{1}^{\text{ds}}$ the elementary $\gamma p \to \phi p$ amplitude, and 
$T_{2}^{\text{ds}}$ the $\phi n \to \phi n$ amplitude. We note that in general 
both 
$T_{1}^{\text{ds}}$ and $T_{2}^{\text{ds}}$ depend on $\vec{q}_{\text{ex}}$ and appear 
inside the $\vec{q}_{\text{ex}}$ integration. 

Since we are interested in finding the effects from double scattering, the 
important point is to pick up the term that leads to largest interference 
with the impulse approximation, accepting, as it is the case, that the largest 
contribution is given by the single scattering. For this purpose we take the 
$\phi n \to \phi n$ amplitude with the same initial and final $\phi$ 
polarization. This selects the $\phi n \to \phi n$ amplitude that leads to the 
same polarization structure as Eq.~(\ref{eq:T1ss}) and hence produces maximum 
interference with the single scattering amplitude. Explicit details on this 
$\phi n \to \phi n$ amplitude are given in Appendix~\ref{sec:phi-n}. Then the 
elementary $\gamma p \to \phi p$ amplitude in Eq.~(\ref{eq:Tds}) is written as, 
\be
T_{1}^{\text{ds}} = - a_{p} 
((q_{\text{ex}} + p_{p})^{2})
\exp (b \tilde{t}^{\, \prime \, \text{ds}} / 2) 
\times \vec{\epsilon} (\gamma ) \cdot \vec{\epsilon} (\phi ) , 
\label{eq:Tds_old}
\ee
with, 
\be
\tilde{t}^{\, \prime \, \text{ds}} 
= (q_{\text{ex}} - k)^{2} - t_{\text{max}} ((q_{\text{ex}} + p_{p})^{2}) . 
\ee
Hence both the factors $a_{p}$ and 
$\exp (b \tilde{t}^{\, \prime \, \text{ds}}/2)$ depend on 
$\vec{q}_{\text{ex}}$. Since the double scattering amplitude is quite small 
compared to the single scattering one, one can safely approximate 
$\tilde{t}^{\, \prime \, \text{ds}}$ taking into account that the deuteron 
wave function $\tilde{\varphi} (p)$ takes the largest component when the 
nucleons are at rest in the rest frame of the deuteron. This 
allows us to write $(q_{\text{ex}} + p_{p})^{2}$ and $(q_{\text{ex}} - k)^{2}$ as, 
\begin{align}
(q_{\text{ex}} + p_{p})^{2} & = (k + p_{1})^{2} 
\simeq (E_{\gamma}^{\text{lab}} + M_{p} - B_{1/2} )^{2} - (E_{\gamma}^{\text{lab}})^{2} , 
\end{align}
\begin{align}
(q_{\text{ex}} - k)^{2} & = 
(q_{\text{ex}}^{0} - E_{\gamma}^{\text{lab}})^{2} - (\vec{p}_{1} - \vec{p}_{p})^{2} 
\nonumber \\ & 
\simeq (q_{\text{ex}}^{0} - E_{\gamma}^{\text{lab}})^{2} - |\vec{p}_{p}|^{2} , 
\end{align}
where $p_{1}^{\mu}$ is the proton momentum inside the deuteron in the laboratory 
frame, and we take 
$\vec{p}_{1} \simeq \vec{0}$ and $p_{1}^{0} = M_{p}-B_{1/2}$, with $B_{1/2}$ the binding energy 
for the proton, which we assume to be half of the 
deuteron binding energy, $B_{1/2}=1.112 \mev$, and 
$\vec{p}_{p}$ is the final proton momentum in the laboratory frame. 
Further, $q_{\text{ex}}^{0}$ is approximated in the laboratory frame as, 
\be
q_{\text{ex}}^{0} \simeq E_{\gamma}^{\text{lab}} + M_{p} - B_{1/2} - E_{p}^{\text{lab}} , 
\label{eq:qexo}
\ee
with $E_{p}^{\text{lab}} = \sqrt{M_{p}^{2} + |\vec{p}_{p}|^{2}}$. Then we have, 
\be
T_{1}^{\text{ds}} \simeq - a_{p} 
( W^{2} )
\exp (b \tilde{t}^{\text{ds}} / 2) 
\times \vec{\epsilon} (\gamma ) \cdot \vec{\epsilon} (\phi ) , 
\label{eq:Tds1}
\ee
with, 
\be
\tilde{t}^{\text{ds}} 
= (q_{\text{ex}}^{0} - E_{\gamma}^{\text{lab}})^{2} - |\vec{p}_{p}|^{2} 
- t_{\text{max}} ( W^{2} ), 
\ee
\be
W^{2} = (E_{\gamma}^{\text{lab}} + M_{p} - B_{1/2} )^{2} - (E_{\gamma}^{\text{lab}})^{2} , 
\label{eq:definition_W}
\ee
with $q_{\text{ex}}^{0}$ given by Eq.~(\ref{eq:qexo}). Since $T_{1}^{\text{ds}}$ in 
Eq.~(\ref{eq:Tds1}) no longer depends on $\vec{q}_{\text{ex}}$, we can 
extract $T_{1}^{\text{ds}}$ outside of the $\vec{q}_{\text{ex}}$ integration in 
Eq.~(\ref{eq:Tds}). 

Next we consider $T_{2}^{\text{ds}}$, which corresponds to the $\phi n
\to \phi n$ amplitude. The details of this amplitude are shown in the
Appendix~\ref{sec:phi-n}.  As will be clear below, it is through the
imaginary part of this amplitude that $T^{\text{ds}}$ interferes
destructively with the single scattering, once the dominant on-shell
part of the intermediate $\phi$ is taken in $T^{\text{ds}}$. In the
Appendix~\ref{sec:phi-n} we show that $\Ima T_{2}^{\text{ds}}$ can be approximated by
taking $\vec{q}_{\text{ex}}=\vec{p}_{\phi}$ if we are concerned about
forward $\phi$ production as one has in the experiment. Yet, since
$\tilde{\varphi} (p)$ is very sensitive to the momentum, we do not
take $\vec{q}_{\text{ex}}=\vec{p}_{\phi}$ in the argument of
$\tilde{\varphi} (p)$ in Eq.~(\ref{eq:Tds}). This approximation allows
us to factorize $T_{2}^{\text{ds}}$ outside the integral of
Eq.~(\ref{eq:Tds}), as we had done with $T_{1}^{\text{ds}}$ before.

Now we have only the $\phi$ meson propagator and the deuteron wave function 
inside the $\vec{q}_{\text{ex}}$ integral of Eq.~(\ref{eq:Tds}). For the $\phi$ 
meson propagator, we take its imaginary part, keeping the $\phi$ 
on-shell, as in Glauber theory, hence, 
\be
\frac{1}{q_{\text{ex}}^{2} - M_{\phi}^{2} + i M_{\phi} \Gamma _{\phi}} 
\simeq - i \pi \delta (q_{\text{ex}}^{2} - M_{\phi}^{2}) . 
\label{eq:approx-prop}
\ee
Using this approximation, one can perform the $\vec{q}_{\text{ex}}$ integration 
in the laboratory frame as, 
\begin{align}
& - i \pi \int \frac{d^{3} q_{\text{ex}}}{(2 \pi )^{3}} 
\delta (q_{\text{ex}}^{2} - M_{\phi}^{2})
\tilde{\varphi} (|\vec{p}_{\phi} + \vec{p}_{n} - \vec{q}_{\text{ex}}|) 
\nonumber \\
& = \frac{- i q}{8 \pi } 
\int _{-1}^{1} d \cos \theta _{q} \,
\tilde{\varphi} \left ( \sqrt{v^{2} + q^{2} - 2 v q \cos \theta _{q}} \right )  
\nonumber \\
& = - i \sum _{j=1}^{11} \frac{C_{j}}{16 \pi v} 
\ln \left ( \frac{(v + q)^{2} + m_{j}^{2}}{(v - q)^{2} + m_{j}^{2}} \right ) , 
\label{eq:WF-prop}
\end{align}
where $q=\sqrt{q_{\text{ex}}^{0 2} - M_{\phi}^{2}}$, $v=|\vec{p}_{\phi}+\vec{p}_{n}|$, 
and $\theta _{q}$ is the angle between $\vec{q}_{\text{ex}}$ and 
$\vec{p}_{\phi}+\vec{p}_{n}$ in the laboratory frame. To obtain 
Eq.~(\ref{eq:WF-prop}), we have used $\vec{p}_{d}=\vec{0}$ in the laboratory 
frame and the explicit form of the deuteron wave function of 
Eq.~(\ref{eq:deutWF}). 

As a consequence, we finally have for the double scattering amplitude, 
\begin{align}
T^{\text{ds}} = & T_{1}^{\text{ds}} \times \text{Im} \, T_{2}^{\text{ds}} (M_{\phi n}^{2})
\times 
\sum _{j=1}^{11} \frac{C_{j}}{16 \pi v} 
\ln \left ( \frac{(v + q)^{2} + m_{j}^{2}}{(v - q)^{2} + m_{j}^{2}} \right ) . 
\label{eq:Tds_final}
\end{align}

Then the sum of the amplitudes of the impulse approximation~(\ref{eq:Tss}) 
and the double scattering~(\ref{eq:Tds_final}) gives us, 
\begin{align}
& T^{\text{ss}} + T^{\text{ds}} = 
- \vec{\epsilon} (\gamma ) \cdot \vec{\epsilon} (\phi ) \times 
\Bigg [ a_{p} 
( M_{\phi p}^{2} ) \exp (b \tilde{t}^{\text{ss}} / 2) 
\tilde{\varphi} (|\vec{p}_{n}|) 
\nonumber \\
& ~ + a_{p} 
( W^{2} ) \exp (b \tilde{t}^{\text{ds}} / 2) 
\nonumber \\
& ~ \phantom{+} \times 
\text{Im} \, T_{2}^{\text{ds}} (M_{\phi n}^{2}) \times 
\sum _{j=1}^{11} \frac{C_{j}}{16 \pi v} 
\ln \left ( \frac{(v + q)^{2} + m_{j}^{2}}{(v - q)^{2} + m_{j}^{2}} \right ) 
\Bigg ] . 
\label{eq:TssTds}
\end{align}
Note that in Eq.~(\ref{eq:TssTds}) we have already made use of the fact that 
in the Appendix~\ref{sec:phi-n} we chose the part of the 
$\phi n \to \phi n$ amplitude with the same initial and final $\phi$ 
polarizations. It is important to note that the double scattering amplitude
$T^{\text{ds}}$ is real, like $T^{\text{ss}}$, because we have chosen 
Eq.~(\ref{eq:approx-prop}) and $\Ima T_{2}^{\text{ds}}$. This allows for 
interference and we find that this interference is destructive, which has the 
physical meaning that the $\phi$ produced in the first step of the double 
scattering can undergo absorption into the $\phi n \to K Y$ channels.

\section{Results}
\label{sec:result}

In this section we will show results for $\sigma _{p}$ for the free $\gamma p 
\to \phi p$ reaction and for the $\gamma d \to \phi p n$, but only for the 
mechanism where the $\gamma$ strikes a proton first, as have discussed 
earlier. We refer to this latter cross section as $\sigma _{p^{\ast}}$, which 
one would like to compare with the $\sigma _{p^{\ast}}$ experimental cross 
section.

\subsection{Differential cross sections}

\begin{figure}[t]
  \centering
    \includegraphics[width=8.6cm]{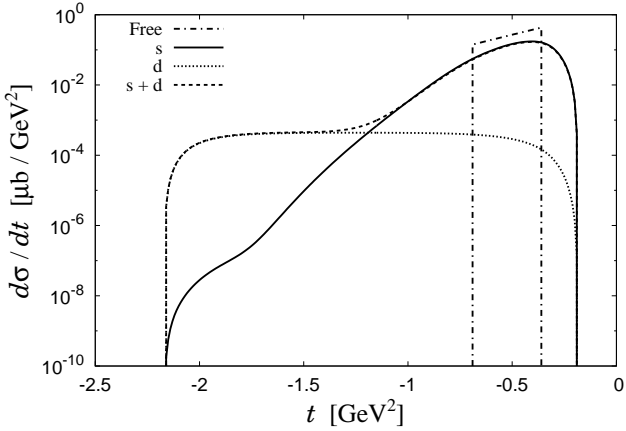} 
    \caption{Differential cross sections $d\sigma _{p}/dt$ and
      $d\sigma _{p^{\ast}}/dt_{\phi}$. We fix $E_{\gamma}^{\text{lab}}
      = 1.6 \gev$. The labels ``Free'', ``s'', ``d'', and ``s$+$d''
      indicate the case of the free proton target, the single, double,
      and single plus double scattering case of the deuteron target,
      respectively. }
  \label{fig:dSdt-16}
\end{figure}

First we show in Fig.~\ref{fig:dSdt-16} the differential cross 
sections $d\sigma _{p}/dt$ and $d\sigma _{p^{\ast}}/dt_{\phi}$ at 
$E_{\gamma}^{\text{lab}}=1.6 \gev$ without any angular cuts. 

As one can see from Fig.~\ref{fig:dSdt-16}, the range of $t_{\phi}$ for the
deuteron target is wider than for the proton. This is a simple 
consequence of having a different reaction, $\gamma p \to \phi p$ or 
$\gamma d \to \phi p n$, where the second one has three particles 
in the final state and different mass for the target. The limits in either 
case were given in Eqs.~(\ref{eq:tmax}) and (\ref{eq:tmin}) for the 
proton and Eqs.~(\ref{eq:tphimin}) and (\ref{eq:tphimax}) for the deuteron. 
At $E_{\gamma}^{\text{lab}}=1.6 \gev$, these values are 
$t_{\text{min}}=-0.69 \gev ^{2}$ and $t_{\text{max}}=-0.36 \gev ^{2}$ for the proton 
and $t_{\phi , \text{min}}=-2.17 \gev ^{2}$ and $t_{\phi , \text{max}}=-0.19 \gev ^{2}$ 
for the deuteron, respectively. 

In addition, we should note that at the minimum and maximum values of $t$ 
for the proton target case the final state phase-space 
of the reaction $\gamma p \to \phi p$ is finite with a sharp drop to zero, 
whereas for the deuteron target case the final state 
phase-space of the reaction $\gamma d \to \phi p n$ goes smoothly to zero. 
This is a consequence of having three particles in the final state for the 
deuteron target case. This is shown in Fig.~\ref{fig:dSdt-16} as a smooth 
decrease of $d\sigma_{p^{\ast}}/dt_{\phi}$ around 
$t_{\phi , \text{min}}=-2.17 \gev ^{2}$ and $t_{\phi , \text{max}}=-0.19 \gev ^{2}$. 

For the deuteron target the double scattering amplitude (see
Fig.~\ref{fig:gamma-d-each}(b)), relatively to single scattering,
contributes more to $d\sigma _{p^{\ast}}/dt_{\phi}$ in the large
$|t_{\phi}|$ region. This is due to the fact that the large momentum
transfer $|t_{\phi}|$ is achieved only by the large Fermi momentum
components in the single scattering. However, in the double scattering
this momentum transfer can be split between two nucleons and it is
easier to accommodate.  
Here we also note that the $t_{\phi}$ dependence 
is different for single and double scatterings, and the factor 
$e^{b\tilde{t}^{\rm ss}}$ strongly suppresses the single scattering 
contribution in the large $t_{\phi}$ region, which is smeared 
by the split of the momentum transfer in the double scattering. 
In the small $|t_{\phi}|$ ($\lesssim 1 \, \text{GeV}^{2}$) region, on
the other hand, the single scattering amplitude dominates $d\sigma
_{p^{\ast}}/dt_{\phi}$. Indeed, here one needs only small Fermi
momentum components for which the deuteron wave function has its
maximum.
%From this discussion one can see that 
%we only need the
%single scattering amplitude for
%the single scattering dominates the differential cross section 
%in the small $|t_{\phi}|$ ($\lesssim 1 \, \text{GeV}^{2}$) region. 
This result is one of the important findings of the present work. We
see that the effect of the double scattering is basically negligible
at $t_{\phi} \simeq t_{\phi , \text{max}}$ by comparing the curve
``s'' and ``s$+$d'' (where the interference appears). The double
scattering alone is less than one per thousand and the interference
around $t_{\text{max}}$ is less than $7 \%$.  The smallness of the
double scattering contribution was hinted in Ref.~\cite{Titov:2007fc}
from the fact that the basic experimental information on this reaction
was reproduced in terms of single scattering alone.  In
Ref.~\cite{Rogers:2005bt} the formalism for double scattering was
developed for coherent $\phi$ photoproduction in the deuteron, but no
explicit evaluation was done.  To the best of our knowledge, this is
the first explicit evaluation of the contribution of double scattering
for this reaction.

\begin{figure*}[!t]
  \centering
  \begin{tabular*}{\textwidth}{@{\extracolsep{\fill}}cc}
    \includegraphics[width=8.4cm]{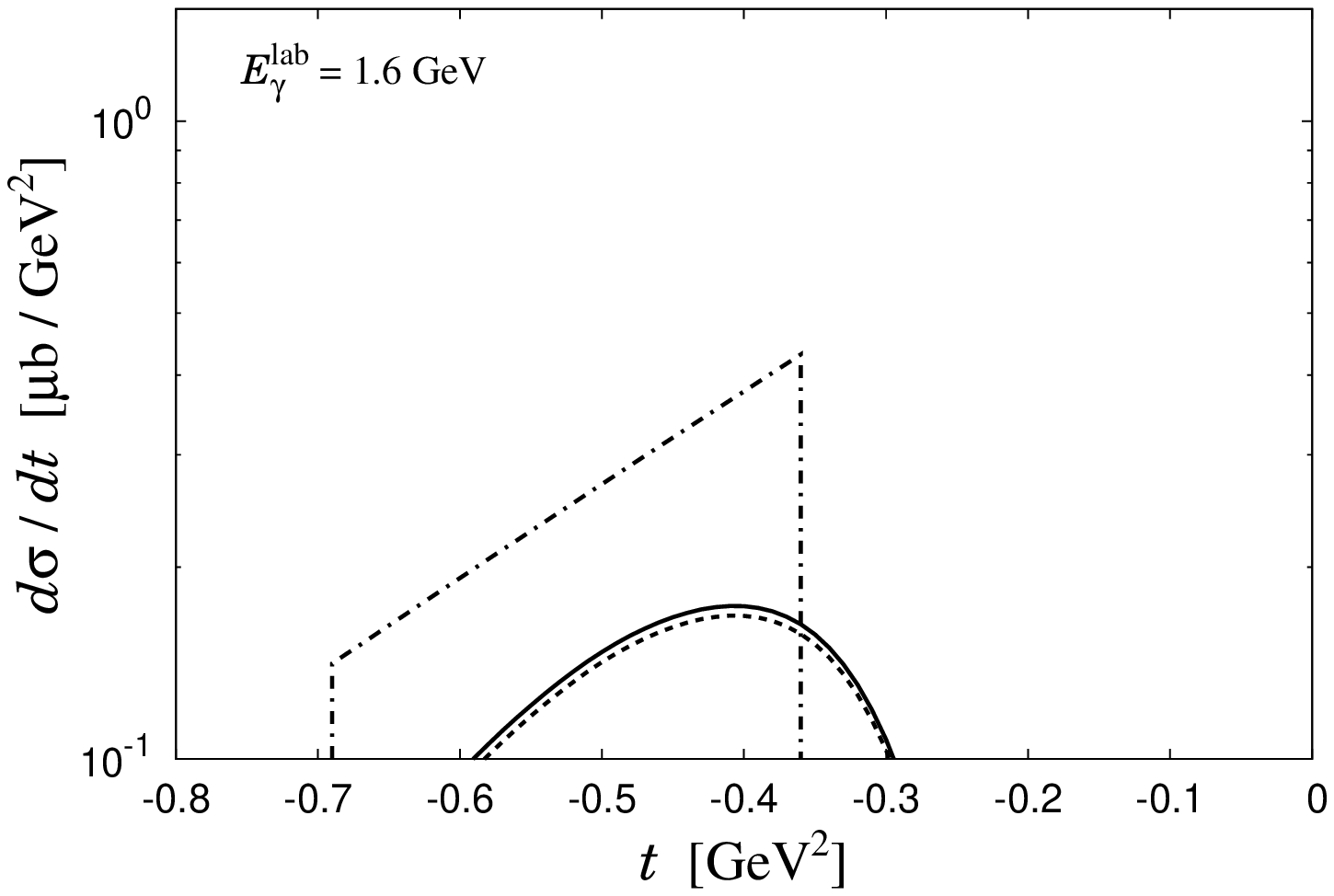} & 
    \includegraphics[width=8.4cm]{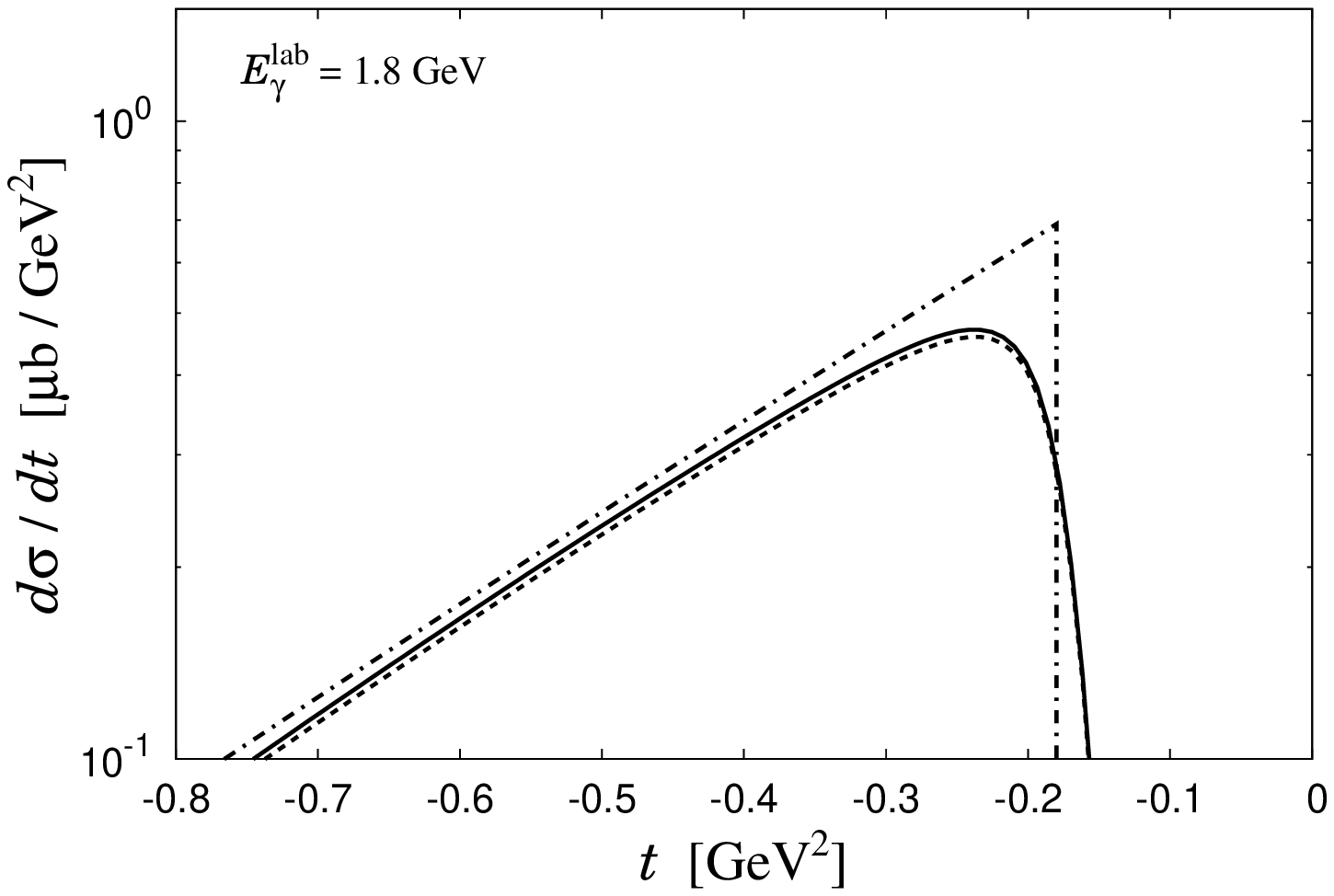} 
    \vspace{-8pt} 
    \\
    \includegraphics[width=8.4cm]{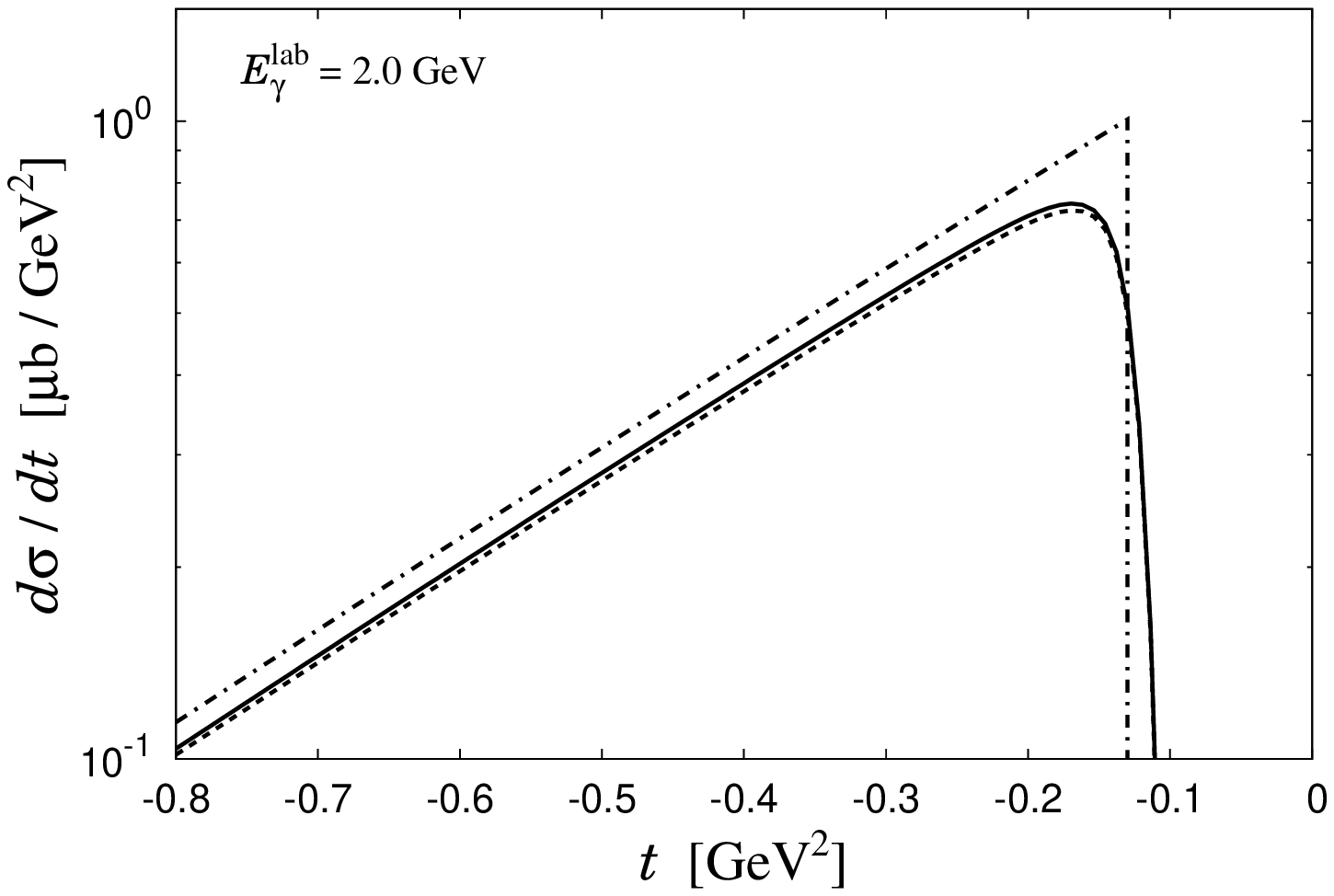} & 
    \includegraphics[width=8.4cm]{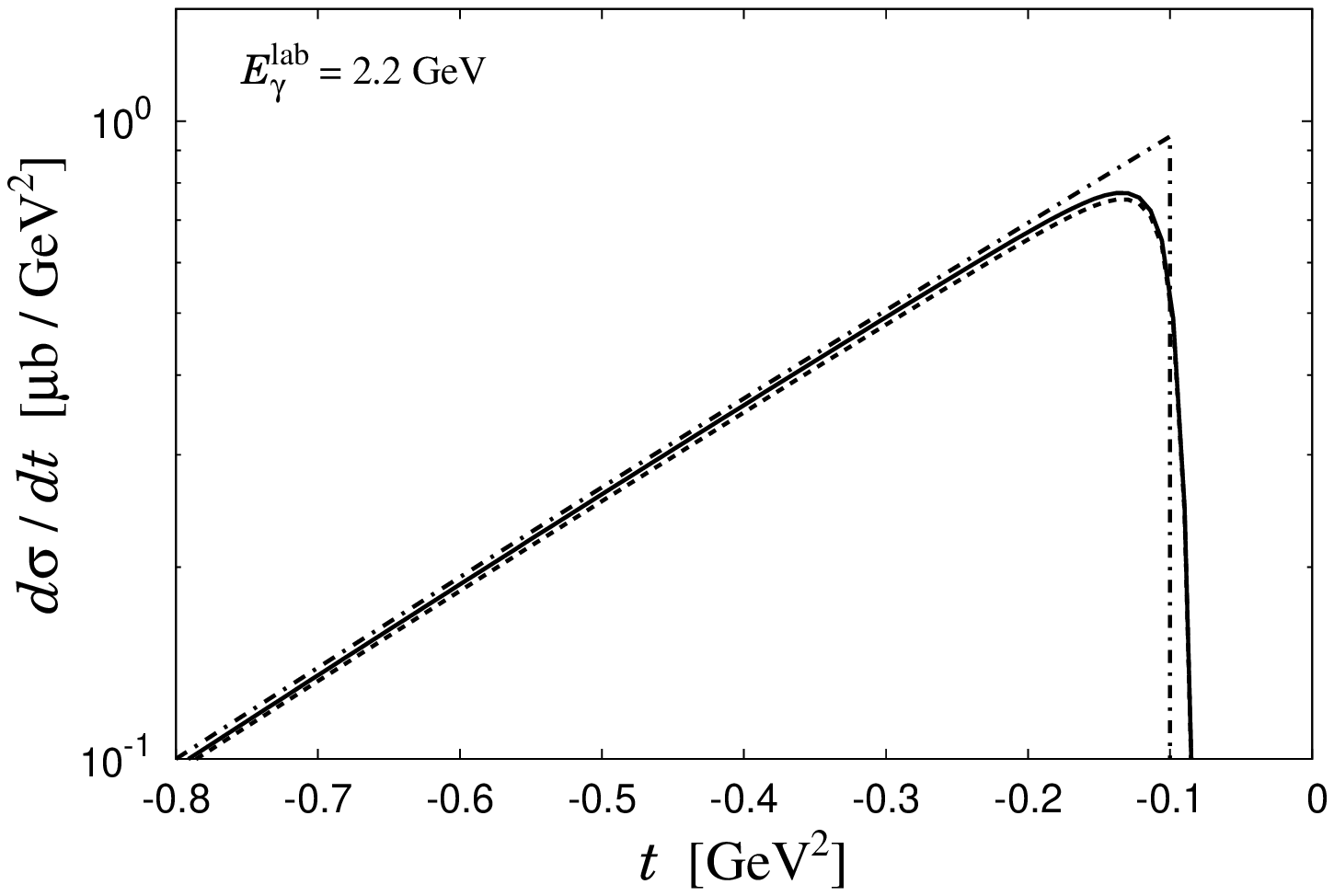} 
  \end{tabular*}
  \caption{ \label{fig:dSdt} Differential cross sections $d\sigma
    _{p}/dt$ and $d\sigma _{p^{\ast}}/dt_{\phi}$ as functions of $t$
    ($t_{\phi}$) for different $E_{\gamma}^{\text{lab}}$. Solid and
    dashed lines indicate the single and single plus double scattering
    case of the deuteron target, and dash-doted line the case of the
    free proton target, respectively. }
\end{figure*}

For each photon energy, we show the differential cross sections in
Fig.~\ref{fig:dSdt}, where we plot $d \sigma /dt$ in the $|t|\le 0.8
\, \text{GeV}^{2}$ region, so as to clarify the behavior of the
differential cross sections around the small $|t|$ region. As one can
see from the figure, $d\sigma _{p^{\ast}}/dt_{\phi}$ shows the smooth
decrease around $t_{\phi , \text{max}}$, which is not seen in $d\sigma
_{p}/dt$ around $t_{\text{max}}$. Also we note that in this $t_{\phi}$
region for the $\gamma d \to \phi p n$ reaction the double scattering
amplitude interferes destructively with the single scattering
amplitude, although $d\sigma _{p^{\ast}}/dt_{\phi}$ is dominated by
the single scattering amplitude.

The results that we find about the accuracy of the single scattering
  to reproduce the photoproduction cross section in deuteron agree
  with the conclusions obtained in Ref.~\cite{haiyan}.

\subsection{Nucleon rescattering effects}

In the previous subsection we have shown the differential cross sections
for the reactions $\gamma p \to \phi p$ and $\gamma d \to \phi p n$.
From our results we have found that the ``in-medium'' $\phi$
propagation in deuteron, which is realized as a $\phi n \to \phi n$
rescattering in the reaction, has only small contributions to the
suppression of the cross section $d\sigma _{p^{\ast}}/dt_{\phi}$
compared to $d\sigma _{p}/dt$.

\begin{figure}[t]
  \centering
    \includegraphics[scale=0.18]{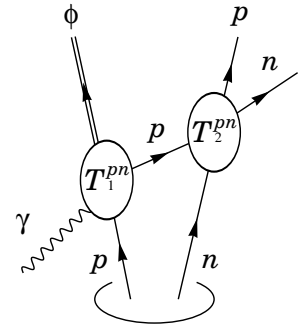} 
  \caption{Diagrams for the nucleon rescattering effects in 
    $\gamma d \to \phi p n$ reaction.}
  \label{fig:gamma-d3}
\end{figure}

Next, let us consider another double-scattering contribution to the
cross section $d\sigma _{p^{\ast}}/dt_{\phi}$, that is, the nucleon
rescattering effects.  In the deuteron target case, this can be 
taken into account by considering a $pn \to pn$ rescattering as a 
final state interaction, which is diagrammatically shown in 
Fig.~\ref{fig:gamma-d3}.  The scattering amplitude can be evaluated 
in an analogous way as the $\phi$ propagation diagrammatically 
shown in Fig.~\ref{fig:gamma-d-each}(b), and can be written as, 
\begin{align}
  T^{pn} = & T_{1}^{pn} \int \frac{d^{3} q_{\rm ex}^{\prime}}{( 2 \pi
    )^{3}} \frac{M_{p}}{E_{\rm ex}^{\prime}} \frac{\tilde{\varphi} (|
    \vec{p}_{p} + \vec{p}_{n} - \vec{q}_{\rm ex}^{\; \prime} |)}
  {q_{\rm ex}^{\prime 0} - E_{\rm ex}^{\prime} + i \epsilon}
  T_{2}^{pn} (M_{pn} , \, \theta _{pn} ) ,
% \nonumber \\ & \times 
\end{align}
where, 
\be
T_{1}^{pn} = - a_{p} ( W^{2} ) \exp (b \tilde{t}^{pn} / 2) 
\times \vec{\epsilon} (\gamma ) \cdot \vec{\epsilon} (\phi ) , 
\ee
\be
\tilde{t}^{pn} = (p_{\phi} - k)^{2} - t_{\text{max}} ( W^{2} ), 
\ee
\be
q_{\text{ex}}^{\prime 0} 
= E_{\gamma}^{\text{lab}} + M_{p} - B_{1/2} - \omega _{\phi}^{\text{lab}} , 
\ee
\be
E_{\rm ex}^{\prime} = \sqrt{M_{p}^{2} + |\vec{q}_{\rm ex}^{\prime}|^{2} } , 
\ee
with $W^{2}$ defined in Eq.~(\ref{eq:definition_W}) and 
the $\phi$ energy in the laboratory frame $\omega _{\phi}^{\text{lab}}$.  
The $pn \to pn$ scattering amplitude, $T_{2}^{pn}$, depends on 
the $p$-$n$ invariant mass $M_{pn}$ as well as the scattering 
angle $\theta _{pn}$ in the $p$-$n$ center-of-mass frame, and will 
be determined later.  For the proton propagator we use an ``on-shell'' 
approximation, 
\begin{align}
& \frac{M_{p}}{E_{\rm ex}^{\prime}}
\frac{1}{q_{\text{ex}}^{\prime 0} - E_{\rm ex}^{\prime} + i \epsilon } 
% \nonumber \\ & 
\to - 2 i \pi M_{p} \delta ((q_{\text{ex}}^{\prime})^{2} - M_{p}^{2} ) ,
\end{align}
as for the $\phi$ propagator in the $\phi$ exchange amplitude. 
Then the scattering amplitude can be rewritten as, 
\begin{align}
  T^{pn} = & \frac{- i M_{p} T_{1}^{pn}}{(2 \pi )^{2}} \int
  _{0}^{\infty} d |\vec{q}_{\rm ex}^{\; \prime}| |\vec{q}_{\rm ex}^{\;
    \prime}|^{2} \delta (|\vec{q}_{\rm ex}^{\; \prime}|^{2} +
  M_{p}^{2} - (q_{\text{ex}}^{\prime 0})^{2} )
  \nonumber \\
  & \times \int d \Omega _{\rm ex}^{\prime} \tilde{\varphi} (| \vec{p}_{p} +
  \vec{p}_{n} - \vec{q}_{\rm ex}^{\; \prime}|) T_{2}^{pn} ( M_{pn} ,
  \, \theta _{pn} )
  \nonumber \\
  = & \frac{- i M_{p} T_{1}^{pn}}{8 \pi ^{2}}
  \sqrt{q_{\text{ex}}^{\prime 0} - M_{p}^{2}} \int d \Omega _{\rm
    ex}^{\prime} \tilde{\varphi} (| \vec{p}_{p} + \vec{p}_{n} - \vec{q}_{\rm
    ex}^{\; \prime} |)
  \nonumber \\
  & \times T_{2}^{pn} ( M_{pn} , \, \theta _{pn} ) .
\end{align} 
Here we emphasize that the solid angle $\Omega _{\rm ex}$ and 
the momenta $\vec{p}_{p}$, $\vec{p}_{n}$, and $\vec{q}_{\rm ex}^{\; \prime}$ 
are evaluated in the laboratory frame, whereas $\theta _{pn}$ 
is evaluated in the $p$-$n$ center-of-mass frame as 
$\cos \theta _{pn} = (\vec{q}_{\rm ex}^{\; \prime} \cdot \vec{p}_{p})_{pn} 
/ (|\vec{q}_{\rm ex}^{\; \prime} | |\vec{p}_{p} | )_{pn} $.  

For the determination of the $pn \to pn$ scattering amplitude, we take
the following procedure.  First, in order to take into account the
angular dependence we parameterize the $pn \to pn$ differential cross
section as follows:
\be \frac{d \sigma _{pn \to pn}}{d \Omega _{pn}} (M_{pn} , \,
\theta _{pn}) = {\cal A} (M_{pn}) + {\cal B} (M_{pn}) \cos ^{2} \theta
_{pn} .  
\ee 
Here spin average and sum are assumed to be taken in the initial and
final states, respectively.  The parameters ${\cal A}$ and ${\cal B}$
are determined so as to reproduce the differential cross section.  In
this study we use experimental data for the total cross section
from~\cite{Amsler:2008zzb} and model calculation for the differential
cross section at $\theta _{pn} = 90$ degrees from~\cite{NNonline}.
Next we evaluate the real part of the $pn \to pn$ scattering amplitude
by assuming that the magnitude of the real part is larger than that of
the imaginary part and that the real part dominates the cross section,
which is certainly the case at low energies, of relevance to the
present problem, where $\text{Im}\, T_{2}^{pn}$ goes to zero:
\be
\text{Re} \, T_{2}^{pn} (M_{pn} , \, \theta _{pn}) 
= \sqrt{\mathcal{A} + {\cal B} \cos ^{2} \theta _{pn}}
\frac{2 \pi M_{pn}}{M_{p} M_{n}} . 
\label{eq:ReTpn}
\ee
The imaginary part of the $pn \to pn$ amplitude, on the other hand, 
is evaluated so as to satisfy the optical theorem and to have the same 
angular dependence as the real part as, 
\begin{align}
& \text{Im} \, T_{2}^{pn} (M_{pn}^{2} , \theta _{pn}) 
\nonumber \\
& = - \sqrt{\frac{{\cal A} + {\cal B} \cos ^{2} \theta _{pn}}
{{\cal A} + {\cal B}}}
\frac{p_{\text{cm}}^{pn} M_{pn}}{2 M_{p} M_{n}} 
\sigma _{pn \to X} (M_{pn}) , 
\label{eq:ImTpn}
\end{align}
where $p_{\text{cm}}^{pn}$ is the center-of-mass momentum of the $p$-$n$
system and the $pn \to X$ total cross
section $\sigma _{pn \to X}$ is obtained by fitting the experimental data
given in~\cite{Amsler:2008zzb}.  

Now let us see how the whole $\gamma d \to \phi p n$ reaction is 
affected by the $pn \to pn$ rescattering effect, for which 
the amplitude can be expressed as, 
\begin{align}
  T^{pn} = & \frac{M_{p} T_{1}^{pn}}{8 \pi ^{2}}
  \sqrt{q_{\text{ex}}^{\prime 0} - M_{p}^{2}} \int d \Omega _{\rm
    ex}^{\prime} \tilde{\varphi} (| \vec{p}_{p} + \vec{p}_{n} - \vec{q}_{\rm
    ex}^{\; \prime} |)
  \nonumber \\
  & \times [ \text{Im} \, T_{2}^{pn} ( M_{pn} , \, \theta _{pn} ) 
- i \text{Re} \, T_{2}^{pn} ( M_{pn} , \, \theta _{pn} ) ] .
\label{eq:Tpn}
\end{align}
First, we note that the first scattering amplitude, $T_{1}^{pn}$,
shows the same $t_{\phi}=(p_{\phi}-k)^{2}$ dependence as the impulse
approximation, $T^{\text{ss}}$.  This indicates that, in contrast to
the $\phi$ rescattering, the $pn$ rescattering effect has a
possibility to become large in the small $|t_{\phi}|$ region.  Next,
the imaginary part of the $pn \to pn$ amplitude, $\text{Im} \,
T_{2}^{pn}$, interferes destructively with the single scattering
contribution, as one can see from comparing Eqs.~(\ref{eq:TssTds}) and
(\ref{eq:Tpn}) (note that $\text{Im} \, T_{2}^{pn} \le 0$ due to the
optical theorem).  This means that in the reaction $\gamma d \to \phi
p n$ the final-state proton and neutron are distorted by the $pn$
rescattering, changing directions of the final proton and neutron with
regard to single scattering.  However, the cross section for the
$\gamma d \to \phi p n$ should not be changed by the inclusion of the
$pn$ rescattering if one observes the final-state proton and neutron
in the whole solid angle (or equivalently one does not observe the
proton and neutron), since the proton and neutron cannot disappear in
the rescattering process, in contrast to the $\phi n \to \phi n$
rescattering where only the large $\phi$ absorption part of the
amplitude was considered (see Appendix).  The recovery of the cross
section after the $pn \to pn$ rescattering is realized by the term
containing the real part of the $pn$ amplitude, $\text{Re} \,
T_{2}^{pn}$, which is the sole pure imaginary term in
Eq.~(\ref{eq:Tpn}) and is added incoherently to the $\gamma d \to \phi
p n$ cross section, compensating the destructive interference of the
term containing $\text{Im} \, T_{2}^{pn}$.

\begin{figure*}[!t]
  \centering
  \begin{tabular*}{\textwidth}{@{\extracolsep{\fill}}cc}
    \includegraphics[width=8.4cm]{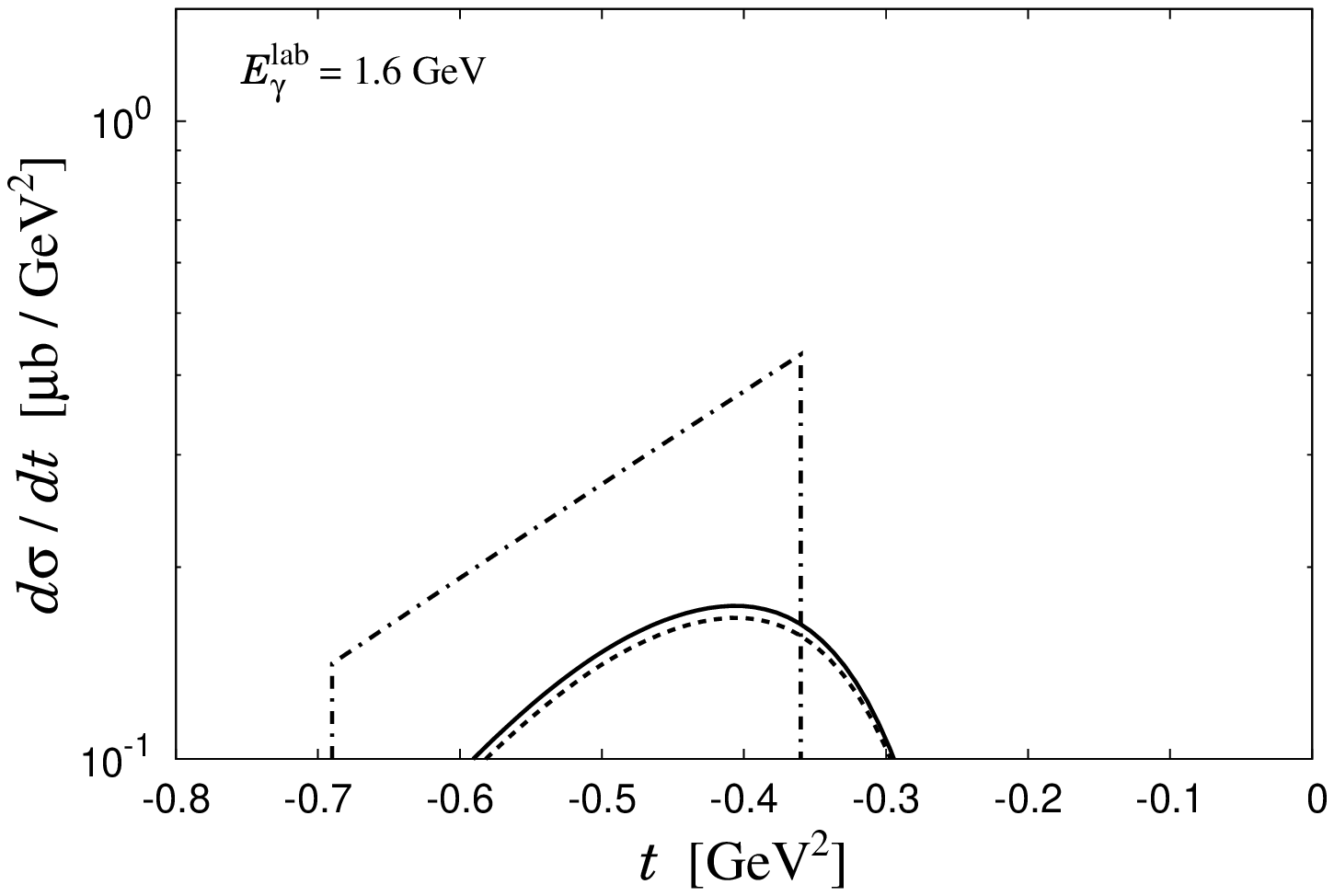} &
    \includegraphics[width=8.4cm]{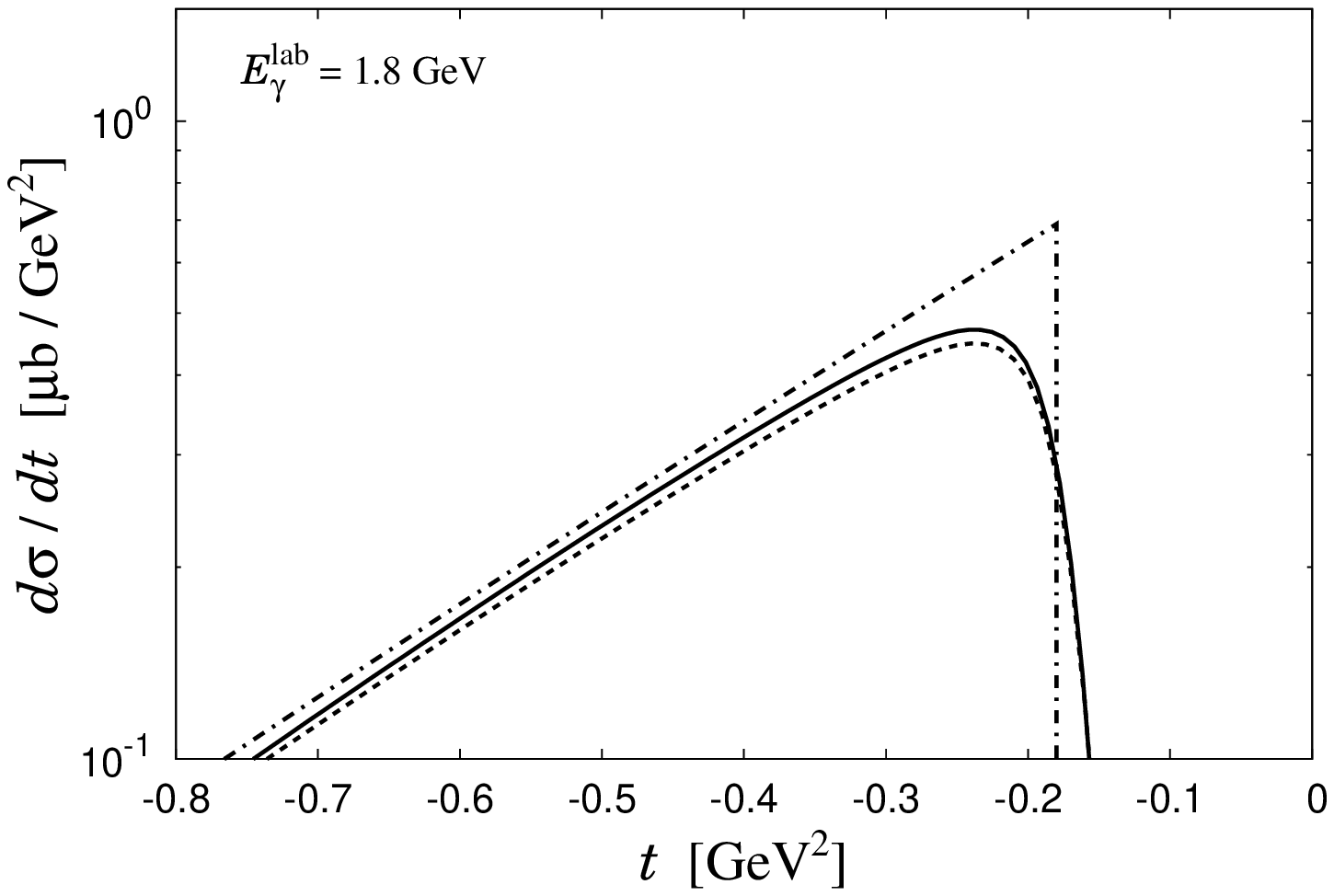} 
    \vspace{-8pt} 
    \\
    \includegraphics[width=8.4cm]{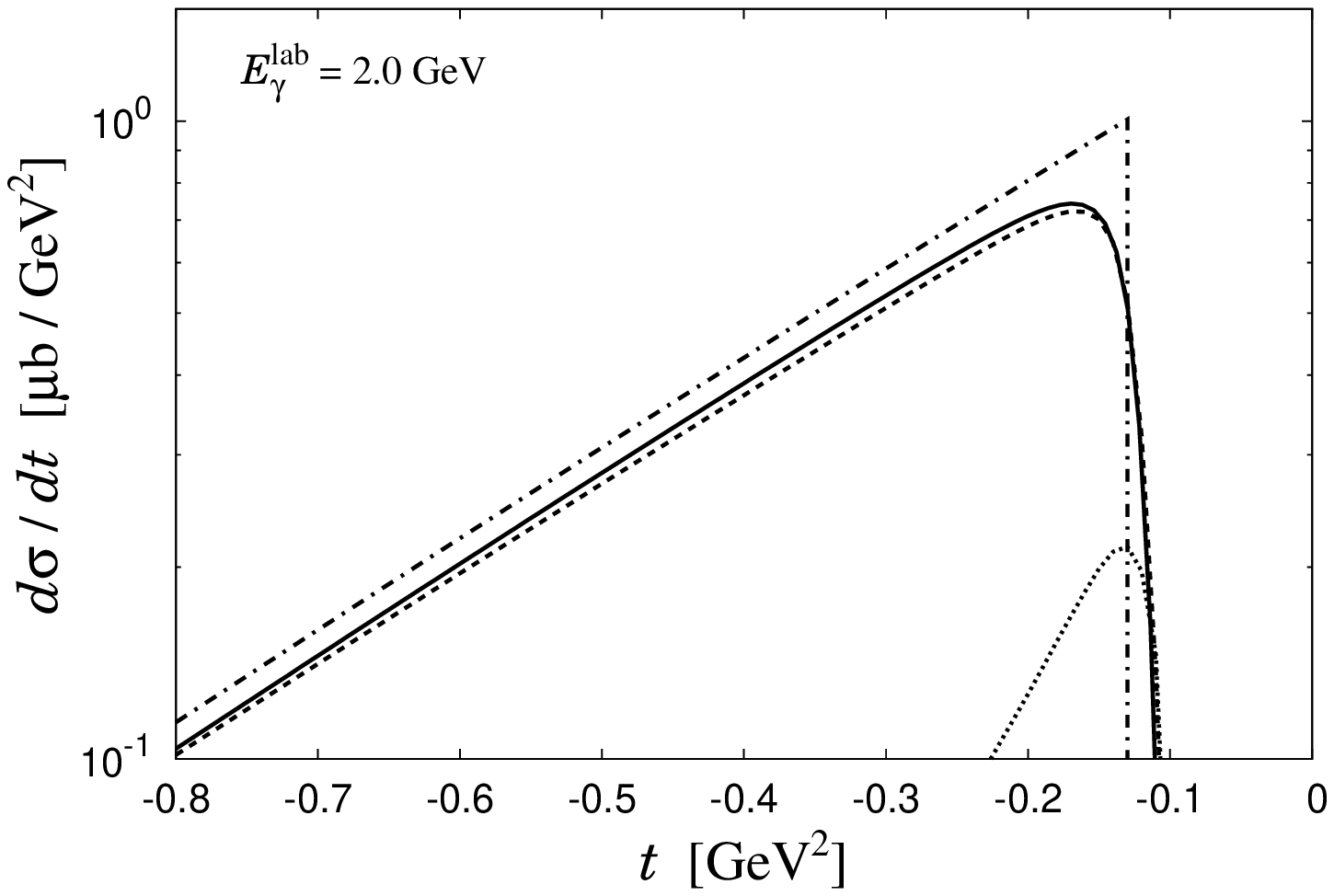} &
    \includegraphics[width=8.4cm]{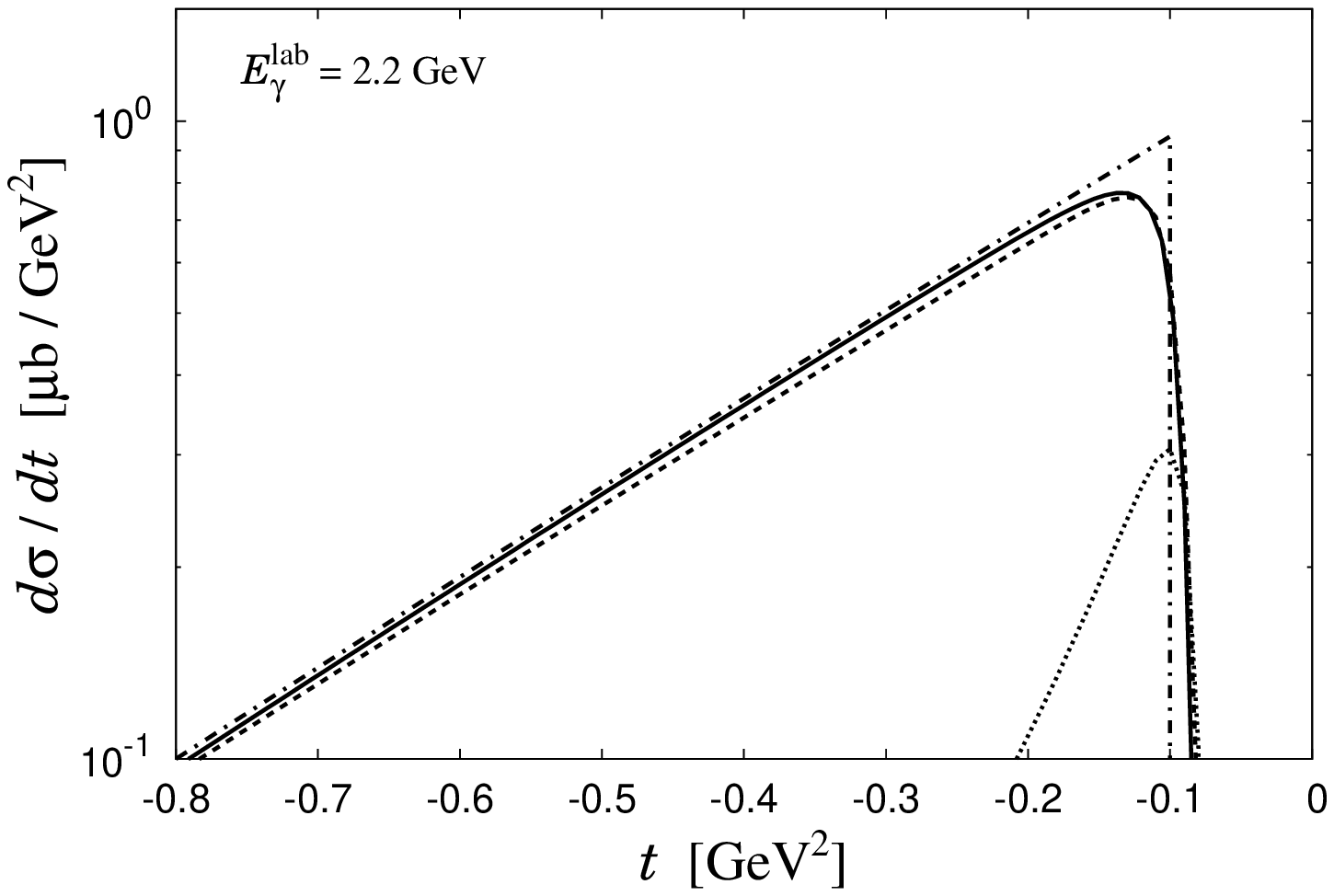} 
  \end{tabular*}
  \caption{ \label{fig:dSdt-pn} Differential cross sections $d\sigma
    _{p}/dt$ and $d\sigma _{p^{\ast}}/dt_{\phi}$ as functions of $t$
    ($t_{\phi}$) for different $E_{\gamma}^{\text{lab}}$. Solid,
    dashed, and dotted lines indicate the single, single plus two
    double scatterings ($\phi$ and proton exchanges), and only
    proton-exchange scattering case of the deuteron target, and
    dash-doted line the case of the free proton target, respectively.}
\end{figure*}

The numerical results of $\gamma d \to \phi p n$ cross section with
the coherent sum of the three amplitudes (one single and two double
scatterings, $\phi$ and proton exchanges), are shown in
Fig.~\ref{fig:dSdt-pn} by the dashed line.  The contribution only from
the proton rescattering is also plotted as the dotted line.  As one
can see from Fig.~\ref{fig:dSdt-pn}, the values of the cross section
are almost unchanged by the inclusion of the proton rescattering
effect.  This takes place due to, as explained before, the competition
between the imaginary part of the $pn \to pn$ amplitude, which produces
destructive interference with the single scattering, and the real part
of the $pn \to pn$ amplitude, which contributes incoherently to 
the cross section.

\begin{figure}[t]
  \centering
    \includegraphics[width=8.6cm]{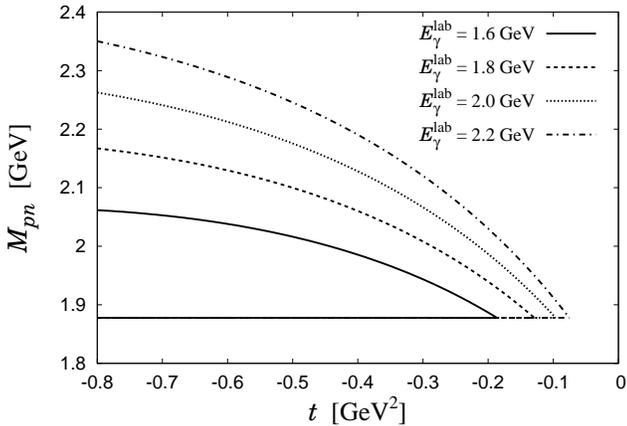}
    \caption{Maximal and minimal values of the achievable $p$-$n$
      invariant mass $M_{pn}$ for the $\gamma d \to \phi p n$ reaction
      in each photon energy. }
  \label{fig:phasespace}
\end{figure}

However, the contribution from the proton rescattering itself is not
negligible compared to the cross section only with the single
scattering, as seen by the dotted line in Fig.~\ref{fig:dSdt-pn}.  It
is important to note that the contribution from the proton
rescattering gets large as $t_{\phi}$ approaches the $t_{\phi ,
  \text{max}}$.  One of the reason is that the amplitude $T^{pn}$
contain a factor $\exp (b \tilde{t}^{pn}/2)$, which shows the same
$t_{\phi}$ dependence as the single scattering amplitude and becomes
large as $t_{\phi}$ goes close to $t_{\phi , \text{max}}$.  In
addition to this, we note that the magnitude of the $pn \to pn$ amplitude
$T_{2}^{pn}$ gets large for $t_{\phi} \to t_{\phi , \text{max}}$.
This is because, in the region $t_{\phi} \simeq t_{\phi , \text{max}}$
the $p$-$n$ invariant mass $M_{pn}$ is very close to the threshold
$M_{p}+M_{n}$, as one can see from Fig.~\ref{fig:phasespace}, in
which the maximal as well as the minimal values of the achievable
$M_{pn}$ for fixed $t_{\phi}$ in each photon energy are plotted.
Since the cross section grows rapidly as the $p$-$n$ invariant mass
gets close to the threshold~\cite{Amsler:2008zzb} and the $pn \to pn$
amplitude is determined from the $pn \to pn$ cross section as in
Eqs.~(\ref{eq:ReTpn}) and (\ref{eq:ImTpn}), the $pn \to pn$ scattering
amplitude also grows rapidly as the invariant mass $M_{pn}$ approaches
the threshold, or equivalently $t_{\phi}$ gets close to $t_{\phi ,
  \text{max}}$.

\begin{figure}[t]
  \centering
  \begin{tabular*}{8.6cm}{@{\extracolsep{\fill}}cc}
    \includegraphics[scale=0.33]{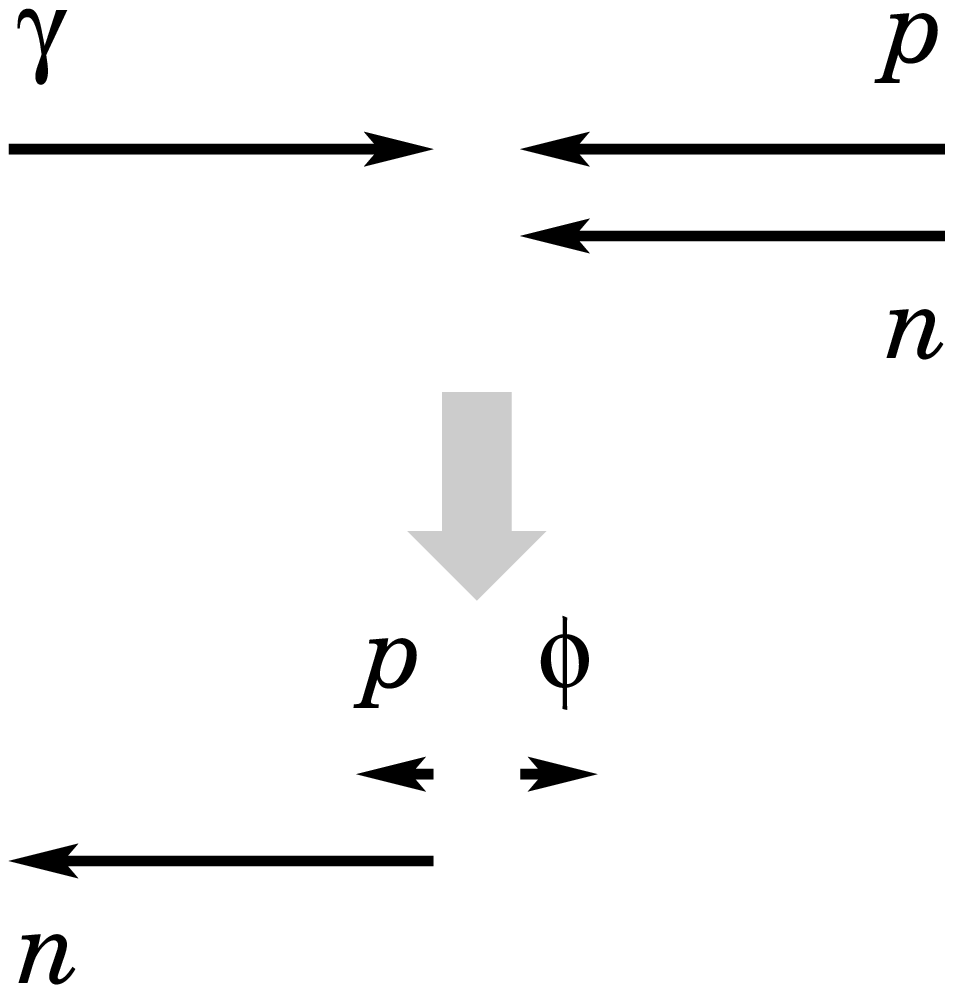} & 
    \includegraphics[scale=0.33]{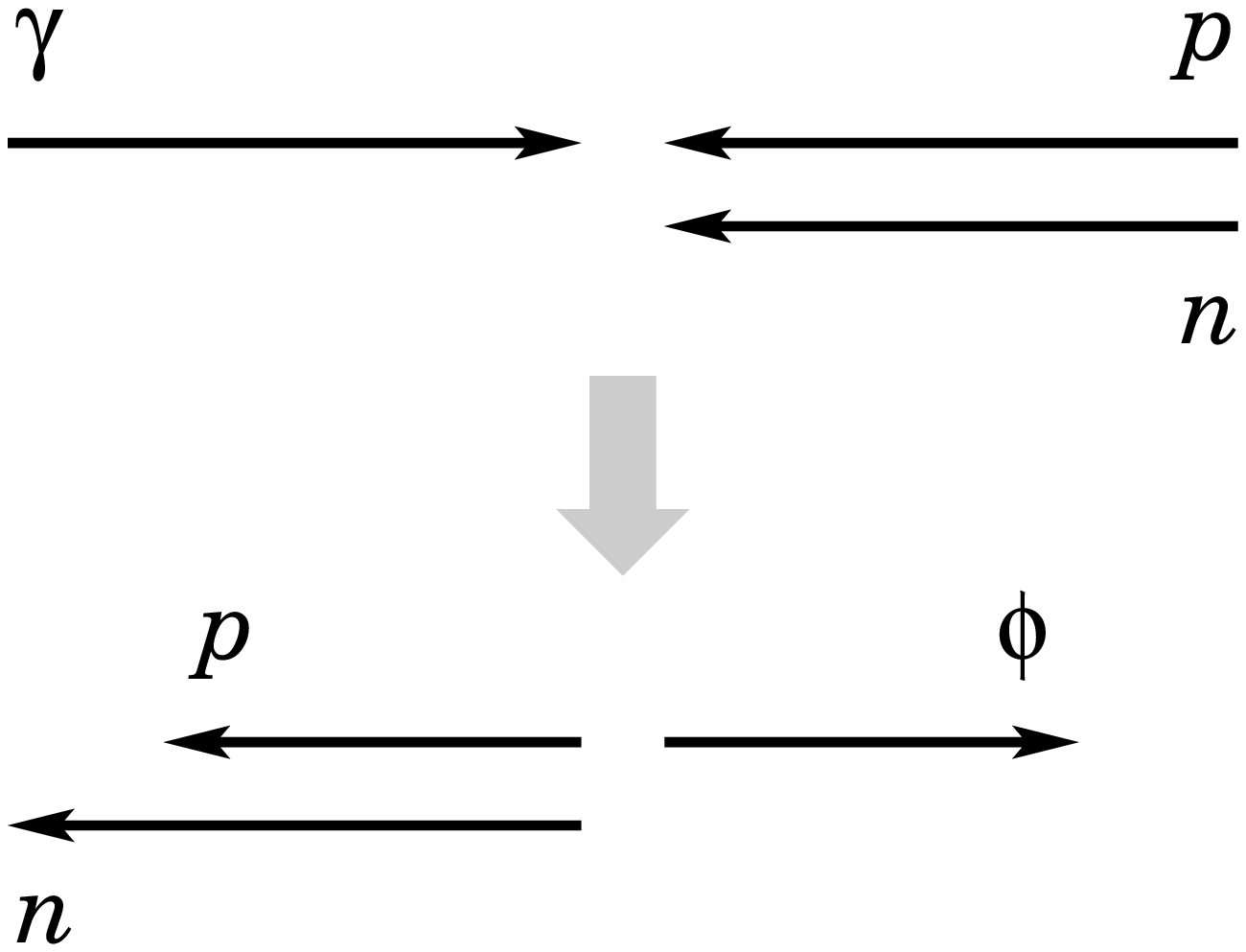} \\
    (a) & (b)
  \end{tabular*}
  \caption{Kinematics for the $\gamma d \to \phi p n$ reaction at
    $t_{\phi} \simeq t_{\phi , \text{max}}$ with (a) photon energy
    close to the $\phi$ photoproduction threshold, and (b)
    appropriately large photon energy.  Length of the arrows
    corresponds to magnitude of momenta. }
  \label{fig:kinematics_threshold}
\end{figure}

\begin{figure*}[t]
  \centering
  \begin{tabular*}{\textwidth}{@{\extracolsep{\fill}}cc}
    \includegraphics[width=8.4cm]{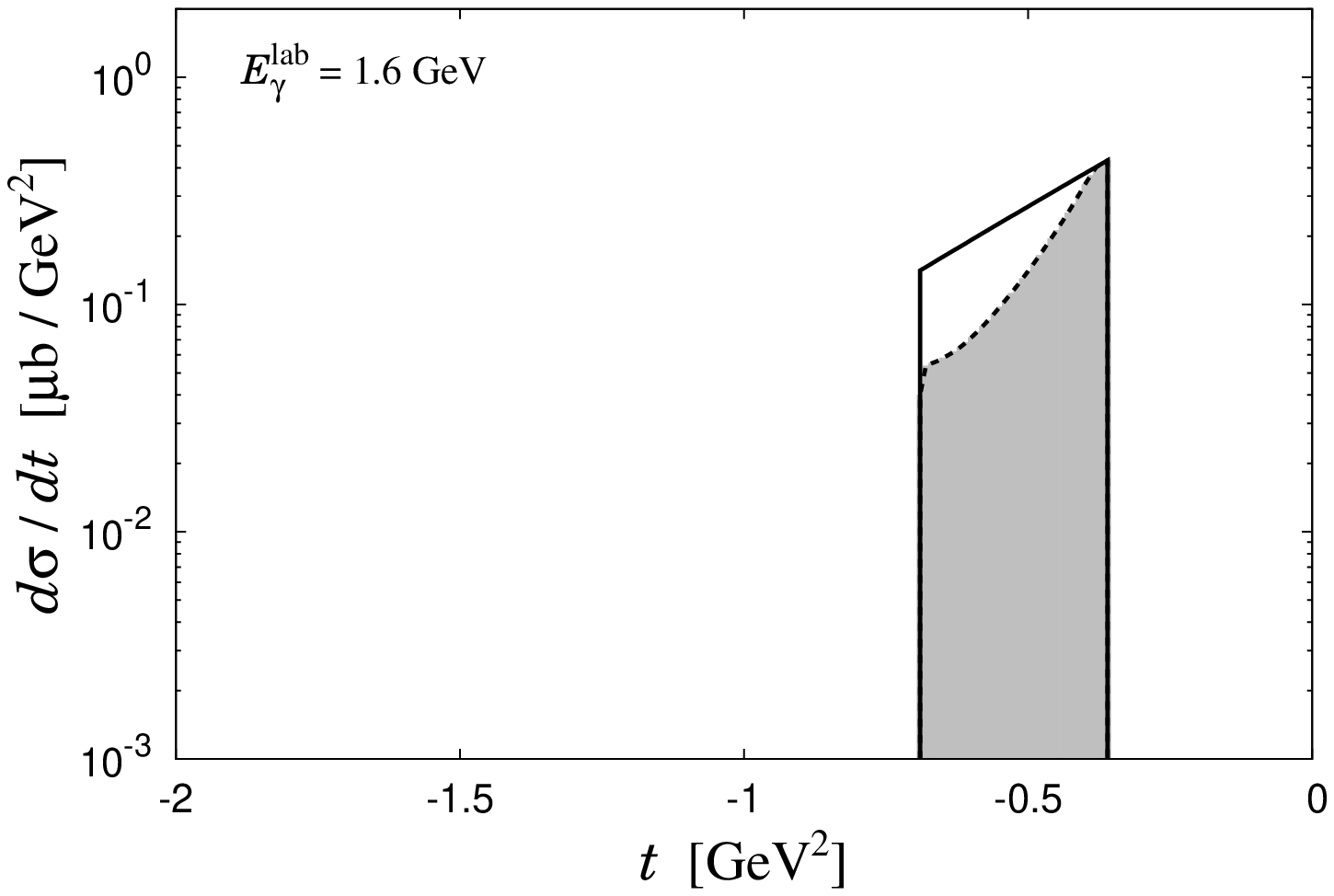} & 
    \includegraphics[width=8.4cm]{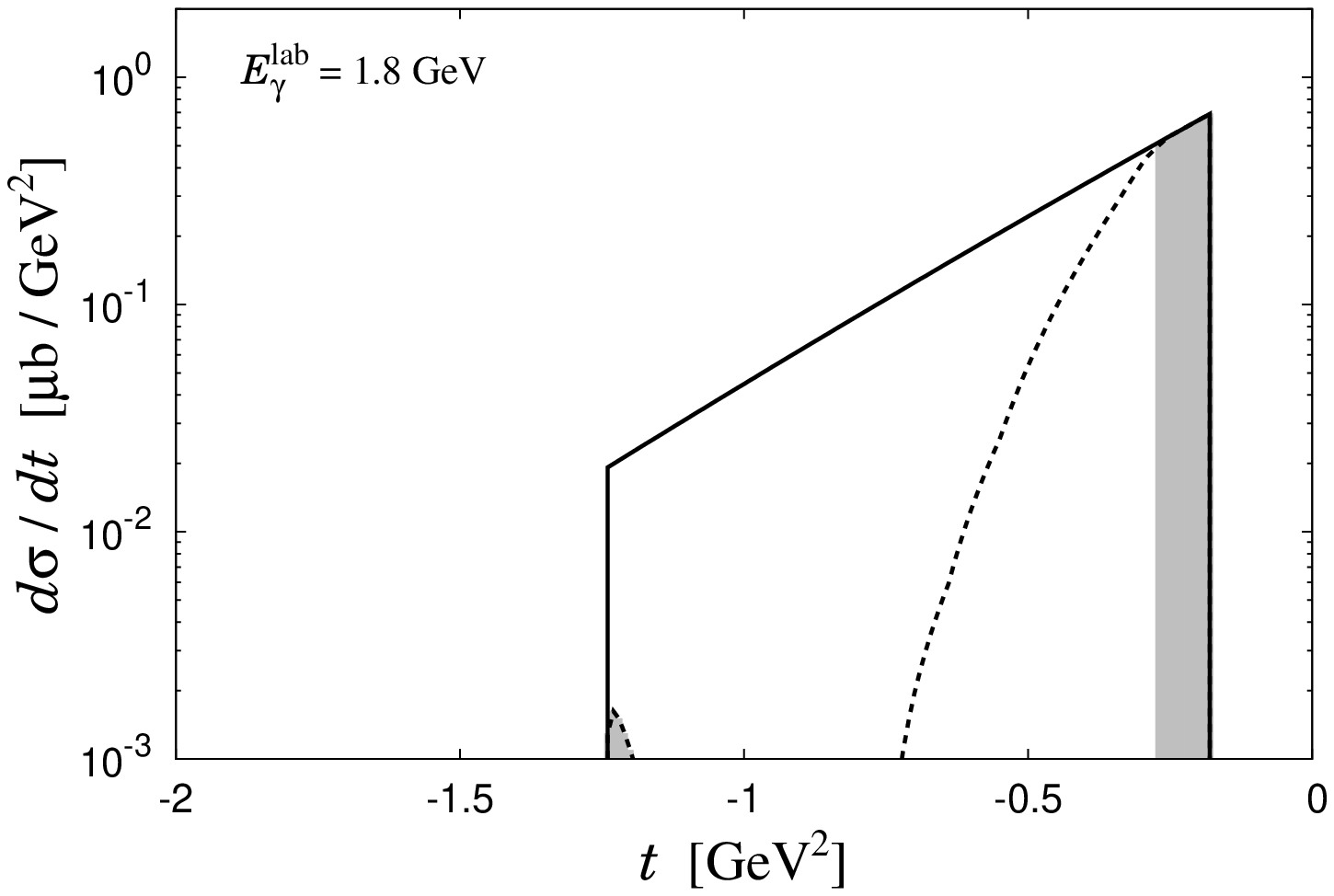} 
    \vspace{-8pt} 
    \\
    \includegraphics[width=8.4cm]{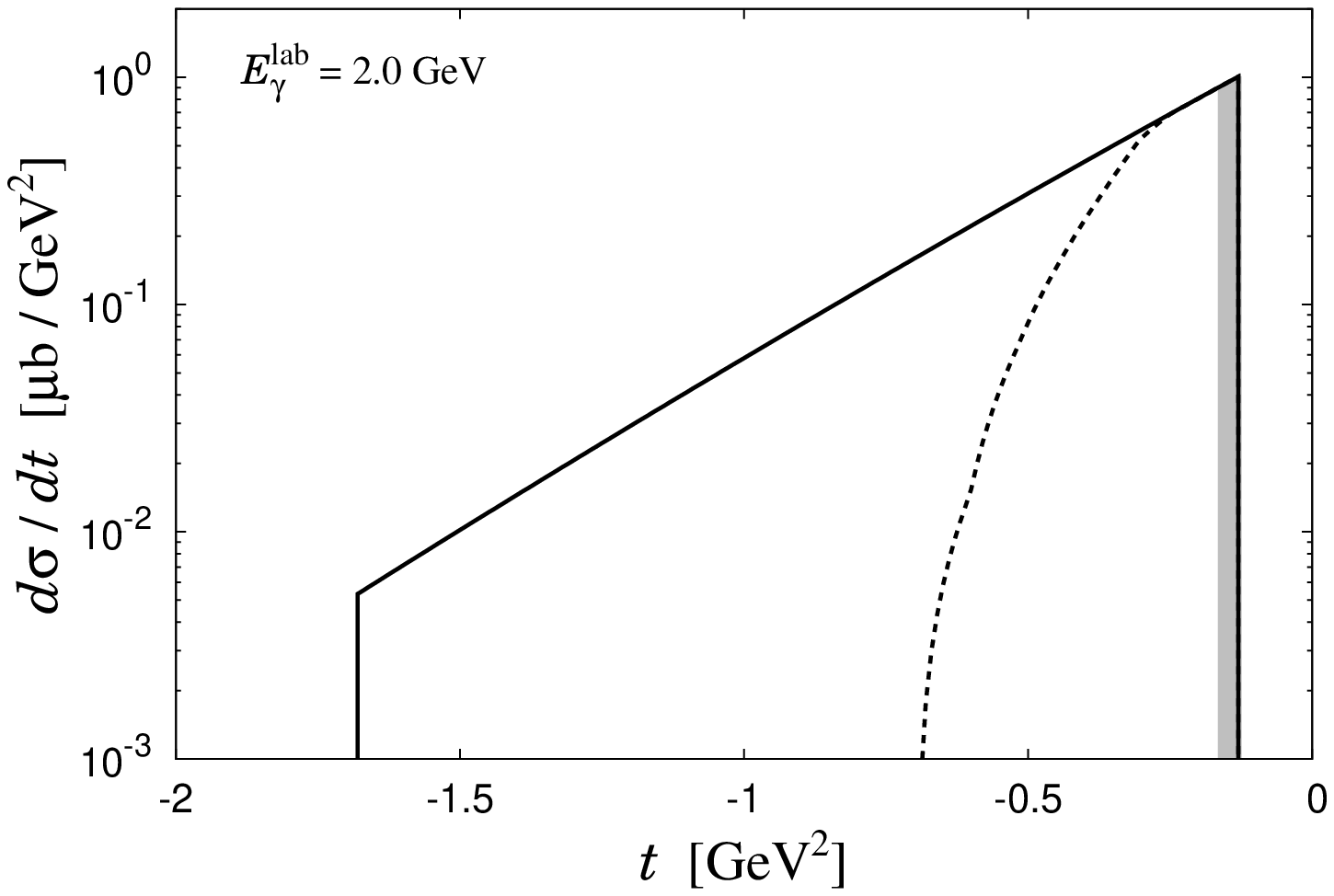} & 
    \includegraphics[width=8.4cm]{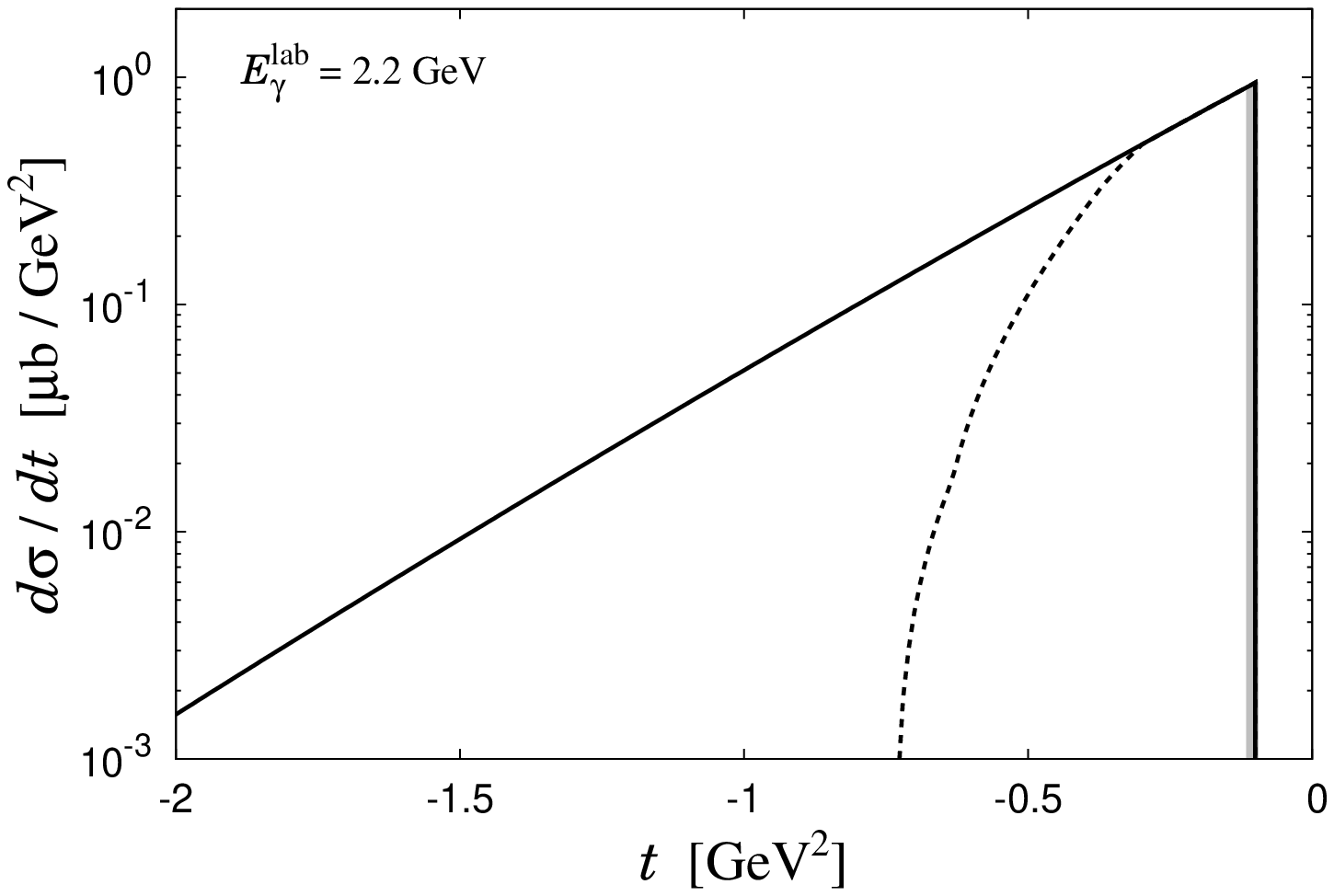} 
  \end{tabular*}
  \caption{ \label{fig:dSdt-p-cut} Differential cross section $d\sigma
    _{p}/dt$ as a function of $t$ for different
    $E_{\gamma}^{\text{lab}}$. The solid and dashed lines, and shaded
    area indicate the case without any angular cuts, with the $\phi$
    cut, and with the $\phi$ and proton cuts, respectively. }
\end{figure*}
  
We also note that the proton rescattering effect increases as the
photon energy increases.  The reason why this occurs is that the
difference between $t_{\phi , \text{max}}$ and $t_{\text{max}}$, the
highest value of $t=(p_{\phi}-k)^{2}$ for the free proton target case,
becomes small (as one can see from Fig.~\ref{fig:dSdt}) and hence
$M_{pn}$ ($t_{\phi}$) can easily go close to the threshold
$M_{p}+M_{n}$ ($t_{\phi , \text{max}}$) without help by the Fermi
motion.  In order to see this, we schematically show the reaction
kinematics for $\gamma p^{\ast} \to \phi p$ with $t_{\phi} \simeq
t_{\text{max}}$ in the photon-bound proton center-of-mass frame in
Fig.~\ref{fig:kinematics_threshold}, where the Fermi motion is
neglected.  For the photon energy close to the $\phi$ photoproduction
threshold [Fig.~\ref{fig:kinematics_threshold}(a)], $M_{pn}$ is large
at $t_{\phi}\simeq t_{\text{max}}$ because after the reaction the
neutron momentum is large while the proton momentum is small in the
photon-bound proton center-of-mass frame.  In this photon energy
region, one can go close to the threshold $M_{pn} \to M_{p}+M_{n}$
($t_{\phi} \to t_{\phi , \text{max}}$) by the help of high momentum 
components of the Fermi motion.  For higher photon energy
[Fig.~\ref{fig:kinematics_threshold}(b)], on the other hand, $M_{pn}$
is very close to the threshold at $t_{\phi} \simeq t_{\text{max}}$
because the neutron and proton momenta after the reaction are very
similar to each other.  Hence, in this photon energy region one can go
close to the threshold $M_{pn} \to M_{p}+M_{n}$ ($t_{\phi} \to t_{\phi
  , \text{max}}$) without large momenta of the Fermi motion compared to the
photon energy close to the $\phi$ photoproduction threshold.  This
means that the proton rescattering contribution becomes large as the
photon energy becomes higher as shown in the dotted line in
Fig.~\ref{fig:dSdt-pn}, recalling that the $pn \to pn$ amplitude gets
much larger if one approaches the threshold, $M_{pn} \to M_{p}+M_{n}$
($t_{\phi} \to t_{\phi , \text{max}}$).

\begin{figure*}[!t]
  \centering
  \begin{tabular*}{\textwidth}{@{\extracolsep{\fill}}cc}
    \includegraphics[width=8.4cm]{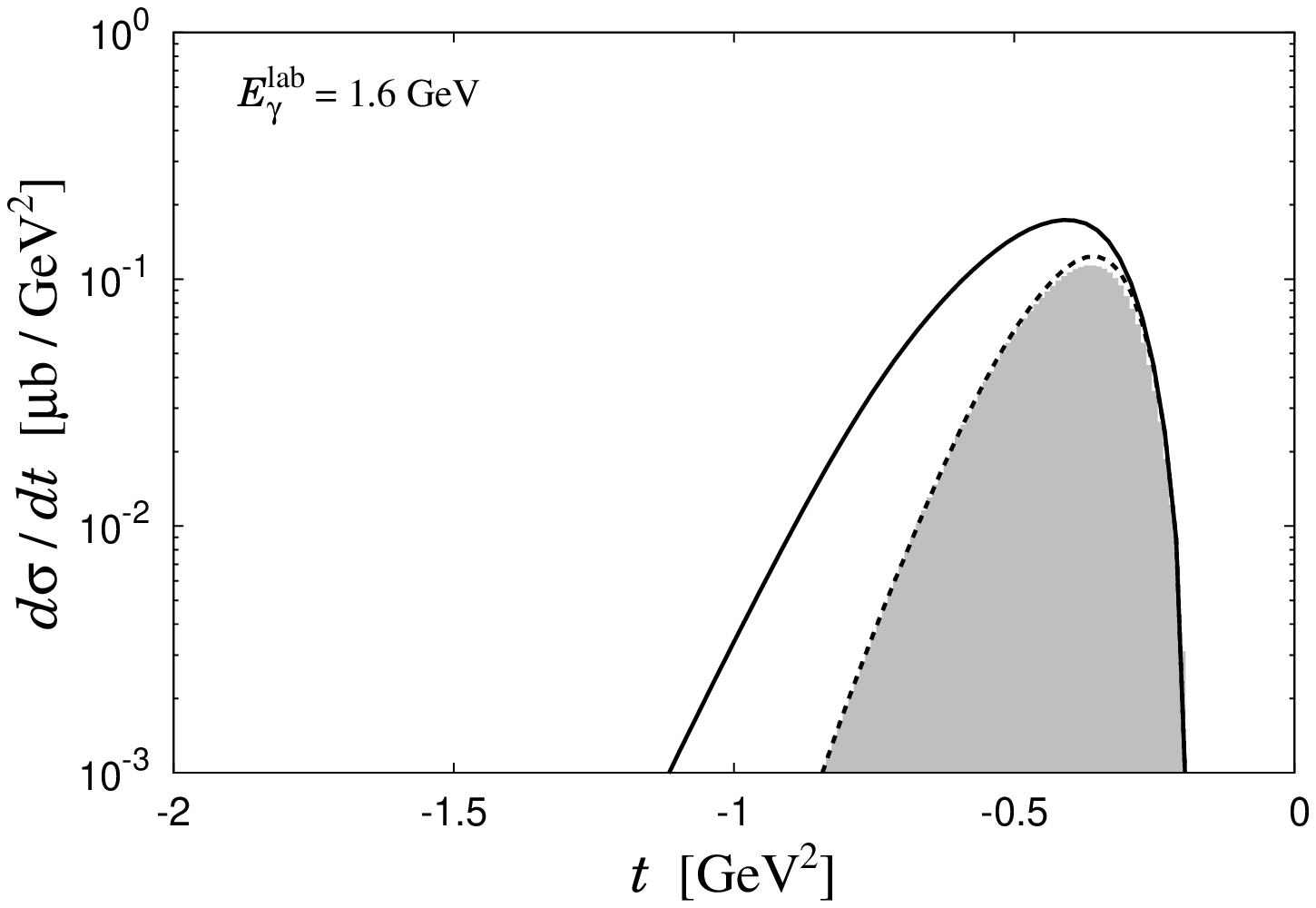} & 
    \includegraphics[width=8.4cm]{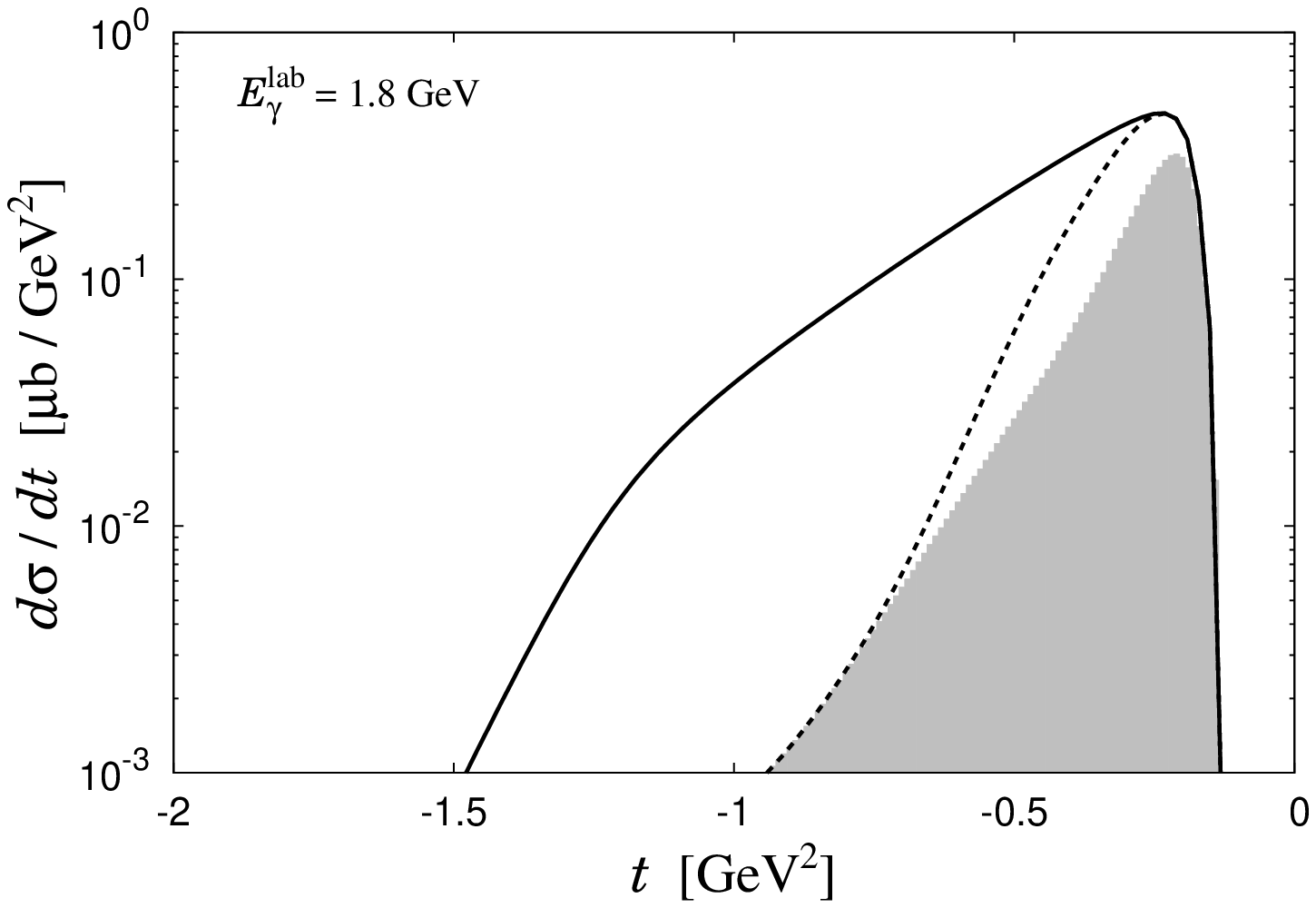} 
    \vspace{-8pt} 
    \\
    \includegraphics[width=8.4cm]{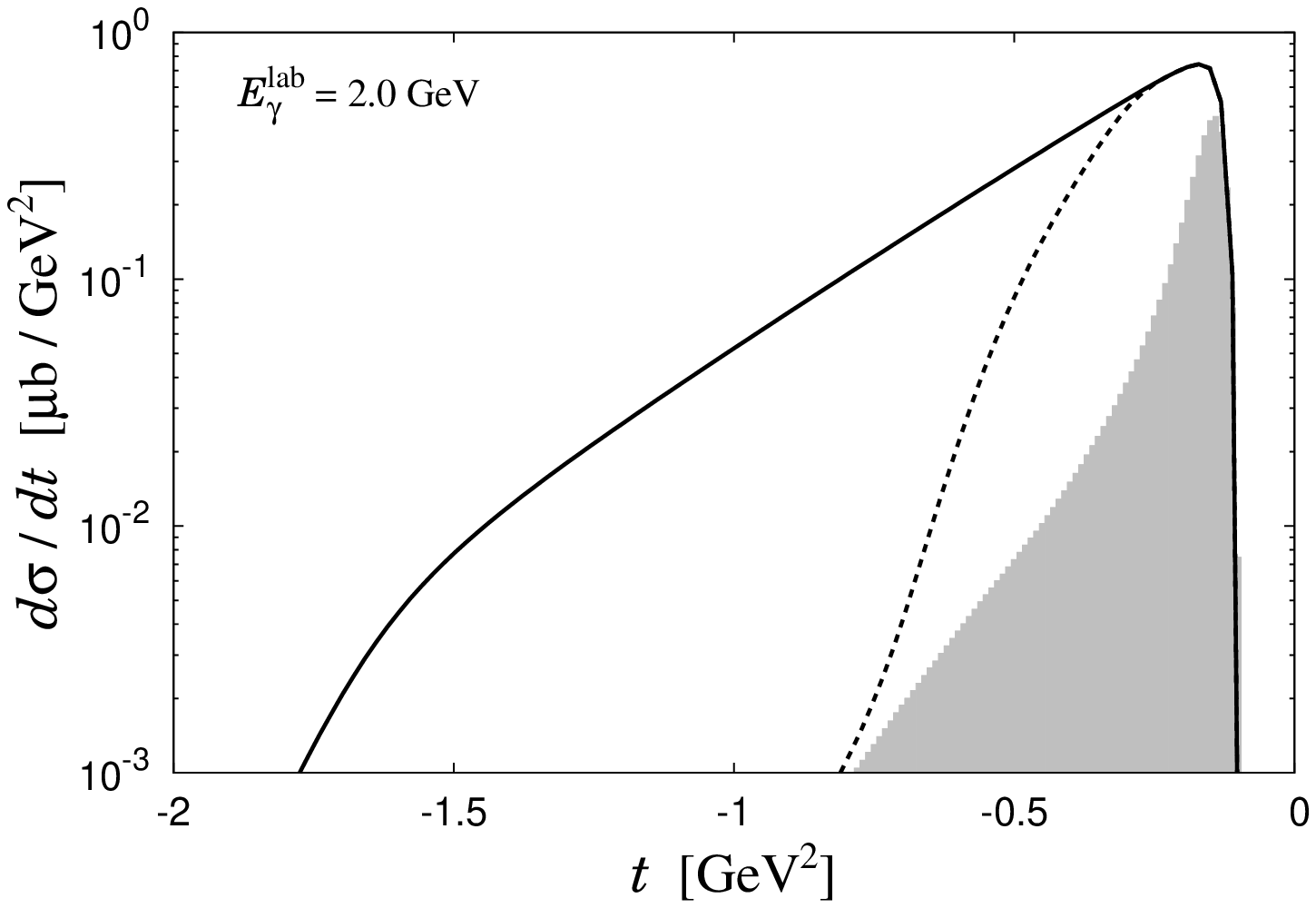} & 
    \includegraphics[width=8.4cm]{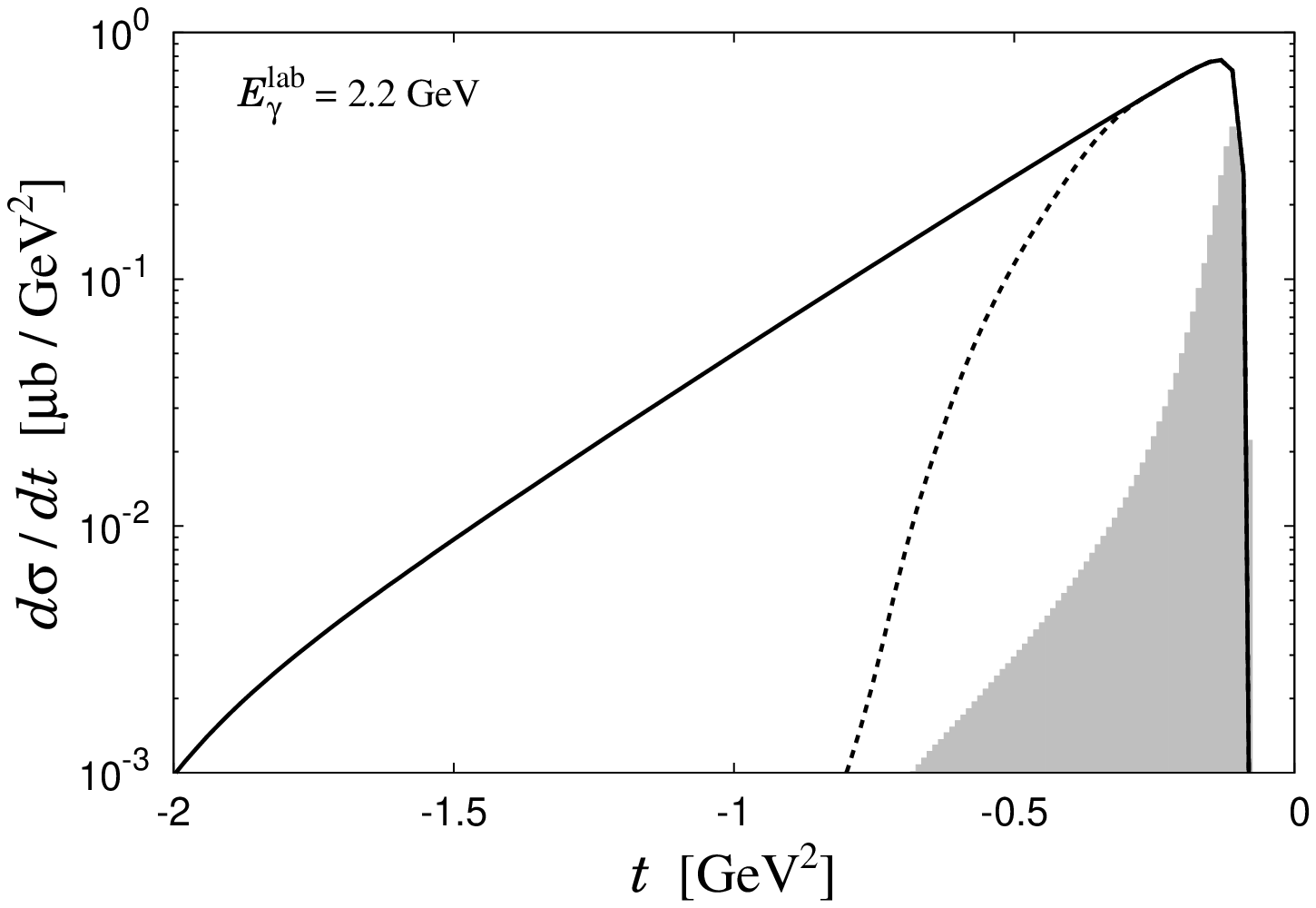} 
    % \vspace{-8pt} 
  \end{tabular*}
  \caption{ \label{fig:dSdt-d-cut} Differential cross section $d\sigma
    _{p^{\ast}}/dt_{\phi}$ for a bound proton in the deuteron as a
    function of $t_{\phi}$ for different
    $E_{\gamma}^{\text{lab}}$. The solid and dashed lines, and shaded
    area indicate the case without any angular cuts, with the $\phi$
    cut, and with the $\phi$ and proton cuts, respectively.}
\end{figure*}

\begin{figure*}[!t]
  \centering
  \begin{tabular*}{\textwidth}{@{\extracolsep{\fill}}cc}
    \includegraphics[width=8.4cm]{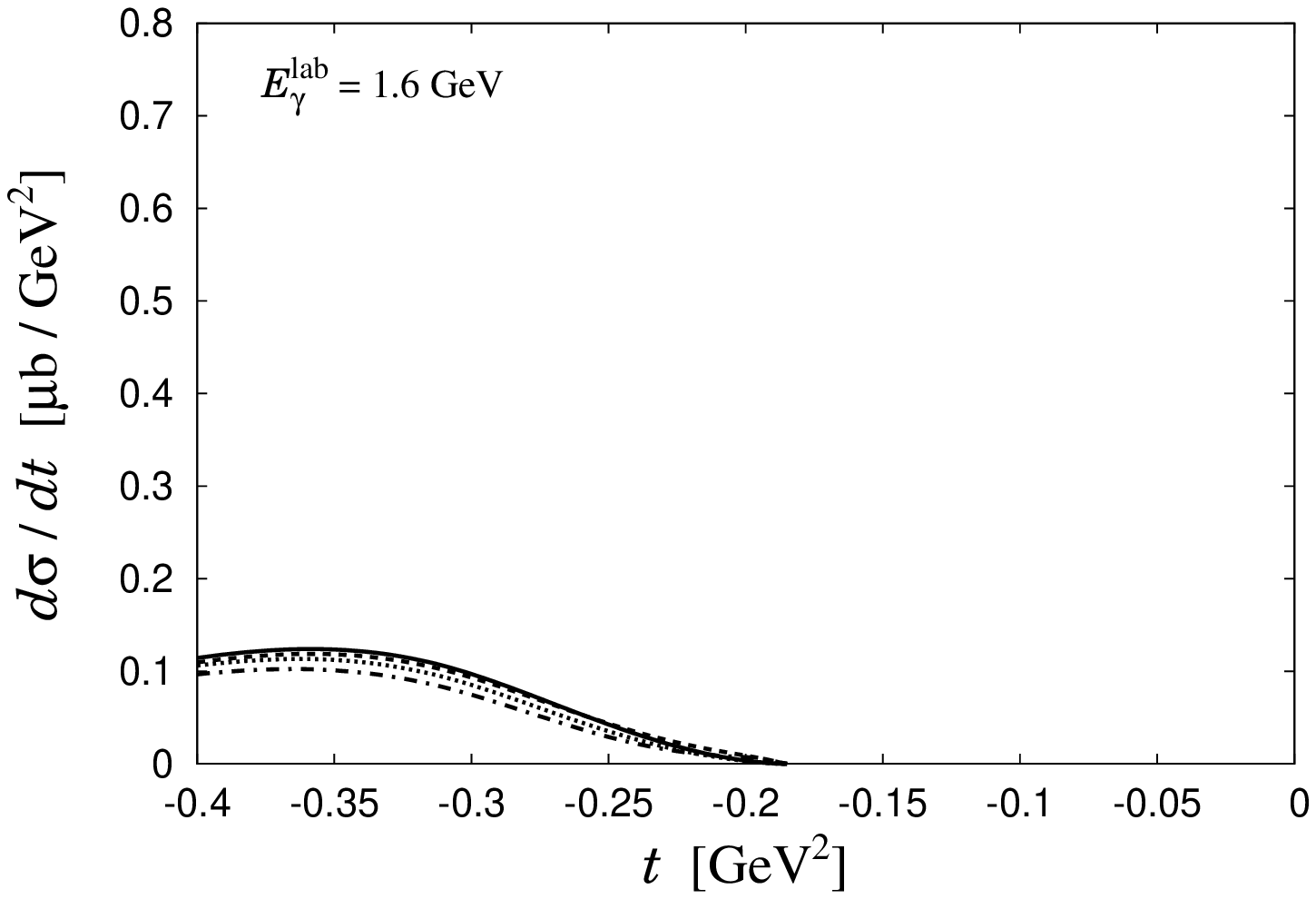} & 
    \includegraphics[width=8.4cm]{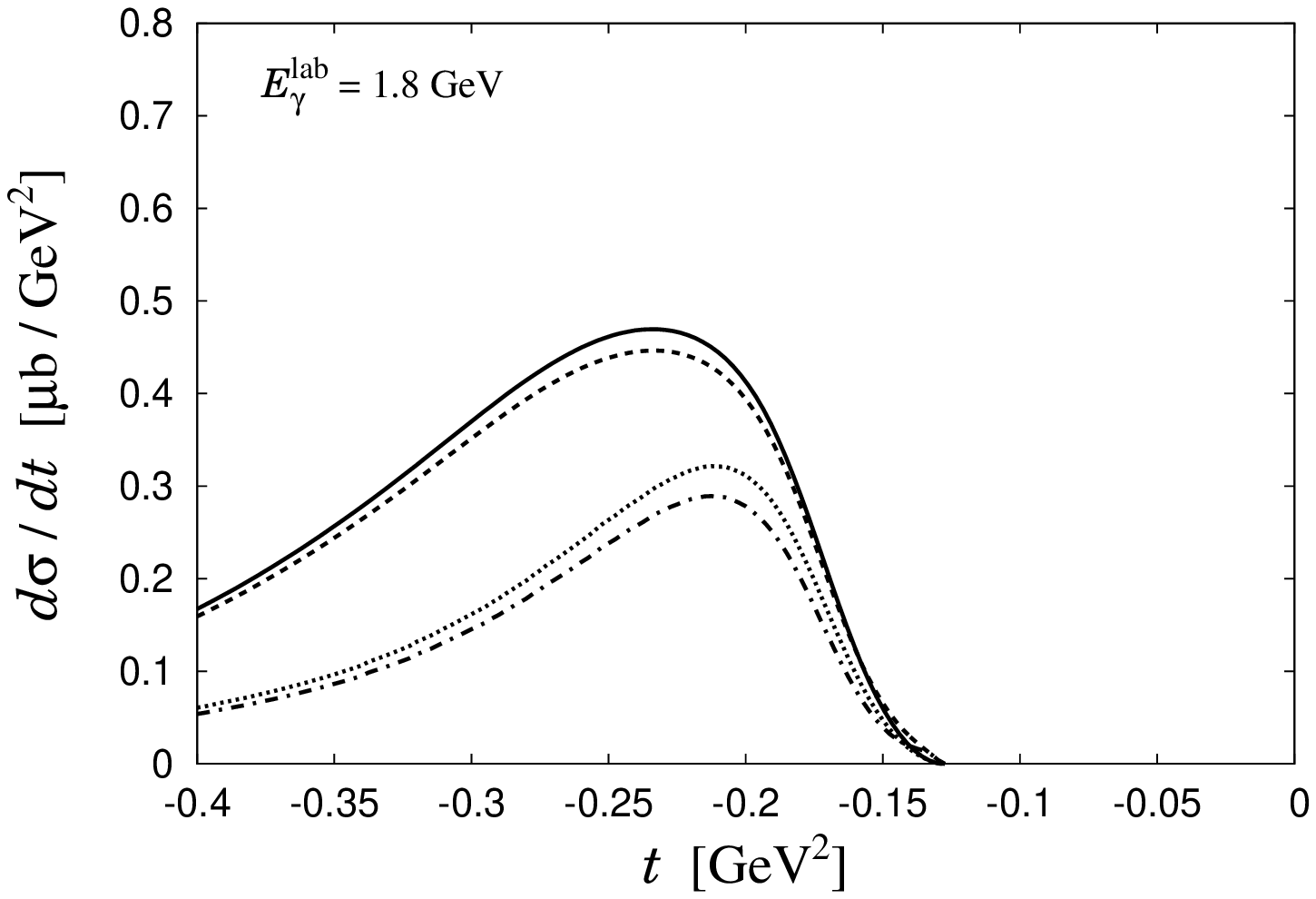} \\
    \includegraphics[width=8.4cm]{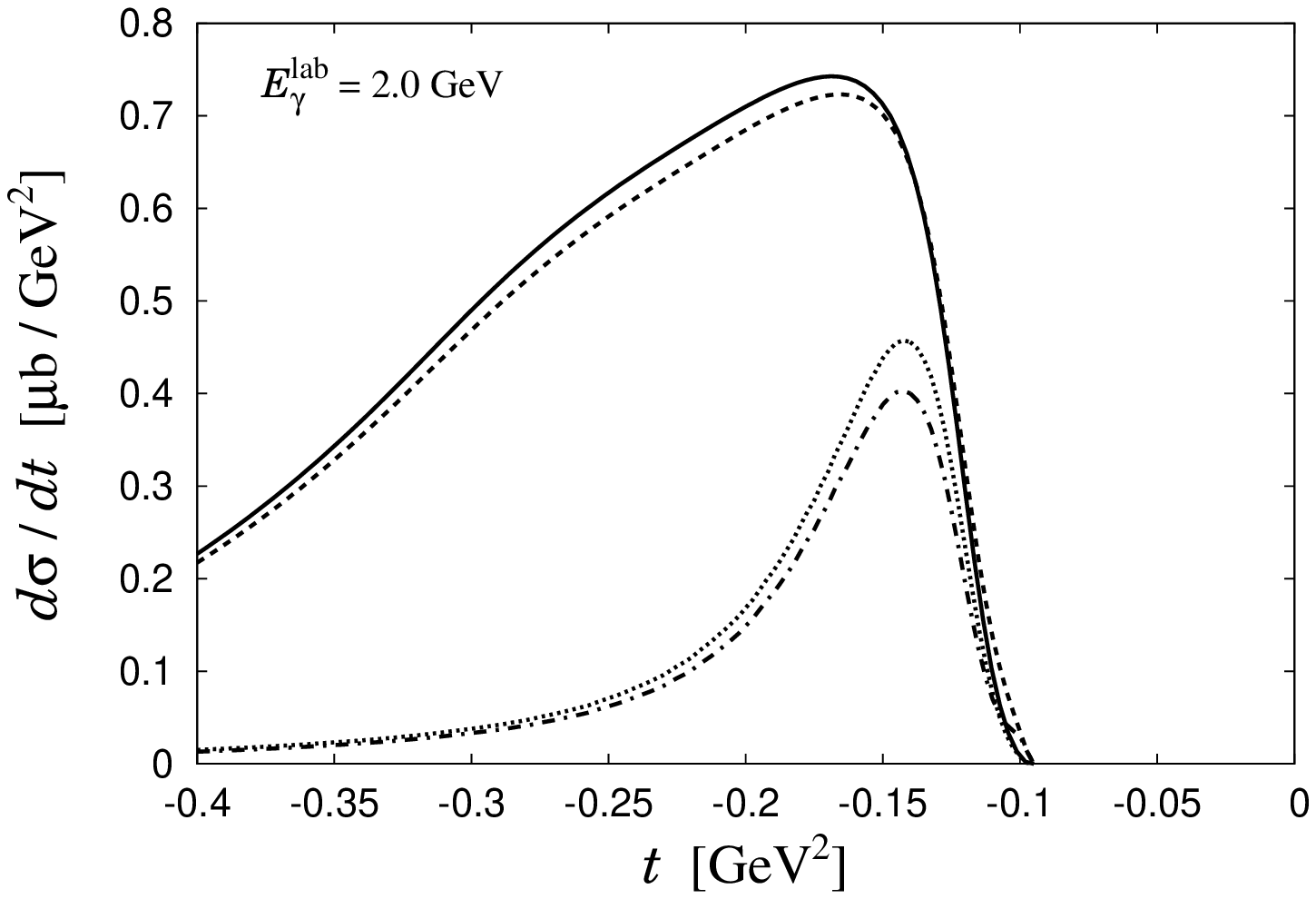} & 
    \includegraphics[width=8.4cm]{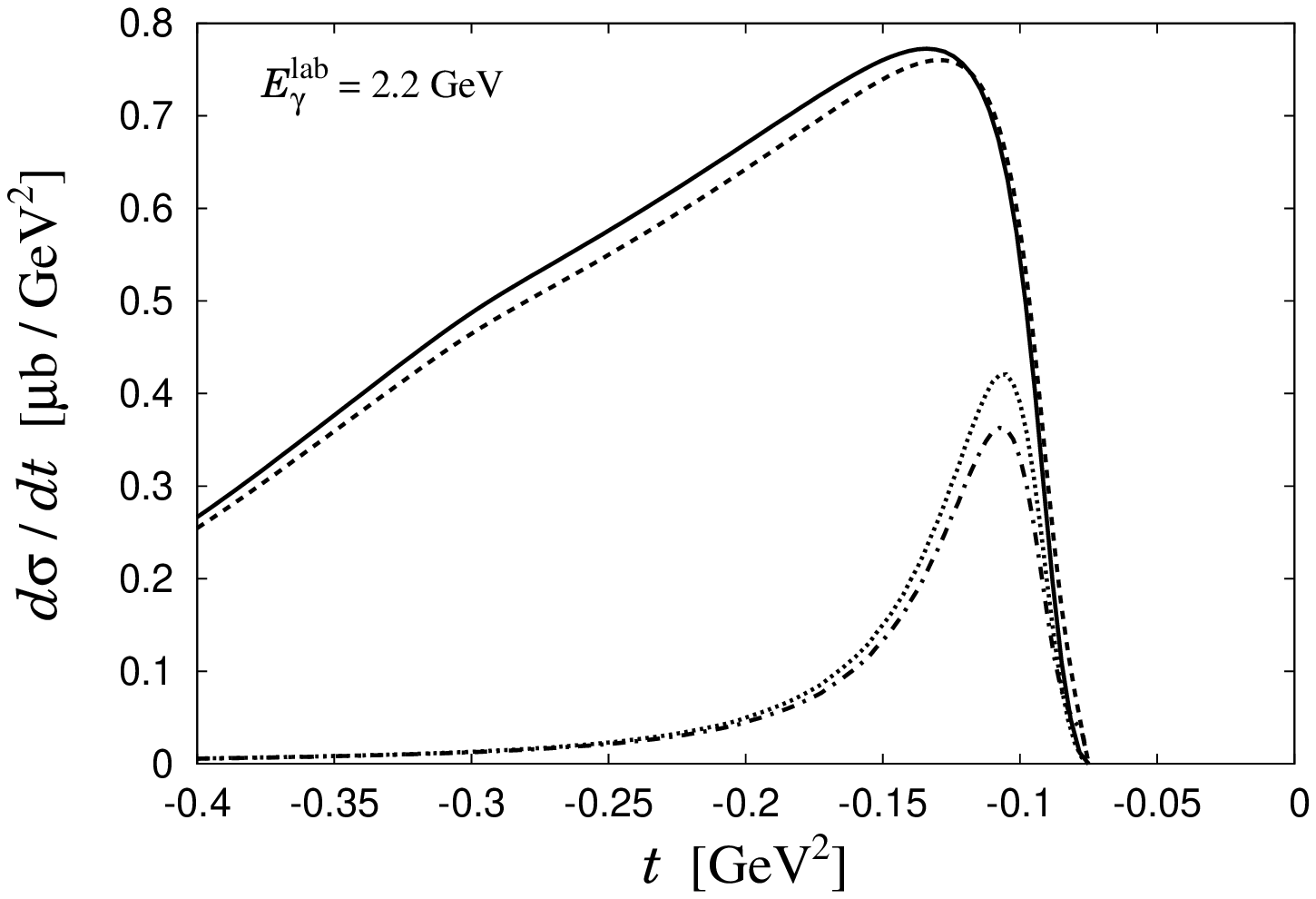} 
  \end{tabular*}
  \caption{ \label{fig:dSdt-d-cut-pn} Differential cross section
    $d\sigma _{p^{\ast}}/dt_{\phi}$ for a bound proton in the deuteron
    as a function of $t_{\phi}$ for different
    $E_{\gamma}^{\text{lab}}$. The solid and dashed lines indicate
    single scattering and single plus two double scattering ($\phi$
    and proton exchanges) contributions, respectively, with only
    $\phi$ cut.  The dotted and dash-dotted lines indicate single
    scattering and single plus two double scattering ($\phi$ and
    proton exchanges) contributions, respectively, with both $\phi$
    and proton cuts.}
\end{figure*}

\subsection{Angular cuts for charged particles}

In the previous subsection we have shown the differential cross
sections for the reactions $\gamma p \to \phi p$ and $\gamma d \to
\phi p n$. In the LEPS experiments these reactions are identified by
detecting charged particles by the spectrometer in the forward angles
in the laboratory frame.  Hence, let us now perform the angular cuts
for the charged particles in the evaluation of the cross section such
that we can compare with the LEPS experiment.

We take the angular cuts so as to keep $\Theta \le 20$ degrees, where
$\Theta$ is the angle between the momenta of the incident photon and
the charged particle in the laboratory frame.

In our study, the charged particles to which we should apply the cut
are the final proton and $K^{+}K^{-}$, which come from $\phi$
decay. For the proton angular cut, we can simply restrict the final
state phase-space so that the final proton comes into the angle
$\Theta \le 20$ degrees. We refer to this cut as the proton cut. For
the $K^{+}$ and $K^{-}$ angular cuts, on the other hand, one needs
some consideration, since we do not explicitly have the final $K^{+}$
and $K^{-}$, but we have the $\phi$. Here we choose the following
method for the $K^{+}$ and $K^{-}$ angular cuts; in each event we
assume that the $K^{+}$ and $K^{-}$ go out with spherical symmetry in
the $\phi$ rest frame with momentum $\simeq 127 \mev /c$, regardless
of the polarization of the $\phi$. Next, we perform the Lorentz boost
from the $\phi$ rest frame to the laboratory frame and evaluate the
probability that both the $K^{+}$ and $K^{-}$ come into the angle
$\Theta \le 20$ degrees. Then we multiply by this probability the
cross sections in order to reproduce the $K^{+}$ and $K^{-}$ angular
cuts. We refer to this as the $\phi$ cut.

Now we show the angle-cut differential cross section for the $\gamma p
\to \phi p$ reaction, $d\sigma _{p}/dt$, in
Fig.~\ref{fig:dSdt-p-cut}. First of all, one can see that the $\phi$
cut suppresses $d\sigma _{p}/dt$ in the large $|t|$ region at any
photon energies.  For the region close to $t=t_{\text{max}}$, however,
$d\sigma _{p}/dt$ is not affected by the $\phi$ cut at any photon
energy, because in this region the $\phi$ goes in the forward
direction in the laboratory frame with sufficiently large
momentum\footnote{Note that both the $K^{+}$ and the $K^{-}$ come into
  the angle $\Theta \le 20$ degrees if the $\phi$ has sufficiently
  large momentum in the forward direction in the laboratory frame,
  since the transverse momenta of the kaons coming from $\phi$ decay
  in the $\phi$ rest frame are restricted to less than $127 \,
  \text{MeV}/c$. }. Hence, the LEPS experiment can detect the whole
$\phi$ photoproduction events in the region close to
$t=t_{\text{max}}$.

Then, one can see that the proton cut does not change $d\sigma
_{p}/dt$ at $E_{\gamma}^{\text{lab}}=1.6 \gev$. This is because this
photon energy is very close to the threshold for the $\phi$
photoproduction so that the final proton, which is almost at rest in
the center-of-mass frame, can come into the $\Theta \le 20$ degrees
without large transverse momentum.

The proton cut produces changes in $d\sigma _{p}/dt$ for photon energy bigger 
than $E_{\gamma}^{\text{lab}}\simeq 1.8 \gev$. The cut produces null cross 
section in the middle of the $|t|$ region, that is the region where the final 
proton does not go forward nor backward in the 
center-of-mass frame. In this region the final proton has large transverse 
momentum, and, hence, the proton goes out of the spectrometer in the LEPS 
experiment. In the case that the $\phi$ goes forward in the center-of-mass 
frame, the proton cut has no effects, since the final proton goes backward in 
the center-of-mass, so the Lorentz boost from the center-of-mass frame to the 
laboratory frame keeps the proton inside the angle $\Theta \le 20$ degrees. 
 We 
should note that as $E_{\gamma}^{\text{lab}}$ increases, the region with null cross 
section that the proton cut produces becomes larger, since the maximal value of 
the transverse momentum of the final proton also increases. 

For the free proton target $\gamma p \to \phi p$ reaction, however, one does 
not need to detect the final proton by the spectrometer; in this reaction one 
can identify the final proton by using the missing mass method. Therefore, the 
proton cut is not implemented in the experiments in this reaction and 
theoretically we should 
calculate $d\sigma _{p}/dt$ only with the $\phi$ cut. As one can see, only with 
the $\phi$ cut $d\sigma _{p}/dt$ is suppressed in the large $|t|$ region, 
whereas it is not affected in the region close to $t=t_{\text{max}}$. Should one 
wish to obtain $d\sigma _{p}/dt$ experimentally for the case without angular 
cuts, one should make acceptance corrections, which were done 
in~\cite{Mibe:2005er}.

Now let us show the effects of the angular cut on the $\gamma d \to
\phi p n$ reaction in Fig.~\ref{fig:dSdt-d-cut}, where only the single
scattering is taken into account. One of the important features from
Fig.~\ref{fig:dSdt-d-cut} is that at any photon energy the $\phi$ cut
does not produce changes in the cross section in the region close to
$t_{\phi} =t_{\phi , \text{max}}$, like in the $\gamma p \to \phi p$
case, whereas the $\phi$ cut changes the cross section in the large
$|t_{\phi}|$ region. This suggests that all of the $K^{+}K^{-}$ from
the $\phi$ can go into the LEPS spectrometer in the region close to
$t_{\phi} =t_{\phi , \text{max}}$.

The proton cut effects, on the other hand, appear above the photon
energy $E_{\gamma}^{\text{lab}} \simeq 1.8 \gev$. This proton cut,
however, does not completely suppress $d\sigma _{p^{\ast}}/dt_{\phi}$,
as it was the case in the $\gamma p \to \phi p$ reaction. The finite
$d\sigma _{p^{\ast}}/dt_{\phi}$ with the proton cut is caused by the
Fermi motion in the deuteron bound system.  This $d\sigma
_{p^{\ast}}/dt_{\phi}$ suppression due to the proton cut gets bigger
as the photon energy increases, the same as for the $\gamma p \to \phi
p$ reaction.

It is important to keep in mind this suppression created by the proton
cut.  Namely, in the $\gamma d \to \phi p n$ reaction in the LEPS
experiment the $\gamma p^{\ast} \to \phi p$ reaction is identified by
detecting also the final proton in addition to the $K^{+}K^{-}$ in the
spectrometer. Since the double scattering mechanism of
Fig.~\ref{fig:gamma-d-each}(b) contributes scarcely to $d\sigma
_{p^{\ast}}/dt_{\phi}$ in the small $|t_{\phi}|$ region (see
Fig.~\ref{fig:dSdt}) and the Fermi motion of the neutron is moderate
in the deuteron, one can identify the $\gamma p^{\ast} \to \phi p$
process in the deuteron by detecting relatively fast protons. However,
in order to compare $d\sigma _{p^{\ast}}/dt_{\phi}$ for
photoproduction on a proton in the deuteron with $d\sigma _{p}/dt$,
one has to consider the proton cut effects on $d\sigma
_{p^{\ast}}/dt_{\phi}$ and has to perform the acceptance correction
for the proton cut in addition to that for the $\phi$ cut.  These
acceptance corrections are done in the experiment, although no details
are given in the paper. In view of the strong $t_{\phi}$ dependence of
the proton cut, details on how the acceptance corrections are done
would be most advisable.

Next let us examine the angular cut effects on the double scattering
contributions.  The effects of the $\phi$ and the proton cuts are
shown in Fig.~\ref{fig:dSdt-d-cut-pn}.  As one can see from
Fig.~\ref{fig:dSdt-d-cut-pn}, the two double scattering ($\phi$ and
proton exchanges) contributions with the $\phi$ cut do not largely
suppress the contribution from the single scattering (solid and dashed
lines).  However, if the proton cut is taken into account the two
double scattering contributions suppress the differential cross
section compared to that with only the single scattering in a larger
amount than if only the $\phi$ cut is considered, especially in
$t_{\phi} \simeq t_{\phi , \text{max}}$ region (dotted and dash-dotted
lines).  Remembering that the $\phi$ rescattering effect is small at
$t_{\phi} \simeq t_{\phi , \text{max}}$, this suppression originates
from the proton rescattering effect.

Since the proton cut is important for the suppression of $d \sigma
_{p^{\ast}} / d t_{\phi}$, the suppression behavior can be interpreted
as follows: namely, the final proton in forward angle is rescattered
by the neutron and the proton direction changes.  Then the number of
protons in the forward direction decreases, and this is taken into
account by the imaginary part of the $pn \to pn$ amplitude
$T_{2}^{pn}$.  The proton in this collision is not lost, but it is
simply redistributed in other directions.  Technically we saw that
this was accomplished by means of the real part of the $pn \to pn$
amplitude $T^{pn}_{2}$.  Then, due to the proton cut in the $t_{\phi}
\simeq t_{\phi , \text{max}}$ region, some of these protons will not
reach the LEPS detector.  Furthermore, the proton rescattering effect
of forward to another angle for the proton is larger than that of
another angle to forward for the proton and we see a net suppression
of forward going protons.

\begin{figure}[!t]
  \centering
    \includegraphics[width=8.6cm]{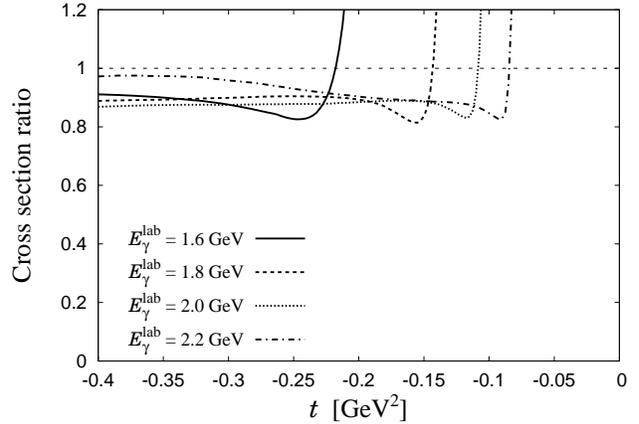} 
    \caption{Ratio of the differential cross sections $d\sigma
      _{p^{\ast}}/dt_{\phi}$ with three contributions (single
      scattering and $\phi$- and proton-exchange amplitudes) to that
      only with single scattering.  Both $\phi$ and proton angle cuts 
      are performed. }
  \label{fig:dSdt-cut-ratio}
\end{figure}

In order to see how much the double scattering effects suppress $d
\sigma _{p^{\ast}} / d t _{\phi}$ compared to the single scattering
contribution, we show the ratio of the differential cross sections
with the three contributions to that with only single scattering in
Fig.~\ref{fig:dSdt-cut-ratio}.  In Fig.~\ref{fig:dSdt-cut-ratio}, both
the $\phi$ and proton cuts are performed.  From the figure, one can
see that the double scattering effects suppress the cross section,
especially in the $t_{\phi}\sim t_{\phi , \text{max}}$ region, about
to $85 \%$ regardless of the initial photon energy.  The ratio is
close to $90 \%$ if a bin of $\Delta t=0.1 \gev ^{2}$ is taken, as
in~\cite{Miyabe:2010}.  This small reduction goes in the direction
reported in~\cite{Chang:2009yq} for the ratio of $\phi$
photoproduction on a bound proton inside the deuteron to the one on a
free proton, but is short of the value around $60 \%$ reported there
for large energies.  The rapid increase of the ratio at $t_{\phi}
\simeq t_{\phi , \text{max}}$ shown in Fig.~\ref{fig:dSdt-cut-ratio}
is caused by the threshold effect for $t_{\phi}$ and gives no trouble
in the differential cross sections themselves.

\subsection{Ratio of the cross sections}

Now that we have all the results, we try to compare the ratio of the 
differential cross sections, 
$(d\sigma _{p^{\ast}}/dt_{\phi})/(d\sigma _{p}/dt)$ from LEPS 
experiments~\cite{Chang:2009yq} with the theoretical results. 

In the LEPS experiment, they found a significant diversion with respect to 
unity (see Fig.~4 of Ref.~\cite{Chang:2009yq}). There they detect the 
$K^{+}K^{-}p$ in the $\gamma d \to \phi p n$ reaction so as to identify the 
$\gamma p^{\ast} \to \phi p$ in the deuteron and evaluate the ratio of this 
cross section to the one on the free proton. 

From the theoretical point of view the comparison of $d\sigma /dt$ for
the proton in the deuteron and the free proton can be obtained from
the ratio of the dashed line and dashed-dotted line in
Fig.~\ref{fig:dSdt-pn} for each energy.  This is leaving apart the
effect of the $\phi$ and proton cuts discussed above, from where one
can expect a reduction of $10 \%$ in the ratio. Of course, one can see
that as one approaches $t_{\text{max}}$, the ratio changes very fast
from values around $0.8$ close to $t_{\text{max}}$, to around $0.4$ at
$t_{\text{max}}$. The ratio becomes infinity if we go a bit beyond
$t_{\text{max}}$, since the reaction is forbidden in the free case,
but is allowed in the deuteron due to the Fermi motion.

However, we cannot compare these results with those obtained in Fig.~4 of 
Ref.~\cite{Chang:2009yq}, the reason being that the experimental paper does not 
provide enough information to allow for a meaningful comparison. We hope that 
an extended version of the concise information given in~\cite{Chang:2009yq} 
can provide these needed details. In order to facilitate this task we write 
below the information that we would need for 
a proper comparison. 

\begin{itemize}
\item[1)] 
How and where is the deuteron wave function taken into account to 
remove effects of Fermi motion in~\cite{Chang:2009yq}, if this is the case? 

\item[2)] 
In Fig.~4 of~\cite{Chang:2009yq} $E_{\gamma}^{\text{eff}}$ is used in the $x$ axis 
without any comment or definition. We assume that this is not a misprint, but 
that indeed the concept of $E_{\gamma}^{\text{eff}}$ introduced 
in~\cite{Nakano:2008ee} is used there\footnote{This seems to be the case 
according to the private communication~\cite{Miyabe:2010}. }. This concept 
relies upon the MMSA (minimum momentum spectator approximation) which 
approximates the spectator nucleon momentum as the minimum one without 
specifying whether it 
is a proton or a neutron (see also the repercussion of its use in the analysis 
of the ``$\Theta ^{+}$'' peak as discussed in~\cite{Torres:2010jh}). 
It is important that the details on the use of this prescription in the present 
analysis are provided.

\item[3)] When dealing with Fermi motion close to $t_{\text{max}}$,
  the binning of $\Delta t$ used to determine $d\sigma _{p^{\ast}}/dt$
  is also important, in view of the fast change of the cross section
  as a function of $t_{\phi}$ close to $t_{\text{max}}$.  On the other hand,
  the use of a large binning can lead to other problems of
  interpretation. Indeed, assume one takes $\Delta t = 0.1 \gev ^{2}$,
  then for events with $t_{\phi}=t_{\text{max}}-\Delta t$, $d\sigma
  _{p^{\ast}}/dt$ has already fallen to $70 \%$ of the value at
  $t_{\text{max}}$, as given by Eq.~(\ref{eq:Amp-phenom}).  Similarly,
  as seen in Fig.~\ref{fig:dSdt-d-cut}, the effect of the proton angle
  cut at the highest energy $E_{\gamma}^{\text{lab}}=2.2 \gev$ leads
  to a reduction of $d \sigma _{p^{\ast}}/dt$ at
  $t_{\phi}=t_{\text{max}}-\Delta t$ to about 
    $31 \%$ of its value at
  $t_{\text{max}}$. The combined effect of the two would be a
  reduction by a factor of about $0.1$ at $\Delta t = 0.1 \gev ^{2}$.

\item[4)]
Although methods could be devised to eliminate the contribution to the ``proton 
in deuteron'' cross section from the unwanted case where the proton is a 
spectator, it should be kept in mind that the Fermi motion provides a 
distribution of momenta to these spectator protons, some of which could be 
observed and be misidentified as participant protons. Details on how this 
problem is avoided would also be most welcome. 

\item[5)]
At some point in the experimental analysis, knowing the proton momentum will be 
important. How this momentum is reconstructed from the observed events in view 
of the expected distortion caused by the target in the detected protons is 
also a relevant information. 
This point becomes more critical once we have shown that the 
proton angular cuts are so relevant for the cross section close to 
$t_{\text{max}}$. Furthermore, 
in view of the losses in the detection of protons, 
when the $K^{+}$ and $K^{-}$ are detected in coincidence, 
one should also clarify the 
statistical situation of these events and how $d \sigma _{p^{\ast}}/dt$ is 
obtained in this case, providing statistical and systematic errors. 

\end{itemize}

When this information is provided, we could continue our work with a
meaningful comparison with the data of~\cite{Chang:2009yq}. At the
present time, this comparison, and the interpretation of the data from
our theoretical perspective is not possible.  Yet, we found that 
the consideration of proton rescattering in the deuteron, together with 
the effect of the proton cut, can produce a moderate reduction of the 
medium to free proton $\phi$ photoproduction cross section of about 
$10 \%$, which goes in the direction reported in~\cite{Chang:2009yq}, 
but falls short of the numbers quoted there.

\section{Summary}
\label{sec:summary}

We have done a study of $\phi$ photoproduction on the proton and
on the deuteron, and in particular on a proton in the deuteron, using
an accurate wave function for the deuteron that accounts for the Fermi
motion, which we found to be very important when studying $d \sigma/dt$
close to $t_{\text{max}}$. We also took into account rescattering of the
$\phi$ including the mechanisms that lead to $\phi$ absorption and
provide realistic values of the $\phi$ width in a nuclear medium.  We
found that the latter mechanisms are very small. The screening of the
$\phi$ production in the deuteron is found to be very small. This could be 
consistent with moderate changes of the screening of the $\phi$ production 
seen in nuclei~\cite{Ishikawa:2004id,Wood:2010ei,Polyanskiy:2010tj}. 
This conclusion is in agreement with the recent experiment at 
Jefferson Lab.~\cite{haiyan} where the extracted $d\sigma /dt$ is 
consistent with predictions based on the quasifree mechanism. 

We studied in detail the effects of the $\phi$ and proton cuts to
accommodate the theoretical results to the measurements of LEPS,
taking into account the LEPS set up which restricts the detected
particles to forward angles. We found that the proton cut, which does
not appear in the free proton case, because its momentum is
reconstructed from the $K^{+}$ and $K^{-}$ momenta, plays an important
role in the scattering on the deuteron since protons are detected in
this case.  
The largest effect, even then rather moderate, was the about $10 \%$
reduction in the ratio of medium to free proton photoproduction cross
sections that we obtain from the consideration of proton rescattering
in the deuteron, together with the $\phi$ and proton cuts of the LEPS
set up implemented in both of them to provide a proper comparison.

Finally, we discussed that a comparison of our results with the
experimental ones provided in~\cite{Chang:2009yq} was not possible at
the present stage due to the absence of relevant experimental
information in~\cite{Chang:2009yq}. We suggested that a more detailed
experimental paper be written and we listed the experimental
information that would be needed for a meaningful comparison in the
future.

\section*{Acknowledgements}

This work is supported in part by
the Grant-in-Aid for Scientific Research
from MEXT and JSPS (Nos.
22105507,  % ??? 
22740161,  % ??? 
and 22-3389), % (Gakushin, Sekihara)
the Bilateral International Exchange Program (BIEP) of the Global COE Program 
``The Next Generation of Physics, Spun from Universality and Emergence'' 
from MEXT of Japan, 
and 
the collaboration agreement between the JSPS of Japan and the CSIC of Spain. 
This work is partly supported by DGICYT contract number FIS2006-03438 and the 
Generalitat Valenciana in the program Prometeo. We acknowledge the support of 
the European Community-Research Infrastructure Integrating Activity ``Study of 
Strongly Interacting Matter'' (acronym HadronPhysics2, Grant Agreement n.
227431) under the Seventh Framework Programme of EU.
One of the authors (T.S.) acknowledges the support by the Grand-in-Aid for 
JSPS fellows. 
This work is partially done under
the Yukawa International Program for Quark-hadron Sciences (YIPQS).

\appendix

\section{$\bm{\phi n \to \phi n}$ scattering amplitude}
\label{sec:phi-n}

\begin{figure}[t]
  \centering
    \includegraphics[scale=0.18]{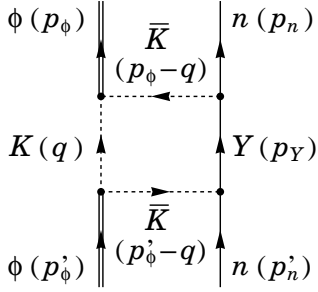} 
  \caption{Box diagram for $\phi n \to \phi n$ reaction 
    together with momentum of each particle. The channels of $K$, $\bar{K}$, 
    and $Y$ are given in Tab.~\ref{tab:phi-n}. }
  \label{fig:phi-n}
\end{figure}

\begin{table}[!t]
  \caption{\label{tab:phi-n} Channels of $K$, $\bar{K}$, and $Y$ appearing in 
    Fig.~\ref{fig:phi-n}. The Clebsch-Gordan coefficients $\alpha$, $\beta$, 
    and $A$ are also shown. }
%  \begin{ruledtabular}
  \begin{tabular}{ccccccc}
    \hline
    \hline
    Channel & $K$ & $\bar{K}$ & $Y$ & $\alpha$ & $\beta$ & $A$ \\ 
    \hline
    1 & $K^{+}$ & $K^{-}$ & $\Sigma ^{-}$ & $0$ & $\sqrt{2}$ & -- \\ 
    2 & $K^{+}$ & $K^{-}$ & $\Sigma ^{\ast -}$ & -- & -- & $-1$ \\ 
    3 & $K^{0}$ & $\bar{K}^{0}$ & $\Lambda$ & $-2/\sqrt{3}$ & $1/\sqrt{3}$ & -- \\
    4 & $K^{0}$ & $\bar{K}^{0}$ & $\Sigma ^{0}$ & $0$ & $-1$ & -- \\ 
    5 & $K^{0}$ & $\bar{K}^{0}$ & $\Sigma ^{\ast 0}$ & -- & -- & $1/\sqrt{2}$ \\ 
    \hline
    \hline
  \end{tabular}
  % \end{ruledtabular}
\end{table}

\begin{figure}[t]
  \centering
  \begin{tabular*}{8.6cm}{@{\extracolsep{\fill}}ccc}
    \includegraphics[scale=0.16]{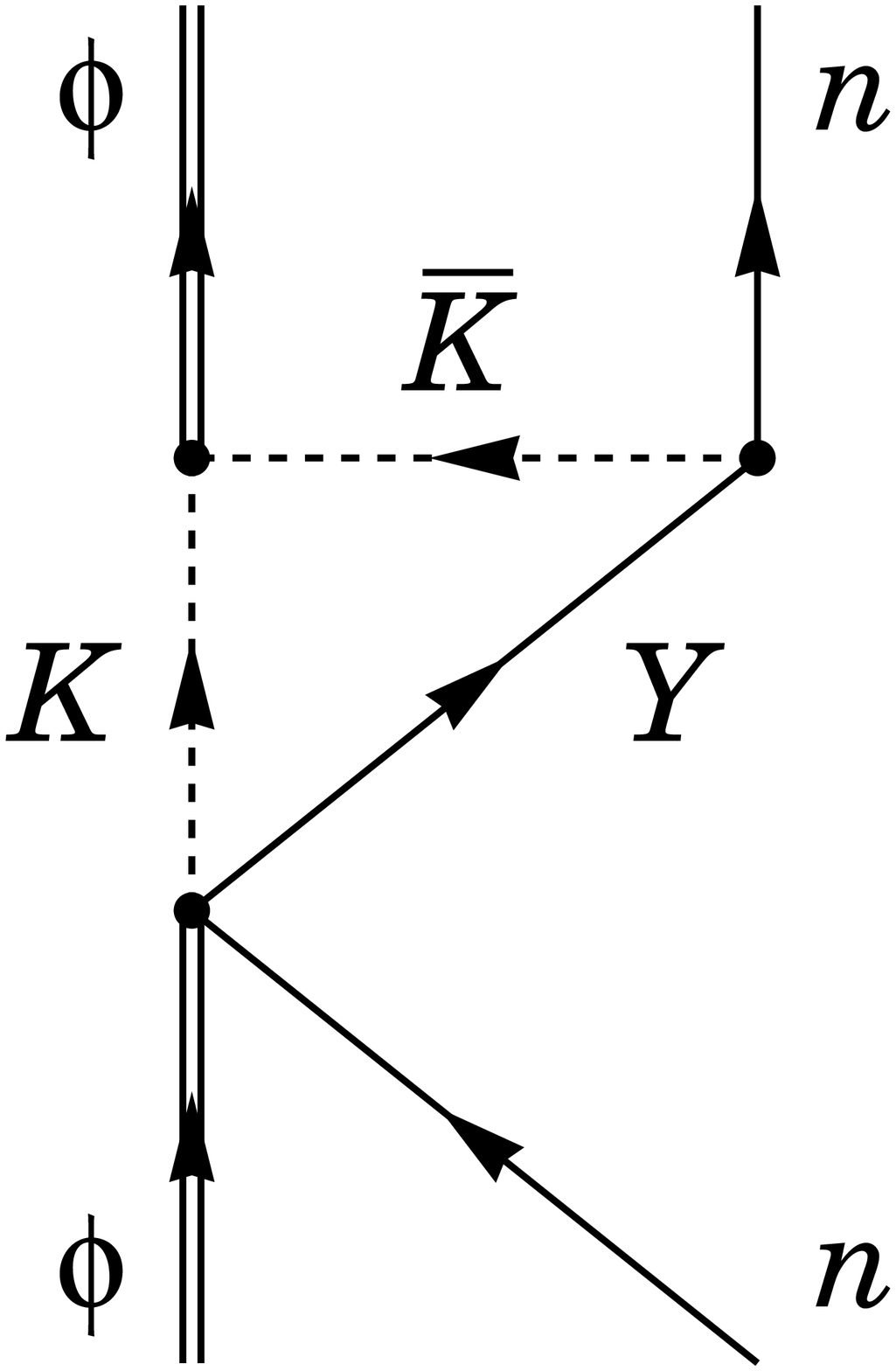} & 
    \includegraphics[scale=0.16]{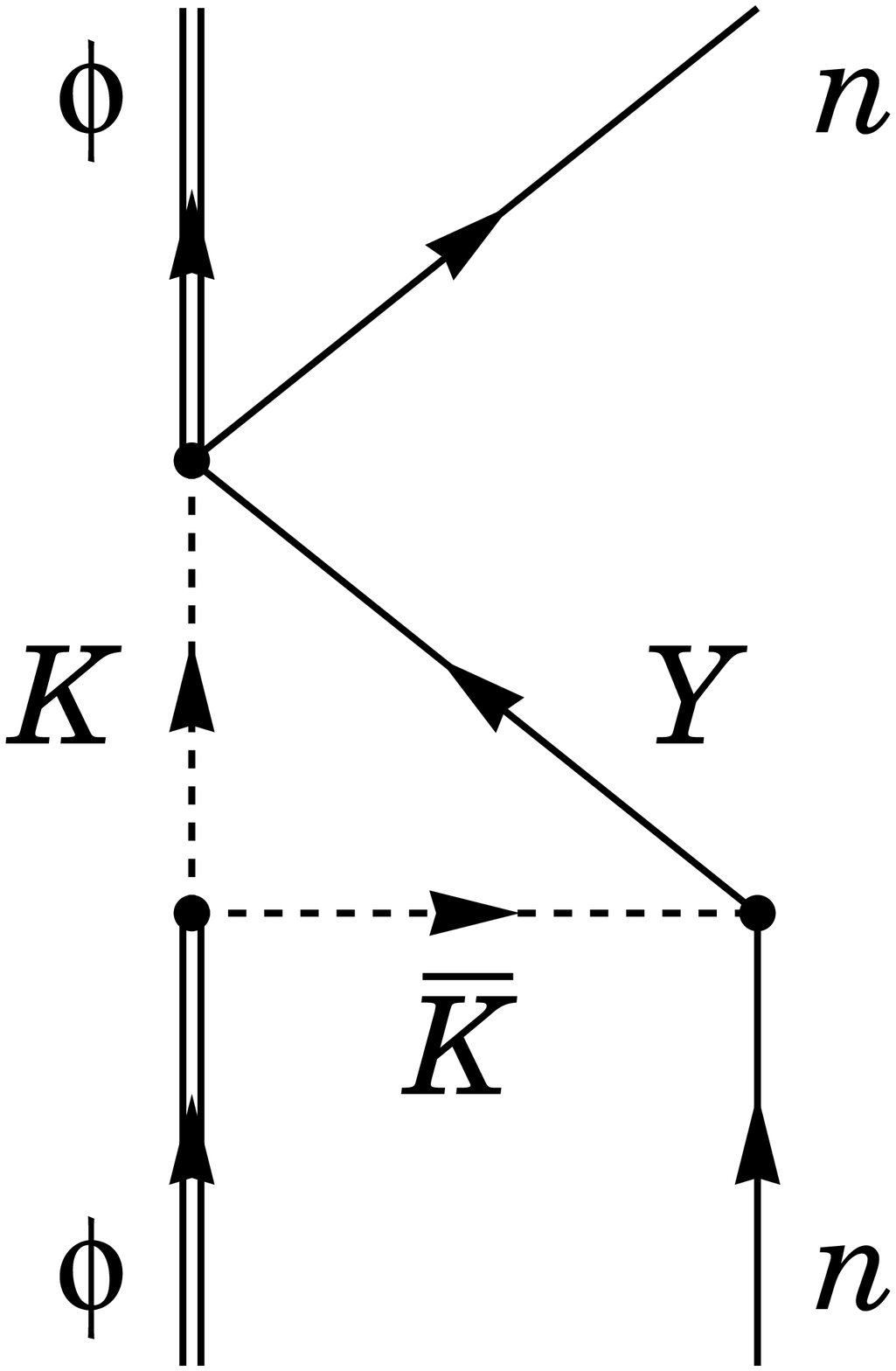} & 
    \includegraphics[scale=0.16]{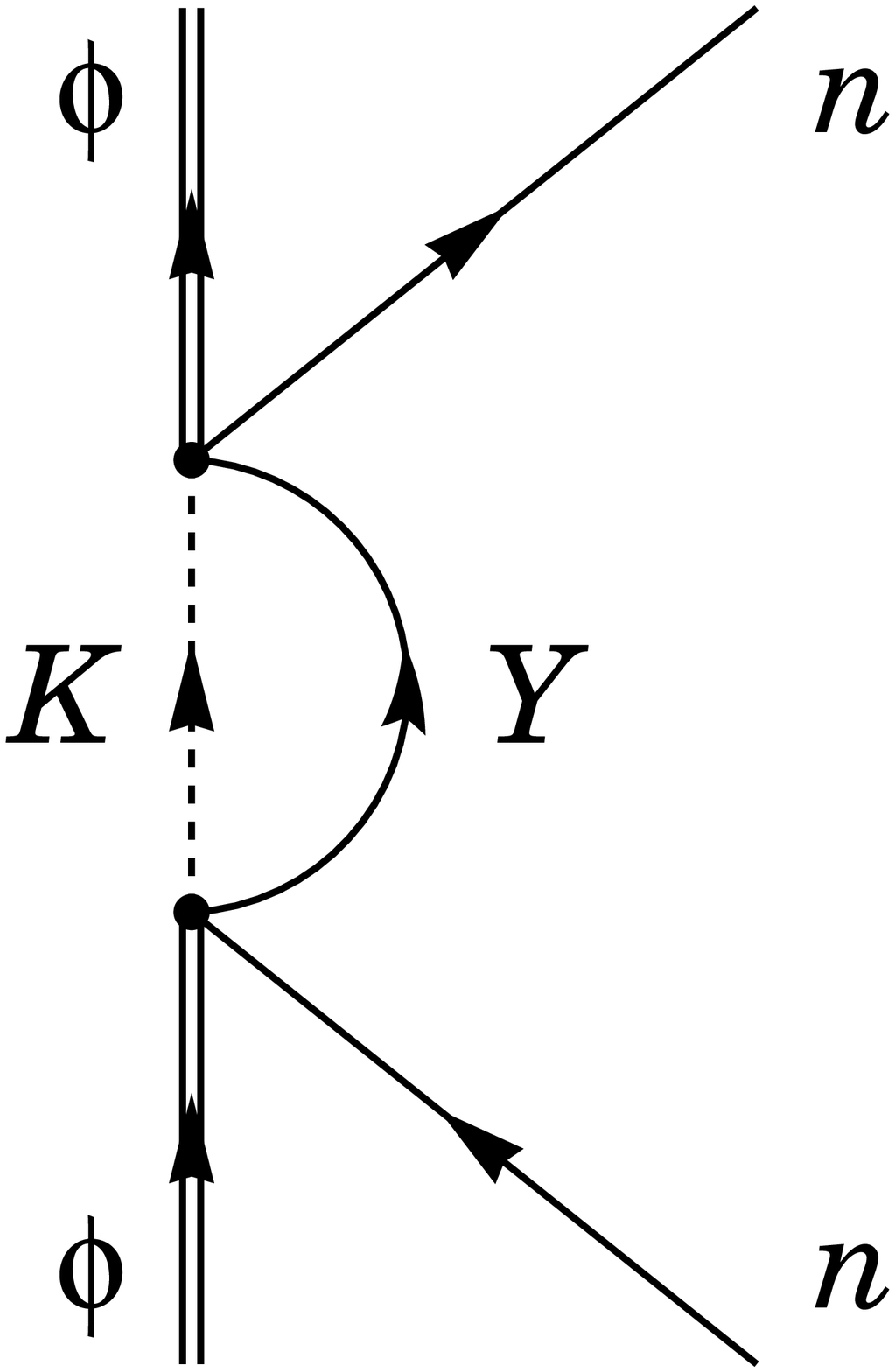} 
  \end{tabular*}
  \caption{Diagrams for the vertex corrections of $\phi n \to \phi n$ reaction. 
    The channels of $K$, $\bar{K}$, and $Y$ are given in Tab.~\ref{tab:phi-n}. }
  \label{fig:phi-n-corr}
\end{figure}

In this Appendix we evaluate the $\phi n \to \phi n$ scattering amplitude. 
Since we are concerned about the absorptive imaginary part of this amplitude, 
we take into account here the same mechanisms that were
considered in the evaluation of the width of the $\phi$ in the medium
in~\cite{Oset:2000eg,Cabrera:2002hc}. These are depicted in 
Figs.~\ref{fig:phi-n} (the box diagram) and \ref{fig:phi-n-corr} (the vertex 
corrections). Here $Y$ denotes the 
$\Lambda$, $\Sigma$ and $\Sigma (1385)$ ($\Sigma ^{\ast}$) hyperons.

First, we consider the box diagram (Fig.~\ref{fig:phi-n}) with $\Lambda$ and 
$\Sigma$ propagation, which is evaluated as, 
\begin{align}
& - i T_{\phi n \to \phi n} ^{\Lambda , \, \Sigma}
\nonumber \\
& = \sum _{\Lambda , \, \Sigma} 
\int \frac{d^{4} q}{(2 \pi )^{4}} 
\frac{i}{q^{2} - m_{K}^{2} + i \epsilon} 
\frac{i}{(p_{\phi} - q)^{2} - m_{\bar{K}}^{2}}  
\nonumber \\
& \phantom{=} 
\times 
\frac{i}{(p_{\phi}^{\prime} - q)^{2} - m_{\bar{K}}^{2}}  
[- i g_{\phi} \epsilon _{\mu}^{\ast} (\phi ) (2 q - p_{\phi})^{\mu} ] 
\nonumber \\
& \phantom{=} 
\times 
\tilde{V} \vec{\sigma} \cdot (\vec{q} - \vec{p}_{\phi}) 
\frac{i M_{Y} / E_{Y}(|\vec{p}_{n}^{\; \prime} + \vec{p}_{\phi}^{\; \prime} - \vec{q}|)}
{p_{n}^{\prime 0} + p_{\phi}^{\prime 0} - q^{0} - 
E_{Y}(|\vec{p}_{n}^{\; \prime} + \vec{p}_{\phi}^{\; \prime} - \vec{q}| ) 
+ i \epsilon}
\nonumber \\
& \phantom{=} 
\times 
[- i g_{\phi} \epsilon _{\nu} (\phi ) (2 q - p_{\phi}^{\prime})^{\nu} ] 
\tilde{V} \vec{\sigma} \cdot (\vec{p}_{\phi}^{\; \prime} - \vec{q}) , 
\label{eq:Amp-Yprop}
\end{align}
where $M_{Y}$ is the mass of the hyperon propagating in the intermediate 
state, and $E_{Y}(p)=\sqrt{M_{Y}^{2}+p^{2}}$. The summation symbol with 
subscripts $\Lambda$ and $\Sigma$ represents the sum of the 
contribution from $\Lambda$ and $\Sigma$ propagation in the intermediate 
states. These channels are explicitly given in~Tab.~\ref{tab:phi-n}. 
The $\phi K \bar{K}$ coupling constant is denoted by $g_{\phi}$ and is fixed as 
$g_{\phi}=4.57$ so as to reproduce the decay width for 
$\phi \to K \bar{K}$. In Eq.~(\ref{eq:Amp-Yprop}) $\vec{\sigma}$ are Pauli 
matrices for baryon spin. The quantity $\tilde{V}$ is the coefficient for the 
meson-baryon-baryon coupling fixed by the flavor $SU (3)$ symmetry as, 
\be
\tilde{V} = \alpha \frac{D + F}{2 f} + \beta \frac{D - F}{2 f} , 
\ee
with empirical parameters $D+F=1.26$, $D-F=0.33$, and $f=1.15 f_{\pi}$ with the 
pion decay constant $f_{\pi}=93.0\, \text{MeV}$. The magnutides $\alpha$ and 
$\beta$ correspond to $SU (3)$ Clebsch-Gordan coefficients and are shown in 
Tab.~\ref{tab:phi-n}. 

In the double scattering diagram of Fig.~\ref{fig:gamma-d-each},
$\vec{p}_{\phi}^{\; \prime}=\vec{q}_{\text{ex}}$. If we consider $\phi$
production forward, as in the experiment, $\vec{p}_{\phi}$ and
$\vec{p}_{\phi}^{\; \prime}$ will be forward and $\vec{p}_{n}$ and
$\vec{p}_{n}^{\; \prime}$ will be small.  Therefore, we can
approximate the amplitude substituting $\vec{q}_{\text{ex}}$ by
$\vec{p}_{\phi}$, in which case $\vec{p}_{n}\simeq \vec{p}_{n}^{\;
  \prime}\simeq \vec{0}$. We also discussed in
Sect.~\ref{subsubsec:amp2} that we were interested in the case where
the polarization of the initial and final $\phi$ were the same, to
optimize the interference of $T^{\text{ds}}$ with
$T^{\text{ss}}$. Hence, we can take the average amplitude over the
$\phi$ polarizations for which we use,
\begin{align}
\overline{\sum _{\lambda _{\phi}}} 
\epsilon _{\mu}^{\ast} (\phi ) (2 q - p_{\phi})^{\mu} 
\epsilon _{\nu} (\phi ) (2 q - p_{\phi})^{\nu}  
= \frac{4}{3} \left [ \frac{(p_{\phi} \cdot q)^{2}}{M_{\phi}^{2}} - q^{2} \right ] . 
\end{align}
Note that the approximation done also forces the neutrons to have the same 
polarization since, 
\be
[ (\vec{q} - \vec{p}_{\phi}) \cdot \vec{\sigma} ]
[ (\vec{p}_{\phi} - \vec{q}) \cdot \vec{\sigma} ] = 
- |\vec{q} - \vec{p}_{\phi}|^{2} , 
\label{eq:Pauli-sigma}
\ee
and there is no spin flip term.  
The independence of this amplitude on the spin of the second particle
is of relevance to our approach.  Indeed, we have used a $\phi$
photoproduction amplitude on the first nucleon which is independent of
the spin of the nucleon.  If we had a spin dependent amplitude the
spin structure of the deuteron should in principle be taken into
account.  However, if there is only spin dependence on the first
nucleon, the sum over initial and final polarization for the case of
initial nucleons uncorrelated by spin, as we have assumed, or
correlated in a spin $=1$ (or zero) state as in the deuteron, give the
same result.

The imaginary part of $T_{\phi n \to \phi n}^{\Lambda , \, \Sigma}$ in 
Eq.~(\ref{eq:Amp-Yprop}) is readily evaluated using 
the Cutkosky rules suited to our normalization as in~\cite{Carrasco:1991mb}, 
\be
T_{\phi n \to \phi n} ^{\Lambda , \, \Sigma} \to 
2 i \text{Im} T_{\phi n \to \phi n} ^{\Lambda , \, \Sigma} , \quad 
G (q) \to 2 i \theta (q^{0}) \text{Im} G (q) , 
\ee
with $G$ the propagators of the particles which are placed on the mass-shell, 
in the 
present case, the $K$ and $Y$ of Fig.~\ref{fig:phi-n}. Hence, we obtain, 
\begin{align}
\text{Im} T_{\phi n \to \phi n} ^{\Lambda , \, \Sigma} 
& = - \frac{1}{2} \sum _{\Lambda , \, \Sigma} \int \frac{d^{3} q}{(2 \pi )^{3}} 
\frac{1}{2 \omega (|\vec{q}|)} 
\frac{M_{Y}}{E_{Y}(|\vec{p}_{\phi} + \vec{p}_{n} - \vec{q}|)} 
\nonumber \\
\times | \mathcal{T}_{1} |^{2}
\times 
& 2 \pi \delta (p_{\phi}^{0} + p_{n}^{0} 
- E_{Y}(|\vec{p}_{\phi} + \vec{p}_{n} - \vec{q}|) - \omega (|\vec{q}|) )
\label{eq:ImTphin1}
\end{align}
with $\omega (q) = \sqrt{m_{K}^{2}+q^{2}}$ and, 
\begin{align}
| \mathcal{T}_{1} |^{2} 
& = \frac{4}{3} 
\left ( \frac{g_{\phi} \tilde{V} |\vec{q} - \vec{p}_{\phi} |}
{M_{\phi}^{2} - 2 p_{\phi} \cdot q}  \right )^{2} \times 
\left [ \frac{(p_{\phi} \cdot q)^{2}}{M_{\phi}^{2}} - m_{K}^{2} \right ] , 
\label{eq:AmpSQphin}
\end{align}
where we have used $q^{2}=m_{K}^{2}=m_{\bar{K}}^{2}$. 

We can see that Eq.~(\ref{eq:ImTphin1}) is the phase-space integral for the 
cross section of $\phi n \to K Y$ with the transition amplitude 
$\mathcal{T}_{1}$ up to a normalization. Performing explicitly the 
$|\vec{q}|$ integration in the $\phi$-$n$ center-of-mass frame we obtain, 
\begin{align}
\text{Im} \, T_{\phi n \to \phi n} ^{\Lambda , \, \Sigma}
& = - \sum _{\Lambda , \, \Sigma} \frac{|\vec{q}| M_{Y}}{8 \pi M_{\phi n}} 
\int _{-1}^{1} d \cos \theta _{K} | \mathcal{T}_{1} |^{2} , 
\label{eq:ImT-final}
\end{align}
with $M_{\phi n} = \sqrt{(p_{\phi} + p_{n})^{2}}$ and $\theta _{K}$ the 
scattering angle between the $\phi$ and the $K$. 
Eq.~(\ref{eq:ImT-final}) is nothing but the expression of the optical theorem 
for the reaction mechanism of Fig.~\ref{fig:phi-n}. 

Next we consider the vertex corrections diagrams of the $\phi n \to \phi n$ 
amplitude shown in Fig.~\ref{fig:phi-n-corr} with the $\Lambda$ and $\Sigma$ 
propagation~\cite{Oset:2000eg}. Following~\cite{Cabrera:2003wb}, 
we obtain for the sum of the three diagrams the same expression for 
$\Ima T_{\phi n \to \phi n}^{\Lambda , \, \Sigma}$ in Eq.~(\ref{eq:ImT-final}), 
substituting $|\mathcal{T}_{1} |^{2}$ by $|\mathcal{T}_{2} |^{2}$ given 
by\footnote{We found that one should use 
$(1 + |\vec{p}_{\phi} |^{2} / 3 M_{\phi}^{2})$ instead of 
$(1 + |\vec{q} |^{2} / 3 M_{\phi}^{2})$ given as the last term of Eq.~(11) 
in~\cite{Cabrera:2003wb}. }, 
\begin{align}
& | \mathcal{T}_{2} |^{2} 
= (g_{\phi} \tilde{V} )^{2} \nonumber \\ 
& \times \Bigg [ 
\frac{4}{3} \frac{1}{{M_{\phi}^{2} - 2 p_{\phi} \cdot q}} 
\left ( |\vec{q}|^{2} - \vec{p}_{\phi} \cdot \vec{q} + 
\frac{p_{\phi} \cdot q}{M_{\phi}^{2}} (|\vec{p}_{\phi}|^{2} 
- \vec{p}_{\phi} \cdot \vec{q}) 
\right ) 
\nonumber \\
& \phantom{\times \Bigg [ ~} 
+ \left ( 1 + \frac{|\vec{p}_{\phi} |^{2}}{3 M_{\phi}^{2}} \right )
\Bigg ] . 
\label{eq:AmpSQcorr}
\end{align}
For the contributions with the $\Lambda$ and $\Sigma$ propagation in 
Figs.~\ref{fig:phi-n} and \ref{fig:phi-n-corr}, summing 
Eqs.~(\ref{eq:AmpSQphin}) and (\ref{eq:AmpSQcorr}), we obtain, 
\be
\text{Im} \, T_{\phi n \to \phi n} ^{\Lambda , \, \Sigma} 
= - \sum _{\Lambda , \, \Sigma} \frac{|\vec{q}| M_{Y}}{8 \pi M_{\phi n}} 
\int _{-1}^{1} d \cos \theta _{K} 
(| \mathcal{T}_{1} |^{2} + | \mathcal{T}_{2} |^{2}) . 
\label{eq:phi-n-LS}
\ee

Now let us consider $\Sigma ^{\ast}$ propagation. 
The amplitude with the $\Sigma^{\ast}$ propagation can be obtained by 
replacement of the coupling constant, spin matrix and mass 
in the amplitudes obtained for the $\Lambda$ and $\Sigma$ propagations. 
The coupling constant for $\Sigma^{\ast}$ is obtained by an $SU(6)$ quark model
and $SU(3)$ flavor symmetry as, 
\be
\tilde{V} \to \tilde{A} = \frac{2 \sqrt{6}}{5} \frac{D + F}{2 f} A , 
\ee
with the Clebsch-Gordan coefficient $A$ given in Tab.~\ref{tab:phi-n}. 
The spin operators in the case of the $\Sigma^{\ast}$ are the transition 
matrices of spin $1/2$ to $3/2$, $\vec{S}^{\dagger}$, which 
should be used instead of the Pauli matrices $\vec{\sigma}$, 
and they satisfy the relation, 
\be
S^{i} S^{\dagger j} = \frac{2}{3} \delta ^{ij} 
- \frac{i}{3} \epsilon _{ijk} \sigma ^{k} , 
\ee
This gives an extra factor $2/3$ since the spin flip part vanishes in the 
equivalent term of Eq.~(\ref{eq:Pauli-sigma}). Finally, we obtain 
for the imaginary part of the $\phi n \to \phi n$ amplitude for the $\Sigma^{\ast}$ propagation  
\be
\text{Im} \, T_{\phi n \to \phi n} ^{\Sigma ^{\ast}}
= - \sum _{\Sigma ^{\ast}} \frac{|\vec{q}| M_{Y}}{12 \pi M_{\phi n}} 
\int _{-1}^{1} d \cos \theta _{K} (|\mathcal{U}_{1}|^{2} + |\mathcal{U}_{2}|^{2}) , 
\label{eq:phi-n-SS}
\ee
with, 
\begin{align}
& |\mathcal{U}_{1}|^{2} 
= \frac{4}{3} 
\left ( \frac{g_{\phi} \tilde{A} |\vec{q} - \vec{p}_{\phi} |}
{M_{\phi}^{2} - 2 p_{\phi} \cdot q}  \right )^{2} \times 
\left [ \frac{(p_{\phi} \cdot q)^{2}}{M_{\phi}^{2}} - m_{K}^{2} \right ] , 
\label{eq:AmpSQ-S}
\\ 
& |\mathcal{U}_{2}|^{2} = (g_{\phi} \tilde{A} )^{2} \nonumber \\ 
& \times \Bigg [ 
\frac{4}{3} \frac{1}{{M_{\phi}^{2} - 2 p_{\phi} \cdot q}} 
\left ( |\vec{q}|^{2} - \vec{p}_{\phi} \cdot \vec{q} + 
\frac{p_{\phi} \cdot q}{M_{\phi}^{2}} (|\vec{p}_{\phi}|^{2} 
- \vec{p}_{\phi} \cdot \vec{q}) 
\right ) 
\nonumber \\
& \phantom{\times \Bigg [ ~} 
+ \left ( 1 + \frac{|\vec{p}_{\phi} |^{2}}{3 M_{\phi}^{2}} \right )
\Bigg ] . 
\label{eq:AmpSQcorr-S}
\end{align}

\begin{figure}[!t]
  \centering
    \includegraphics[width=8.6cm]{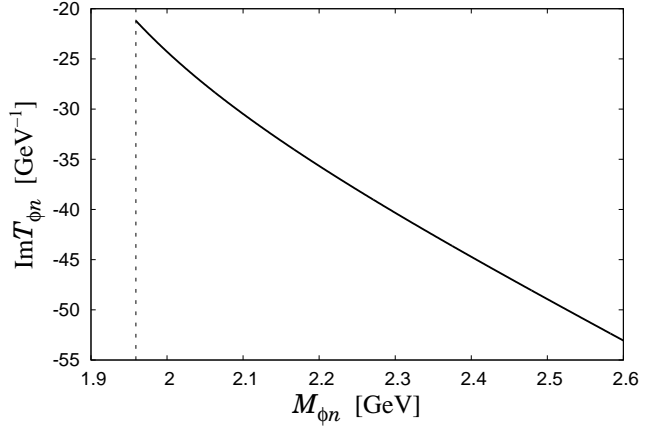}
    \caption{$\text{Im} \, T_{\phi n \to \phi n}$ as a function of
      $M_{\phi n}$.  Vertical dashed line represents the $\phi$-$n$
      threshold. }
  \label{fig:Im_phin}
\end{figure}

As a consequence, the final form for the imaginary part of the 
$\phi n \to \phi n$ amplitude can be written as 
\be
\text{Im} \, T_{\phi n \to \phi n} (M_{\phi n} )
= \text{Im} \, T_{\phi n \to \phi n} ^{\Lambda , \, \Sigma}
+ \text{Im} \, T_{\phi n \to \phi n} ^{\Sigma ^{\ast}} , 
\ee
which can be readily evaluated using Eqs.~(\ref{eq:phi-n-LS}) and
(\ref{eq:phi-n-SS}).
In Fig.~\ref{fig:Im_phin} we show the results that we get for
$\text{Im} \, T_{\phi n \to \phi n}$ as a function of $M_{\phi n}$.
In order to get a feeling for the results in Fig.~\ref{fig:Im_phin},
rather than calculating an inelastic cross section which goes to
infinity at the $\phi$-$n$ threshold, we give the $\phi$ width in a
nuclear medium $\Gamma = - \text{Im}\, T_{\phi n \to \phi n} \rho /
\omega _{\phi}$, with $\rho$ the nuclear matter density.  For normal
nuclear matter density, $\rho _{0} = 0.17 \fm ^{-3}$, this gives
$\Gamma \simeq 27 \mev$ at threshold, in agreement with the results
obtained in Refs.~\cite{Torres:2010jh,Oset:2000eg}.


\begin{thebibliography}{99}


%\cite{Ballam:1972eq}
\bibitem{Ballam:1972eq}
  J.~Ballam {\it et al.},
  %``Vector Meson Production By Polarized Photons At 2.8-Gev, 4.7-Gev, And
  %9.3-Gev,''
  Phys.\ Rev.\  D {\bf 7} (1973) 3150.
  %%CITATION = PHRVA,D7,3150;%%
  
 %\cite{Besch:1974rp}
\bibitem{Besch:1974rp}
  H.~J.~Besch, G.~Hartmann, R.~Kose, F.~Krautschneider, W.~Paul and U.~Trinks,
  %``Photoproduction Of Phi Mesons On Protons At 2.0 Gev,''
  Nucl.\ Phys.\  B {\bf 70} (1974) 257.
  %%CITATION = NUPHA,B70,257;%%
  
  
 %\cite{Behrend:1978ik}
\bibitem{Behrend:1978ik}
  H.~J.~Behrend {\it et al.},
  %``Elastic And Inelastic Phi - Photoproduction,''
  Nucl.\ Phys.\  B {\bf 144} (1978) 22.
  %%CITATION = NUPHA,B144,22;%%
  
  %\cite{Barber:1981fj}
\bibitem{Barber:1981fj}
  D.~P.~Barber {\it et al.},
  %``A STUDY OF ELASTIC PHOTOPRODUCTION OF LOW MASS $K^{+} K^{-}$ PAIRS FROM
  %HYDROGEN IN THE ENERGY RANGE 2.8-GeV TO 4.8-GeV,''
  Z.\ Phys.\  C {\bf 12} (1982) 1.
  %%CITATION = ZEPYA,C12,1;%%

%\cite{Anciant:2000az}
\bibitem{Anciant:2000az}
  E.~Anciant {\it et al.}  [CLAS Collaboration],
  %``Photoproduction of Phi(1020) mesons on the proton at large momentum
  %transfer,''
  Phys.\ Rev.\ Lett.\  {\bf 85} (2000) 4682
  [arXiv:hep-ex/0006022].
  %%CITATION = PRLTA,85,4682;%%
  
  
%\cite{Barth:2003bq}
\bibitem{Barth:2003bq}
  J.~Barth {\it et al.},
  %``Low-energy photoproduction of Phi mesons,''
  Eur.\ Phys.\ J.\  A {\bf 17} (2003) 269.
  %%CITATION = EPHJA,A17,269;%%


%\cite{Bauer:1977iq}
\bibitem{Bauer:1977iq}
  T.~H.~Bauer, R.~D.~Spital, D.~R.~Yennie and F.~M.~Pipkin,
  %``The Hadronic Properties of the Photon in High-Energy Interactions,''
  Rev.\ Mod.\ Phys.\  {\bf 50} (1978) 261
  [Erratum-ibid.\  {\bf 51} (1979) 407].
  %%CITATION = RMPHA,50,261;%%

%\cite{Titov:1996bg}
\bibitem{Titov:1996bg}
  A.~I.~Titov, S.~N.~Yang and Y.~s.~Oh,
  %``Electroproduction of a Phi meson from the proton and the strangeness  in
  %the nucleon,''
  Nucl.\ Phys.\  A {\bf 618} (1997) 259
  [arXiv:nucl-th/9612059].
  %%CITATION = NUPHA,A618,259;%%
  
  %\cite{Titov:1997qz}
\bibitem{Titov:1997qz}
  A.~I.~Titov, Y.~s.~Oh and S.~N.~Yang,
  %``Polarization observables in Phi meson photoproduction and the  strangeness
  %content of the proton,''
  Phys.\ Rev.\ Lett.\  {\bf 79} (1997) 1634
  [arXiv:nucl-th/9702015].
  %%CITATION = PRLTA,79,1634;%%
  
  %\cite{Titov:1998bw}
\bibitem{Titov:1998bw}
  A.~I.~Titov, Y.~s.~Oh, S.~N.~Yang and T.~Morii,
  %``Photoproduction of Phi meson from proton: Polarization observables and  the
  %strangeness in the nucleon,''
  Phys.\ Rev.\  C {\bf 58} (1998) 2429
  [arXiv:nucl-th/9804043].
  %%CITATION = PHRVA,C58,2429;%%
  
 %\cite{Titov:1998tx}
\bibitem{Titov:1998tx}
  A.~I.~Titov, T.~S.~H.~Lee and H.~Toki,
  %``An isotopic effect in Phi photoproduction at a few GeV,''
  Phys.\ Rev.\  C {\bf 59} (1999) 2993
  [arXiv:nucl-th/9812074].
  %%CITATION = PHRVA,C59,2993;%%
  
  %\cite{Titov:1999eu}
\bibitem{Titov:1999eu}
  A.~I.~Titov, T.~S.~Lee, H.~Toki and O.~Streltsova,
  %``Structure of the fgr photoproduction amplitude at a few GeV,''
  Phys.\ Rev.\  C {\bf 60} (1999) 035205.
  %%CITATION = PHRVA,C60,035205;%%
  
%\cite{Kisslinger:1999jk}
\bibitem{Kisslinger:1999jk}
  L.~S.~Kisslinger and W.~h.~Ma,
  %``Pomeron and reggeized glueball/sigma,''
  Phys.\ Lett.\  B {\bf 485} (2000) 367
  [arXiv:hep-ph/9905479].
  %%CITATION = PHLTA,B485,367;%%

  %\cite{LlanesEstrada:2000jw}
\bibitem{felipe}
  F.~J.~Llanes-Estrada, S.~R.~Cotanch, P.~J.~de A. Bicudo, J.~E.~F.~Ribeiro and A.~P.~Szczepaniak,
  %``QCD glueball Regge trajectories and the Pomeron,''
  Nucl.\ Phys.\  A {\bf 710} (2002) 45
  [arXiv:hep-ph/0008212].
  %%CITATION = NUPHA,A710,45;%%
  
%\cite{Mibe:2005er}
\bibitem{Mibe:2005er}
  T.~Mibe {\it et al.}  [LEPS Collaboration],
  %``Diffractive Phi-meson photoproduction on proton near threshold,''
  Phys.\ Rev.\ Lett.\  {\bf 95} (2005) 182001
  [arXiv:nucl-ex/0506015].
  %%CITATION = PRLTA,95,182001;%%
 
%\cite{Titov:2008zz}
\bibitem{Titov:2008zz}
  A.~I.~Titov, T.~Nakano, S.~Date and Y.~Ohashi,
  %``Phi-meson photoproduction at threshold and phi - N scattering length,''
  Mod.\ Phys.\ Lett.\  A {\bf 23} (2008) 2301.
  %%CITATION = MPLAE,A23,2301;%%
  
%\cite{Chang:2009yq}
\bibitem{Chang:2009yq}
  W.~C.~Chang {\it et al.}  [LEPS Collaboration],
  %``Reduction of the incoherent $\gamma d \to \phi p n$ photoproduction near
  %threshold,''
  Phys.\ Lett.\  B {\bf 684} (2010) 6
  [arXiv:0907.1705 [nucl-ex]].
  %%CITATION = PHLTA,B684,6;%%


%\cite{Titov:2007fc}
\bibitem{Titov:2007fc}
  A.~I.~Titov and B.~Kampfer,
  %``Photoproduction of Phi meson off deuteron near threshold,''
  Phys.\ Rev.\  C {\bf 76} (2007) 035202
  [arXiv:0705.2010 [nucl-th]].
  %%CITATION = PHRVA,C76,035202;%%
  


%\cite{Ishikawa:2004id}
\bibitem{Ishikawa:2004id}
  T.~Ishikawa {\it et al.},
  %``Phi photo-production from Li, C, Al, and Cu nuclei at E(gamma) =  1.5-GeV -
  %2.4-GeV,''
  Phys.\ Lett.\  B {\bf 608} (2005) 215
  [arXiv:nucl-ex/0411016].
  %%CITATION = PHLTA,B608,215;%%

%\cite{Hernandez:1992rv}
\bibitem{Hernandez:1992rv}
  E.~Hernandez and E.~Oset,
  %``Many body modes of anti-proton annihilation tested with anti-proton
  %production,''
  Z.\ Phys.\  A {\bf 341} (1992) 201.
  %%CITATION = ZEPYA,A341,201;%%

  %\cite{Rapp:1999ej}
\bibitem{Rapp:1999ej}
  R.~Rapp and J.~Wambach,
  %``Chiral symmetry restoration and dileptons in relativistic heavy-ion
  %collisions,''
  Adv.\ Nucl.\ Phys.\  {\bf 25} (2000) 1
  [arXiv:hep-ph/9909229].
  %%CITATION = ANUPB,25,1;%%

%\cite{Hayano:2008vn}
\bibitem{Hayano:2008vn}
  R.~S.~Hayano and T.~Hatsuda,
  %``Hadron properties in the nuclear medium,''
  Rev.\ Mod.\ Phys.\ \ {\bf 82} (2010) 2949
  [arXiv:0812.1702 [nucl-ex]].
  %%CITATION = RMPHA,82,2949;%%

  
%\cite{Magas:2004eb}
\bibitem{Magas:2004eb}
  V.~K.~Magas, L.~Roca and E.~Oset,
  %``The Phi meson width in the medium from proton induced Phi production in
  %nuclei,''
  Phys.\ Rev.\  C {\bf 71} (2005) 065202
  [arXiv:nucl-th/0403067].
  %%CITATION = PHRVA,C71,065202;%%

%\cite{arXiv:1011.1305}
\bibitem{haiyan} 
  X.~Qian, W.~Chen, H.~Gao, K.~Hicks, K.~Kramer, J.~M.~Laget, T.~Mibe and Y.~Qiang {\it et al.},
  % ``Near-threshold Photoproduction of Phi Mesons from Deuterium,''
  Phys.\ Lett.\ B\ {\bf 696} (2011) 338 
  [arXiv:1011.1305 [nucl-ex]].
  %% CITATION = PHLTA,B696,338;%%

%\cite{Rogers:2005bt}
\bibitem{Rogers:2005bt}
  T.~C.~Rogers, M.~M.~Sargsian and M.~I.~Strikman,
  %``Coherent Vector Meson Photo-Production from Deuterium at Intermediate
  %Energies,''
  Phys.\ Rev.\  C {\bf 73} (2006) 045202
  [arXiv:hep-ph/0509101].
  %%CITATION = PHRVA,C73,045202;%%

%\cite{Amsler:2008zzb}
\bibitem{Amsler:2008zzb}
  C.~Amsler {\it et al.}  [Particle Data Group],
  %``Review of particle physics,''
  Phys.\ Lett.\  B {\bf 667} (2008) 1.
  %%CITATION = PHLTA,B667,1;%%

%\cite{Jido:2009jf}
\bibitem{Jido:2009jf}
  D.~Jido, E.~Oset and T.~Sekihara,
  %``Kaonic production of Lambda(1405) off deuteron target in chiral dynamics,''
  Eur.\ Phys.\ J.\  A {\bf 42} (2009) 257. 
  %[arXiv:0904.3410 [nucl-th]].
  %%CITATION = EPHJA,A42,257;%%

%\cite{Lacombe:1981eg}
\bibitem{Lacombe:1981eg}
  M.~Lacombe, B.~Loiseau, R.~Vinh Mau, J.~Cote, P.~Pires and R.~de Tourreil,
  %``Parametrization of the deuteron wave function of the Paris n-n potential,''
  Phys.\ Lett.\  B {\bf 101} (1981) 139.
  %%CITATION = PHLTA,B101,139;%%

%\cite{Machleidt:2000ge}
\bibitem{Machleidt:2000ge}
  R.~Machleidt,
  %``The high-precision, charge-dependent Bonn nucleon-nucleon potential
  %(CD-Bonn),''
  Phys.\ Rev.\  C {\bf 63} (2001) 024001. 
  %[arXiv:nucl-th/0006014].
  %%CITATION = PHRVA,C63,024001;%%

\bibitem{NNonline}
  NN-OnLine, 
  http://nn-online.org/

%\cite{Miyabe:2010}
\bibitem{Miyabe:2010}
  M.~Miyabe, 
  %``Private communication,''
  Doctor thesis (2010); in private communication. 


%\cite{Nakano:2008ee}
\bibitem{Nakano:2008ee}
  T.~Nakano {\it et al.}  [LEPS Collaboration],
  %``Evidence of the \Theta^+ in the \gamma d \to K^+K^-pn reaction,''
  Phys.\ Rev.\  C {\bf 79} (2009) 025210. 
  %[arXiv:0812.1035 [nucl-ex]].
  %%CITATION = PHRVA,C79,025210;%%

%\cite{Torres:2010jh}
\bibitem{Torres:2010jh}
  A.~M.~Torres and E.~Oset,
  %``Study of the $\gamma d\to K^{+}K^{-}np$ reaction and an alternative
  %explanation for the '$\Theta^{+}(1540)$ pentaquark' peak,''
  Phys.\ Rev.\  C {\bf 81} (2010) 055202
  [arXiv:1003.1098 [nucl-th]].
  %%CITATION = PHRVA,C81,055202;%%

%\cite{Oset:2000eg}
\bibitem{Oset:2000eg}
  E.~Oset and A.~Ramos,
  %``Phi decay in nuclei,''
  Nucl.\ Phys.\  A {\bf 679} (2001) 616. 
  %[arXiv:nucl-th/0005046].
  %%CITATION = NUPHA,A679,616;%%

%\cite{Cabrera:2002hc}
\bibitem{Cabrera:2002hc}
  D.~Cabrera and M.~J.~Vicente Vacas,
  %``Phi meson mass and decay width in nuclear matter,''
  Phys.\ Rev.\  C {\bf 67} (2003) 045203. 
  %[arXiv:nucl-th/0205075].
  %%CITATION = PHRVA,C67,045203;%%

%\cite{Carrasco:1991mb}
\bibitem{Carrasco:1991mb}
  R.~C.~Carrasco, E.~Oset and L.~L.~Salcedo,
  %``Inclusive (gamma, pi) reactions in nuclei,''
  Nucl.\ Phys.\  A {\bf 541} (1992) 585.
  %%CITATION = NUPHA,A541,585;%%

%\cite{Cabrera:2003wb}
\bibitem{Cabrera:2003wb}
  D.~Cabrera, L.~Roca, E.~Oset, H.~Toki and M.~J.~Vicente Vacas,
  %``Mass dependence of inclusive nuclear Phi photoproduction,''
  Nucl.\ Phys.\  A {\bf 733} (2004) 130. 
  %[arXiv:nucl-th/0310054].
  %%CITATION = NUPHA,A733,130;%%

%\cite{Wood:2010ei}
\bibitem{Wood:2010ei}
  M.~H.~Wood {\it et al.}  [CLAS Collaboration],
  %``Absorption of the $\omega$ and $\phi$ Mesons in Nuclei,''
  Phys.\ Rev.\ Lett.\  {\bf 105} (2010) 112301
  [arXiv:1006.3361 [nucl-ex]].
  %%CITATION = PRLTA,105,112301;%%

%\cite{Polyanskiy:2010tj}
\bibitem{Polyanskiy:2010tj}
  A.~Polyanskiy, M.~Hartmann, Y.~.T.~Kiselev, E.~Y.~.Paryev, M.~Buscher, D.~Chiladze, S.~Dymov and A.~Dzyuba {\it et al.},
  %``Measurement of the in-medium phi-meson width in proton-nucleus collisions,''
  Phys.\ Lett.\ B\ {\bf 695} (2011) 74
  [arXiv:1008.0232 [nucl-ex]].
  %%CITATION = PHLTA,B695,74;%%

\end{thebibliography}
\end{document}